\documentclass[lettersize,journal]{IEEEtran}

\usepackage{graphicx,xcolor}
\usepackage{amsmath}
\usepackage{newtxtext}
\usepackage[varg]{newtxmath}
\usepackage{enumerate}
\usepackage{framed}  
\usepackage{makecell}
\usepackage{url}    
\usepackage{booktabs}
\usepackage{enumerate}
\usepackage{bbding}
\usepackage{pifont}
\usepackage{subcaption}
\usepackage{algorithmic}
\usepackage[ruled,linesnumbered]{algorithm2e}
\usepackage{CJKutf8}
\usepackage{multirow}
\usepackage{subfiles}

\definecolor{blue}{RGB}{0, 111, 184}
\definecolor{brown}{RGB}{112, 38, 1}
\definecolor{zi}{RGB}{73,59,148}
\newcommand{\brown}[1]{\textcolor{brown}{#1}}
\newcommand{\purple}[1]{\textcolor{zi}{#1}}
\newcommand{\blue}[1]{\textcolor{blue}{#1}}

\SetKwInput{KwInput}{Input}                
\SetKwInput{KwOutput}{Output}              

\usepackage[T1]{fontenc}

\begin{document}
	
	\title{Voltran: Unlocking Trust and Confidentiality in Decentralized Federated Learning Aggregation}

	\author{Hao~Wang,
		\ Yichen~Cai,
		\ Jun Wang,
		\ Chuan~Ma,
		\ Chunpeng~Ge{*},
		\ Xiangmou~Qu,
		\ and Lu~Zhou
}

%

\maketitle

\begin{abstract}
	The decentralized Federated Learning (FL) paradigm built upon blockchain architectures leverages distributed node clusters to replace the single server for executing FL model aggregation. This paradigm tackles the vulnerability of the centralized malicious server in vanilla FL and inherits the trustfulness and robustness offered by blockchain. However, existing blockchain-enabled schemes face challenges related to inadequate confidentiality on models and limited computational resources of blockchains to perform large-scale FL computations. In this paper, we present Voltran, an innovative hybrid platform designed to achieve trust, confidentiality, and robustness for FL based on the combination of the Trusted Execution Environment (TEE) and blockchain technology. We offload the FL aggregation computation into TEE to provide an isolated, trusted and customizable off-chain execution, and then guarantee the authenticity and verifiability of aggregation results on the blockchain. Moreover, we provide strong scalability on multiple FL scenarios by introducing a multi-SGX parallel execution strategy to amortize the large-scale FL workload. We implement a prototype of Voltran and conduct a comprehensive performance evaluation. Extensive experimental results demonstrate that Voltran incurs minimal additional overhead while guaranteeing trust, confidentiality, and authenticity, and it significantly brings a significant speed-up compared to state-of-the-art ciphertext aggregation schemes.
\end{abstract}


\section{Introduction}
\IEEEPARstart{F}{ederated Learning} (FL) is a decentralized machine learning methodology that involves training deep neural networks on individual devices to produce local models, which are then aggregated to construct a global model on a central server. Its ability to keep clients' individual data locally has garnered significant attention and is extensively utilized in multiple data-sensitive domains, such as Google Keyboard \cite{DaanvanEsch2019WritingAT}, e-healthcare \cite{nguyen2022federated}, and economic applications \cite{QiangYang2019FederatedML}.

However, current FL paradigms rely on a centralized server to execute model aggregation. This paradigm renders them vulnerable to malfunction or active attacks that may incur interruption or termination of tasks \cite{li2020blockchain, lalitha2018fully}. Moreover, a centralized server has complete dominance over the aggregation process, thus potentially being able to tamper with data or bias model training results maliciously. To mitigate these issues brought by the centralized server, the utilization of blockchain technology helps establish a decentralized framework for FL by its inherent distributed architecture. Specifically, smart contracts supported by the blockchain can be leveraged to implement secure aggregation algorithms to prevent malicious FL attacks such as poisoning. Moreover, the financial attribute of blockchains can handle the lack of incentives for the FL clients for a fair billing marketplace.

Despite the considerable alignment between blockchain and FL, there are two main deficiencies needed to be addressed: a). \textit{confidentiality}, adversaries can retrieve sensitive information from the model parameters (e.g., gradient information can be used to infer the private clients' training data). Existing blockchain-enabled schemes do not perform well in maintaining confidentiality due to the publicly visible nature of blockchain \cite{ChuanMa2020WhenFL,wang2021blockchain,qu2022blockchain}. These solutions either directly carry out plaintext model transmission or use differential privacy (DP) to inject noise into models, which leads to a dilemma between accuracy and privacy; b). \textit{practicality}, the blockchain takes the form of full-repetitive computations on all nodes to ensure reliability, which causes little computational power and storage capacity to handle large-scale computing tasks. FL aggregation is a computation-intensive task with large-scale data, intricate calculations, and high-efficiency requirements. Low aggregation efficiency will affect model convergence, thereby impacting the real-time demand and accuracy of models.  Therefore, it is impractical for FL to perform direct on-chain computations due to its high computational costs and low throughput. Overall, it is necessary to design a solution that can solve these two problems for blockchain-based FL.

Fortunately, the Trusted Execution Environment (TEE), such as Intel Software Guard eXtensions (SGX), can be applied to complement blockchain-based FL systems. It provides an isolated region to guarantee confidentiality and integrity of codes and data running in it. Hence, blockchain nodes can execute the FL aggregation process within TEEs to address confidentiality and trustfulness concerns. Local models are securely transmitted into TEE based on cryptographic primitives for off-chain aggregation execution, and then results are uploaded on the blockchain after verification for storage. Meanwhile, by offloading aggregation computation into TEE, computational workloads for blockchain nodes are greatly decreased, thus addressing the issue of practicality by low computational power.

\textbf{Challenges.} Building such a hybrid system is not straightforward and requires addressing several challenges: a). \textit{securely hybridization}, which means that it needs to design a comprehensive protocol to ensure the system achieves data confidentiality and computation result correctness and authenticity in FL aggregation computation while maintaining the privacy-protection of clients' raw data in vanilla FL; b). \textit{throughput}, which means that inefficient blockchains may become a performance bottleneck, requiring a shift in the existing blockchain computing paradigm to enhance its throughput to meet FL performance requirements; c). \textit{capacity}, which means large-scale FL models may exceed the capacity of both blockchain and TEE (e.g., Intel SGX 1 is limited to 128 MB), necessitating considerations on how to make the system capable of handling such large volumes of data. We will further elaborate on the specific challenges and countermeasures in Section IV. Furthermore, focusing on FL scenarios, the system needs to be general to adapt to various FL aggregation algorithms. All the considerations above motivate us to propose a high-level blockchain-TEE hybridized system for FL aggregation.

In this work, we propose Voltran, a trusted, decentralized and privacy-preserving FL aggregation platform that enables high integratability on secure aggregation strategies. The core of our idea is a secure and well-structured integration of blockchain and TEE technology with FL's real-world scenarios. This is not trivial work because combining these three parties requires solving complex hybrid processes and technical challenges. Voltran uses blockchain as the underlying architecture that abolishes the vanilla FL paradigm of a single server. Instead, we conduct the aggregation computation on the distributed blockchain nodes by perform smart contracts. Different from the traditional on-chain smart contract execution form, Voltran extends a new form of smart contracts with Intel SGX by offloading the contract execution off chain in TEEs and performing on-chain verification of the correct execution. Based on this paradigm, Voltran can support large-scale computation-intense task such as FL aggregation by implementing aggregation algorithms into our \textit{new-style} contracts (i.e., into TEEs). Furthermore, we consider the limited size of TEE memory and design a multi-SGX parallel execution strategy, placing models from different clients into multiple SGXs for sub-aggregation based on their different weights, to amortize the computation and communication overhead. To our knowledge, we are the first FL aggregation scheme while considering these specific challenges and providing solutions.

Another consequent advantage brought from our combination is that Voltran implements computations directly in plaintext with confidentiality by performing aggregation in TEE, resulting in minimal performance overhead and high scalability. It brings immediate convenience and realizability to defend against aggregation attacks. Previous security aggregation algorithms using cryptographic primitives, such as homomorphic encryption \cite{miao2022privacy} and secure multi-party computing \cite{liu2022efficient}, perform aggregation on encrypted data. These schemes bring massive computation and communication overhead, which makes them challenging to implement in real-world applications. In contrast, Voltran can support plaintext aggregation that achieves multiple customized efficient and secure aggregation schemes based on plaintext \cite {PevaBlanchard2017MachineLW, yin2018byzantine, chen2017distributed}.

\textbf{Contributions.} In summary, we make the following contributions:
\begin{enumerate}
	\item \textit{Platform:} We propose Voltran, an innovative platform for federated learning aggregation. It combines the trusted hardware Intel SGX with blockchain architecture to provide confidentiality and authenticity for FL data, as well as decentralization and robustness against the centralized server. Due to the created isolated region, Voltran can execute FL aggregation in plaintext so that can support the integration of existing secure aggregation algorithms to resist against model attacks.
	
	\item \textit{Implementation:} We carefully consider the practical implementation of our hybrid TEE-blockchain system in FL scenarios and comprehensively address the associated challenges. We propose a secure data transmission mechanism based Intel remote attestation and cryptographic primitives. Moreover, we take into account the memory limitation of SGX and the maximum single transaction capacity of the blockchain and propose a multi-SGX parallel processing strategy to amortize the computation and communication overhead to multiple nodes.

	\item \textit{Evaluation:} To evaluate Voltran's performance, we conduct diverse experiments on diverse FL tasks. We compare our approach to state-of-the-art privacy-preserving FL schemes, and results demonstrate a significant runtime speed-up, e.g., almost 200$\times$ compared to the SMPC-based scheme \cite{liu2022efficient}. Additionally, we evaluate the performance of our multiple SGX execution mode against a single SGX mode in large-scale model tasks, revealing a notable reduction in execution time. Moreover, we demonstrate our framework's scalability by implementing off-the-shelf aggregation schemes and confirm that Voltran can integrate them without any loss in performance.
\end{enumerate}

\textbf{Organization.} The rest of this paper is organized as follows. Section \ref{preliminaries} presents the preliminaries used in this paper and the security goals. Section \ref{overview} describes the overview of our platform, including the system overview, threat model and workflow. Section \ref{blocks} presents the technical details of Voltran on implementation. Section \ref{protocol} introduces the protocol and security analysis of our framework. Section \ref{implement} displays the implementation, including the experimental setup and information of FL tasks. Section \ref{evaluation} presents the evaluation and discussion on our experiments.

\section{Preliminaries}\label{preliminaries}

\subsection{Federated Learning}
In an FL task, there are \textit{n} clients, each of which possesses the private dataset $D_i$, $i = 1, ... ,n$. The machine learning model is trained locally by clients and aggregated iteratively into a joint global model by a centralized server. A task may contain a number of rounds for exchanging models between clients and the server. In a round \textit{r}, the server randomly chooses several participants to join and sends the global model $w_{g}^r$ to them. Then, the client \textit{i} uses its private dataset $D_i$ to train the local model $\overline{w}_i^r$ based on $w_{g}^r$ and send $\overline{w}_i^r$ to the server. The server performs an aggregation algorithm, e.g., $w_{g}^{r+1} = \frac{D_i}{\sum_{i=1}^n D_i}\sum_{i=1}^n \overline{w}_i^r$ according to \textit{FedAvg} \cite{mcmahan2017communication}.

\subsection{TEE and Intel SGX}

Developing applications that emphasize data confidentiality poses numerous challenges. The inadvertent disclosure of sensitive information can occur even with a minor vulnerability in privileged code running on the platform. To address this, the Trusted Execution Environment (TEE) \cite{sabt2015trusted} provides a secure area within the central processor, ensuring the confidentiality and integrity of the code and data loaded into it. GlobalPlatform \cite{GlobalPlatform} gives a definition of TEE: the TEE is an execution environment that runs alongside but is isolated from the device's main operating system. It protects its assets against general software attacks. It can be implemented using multiple technologies, and its level of security varies accordingly. 

Intel's SGX \cite{Intel, FrankMckeen2013InnovativeIA, MatthewEHoekstra2013UsingII} introduce a set of unique instructions that offer hardware-level protections for user-level codes. This empowers software developers with the ability to exert control over the security of sensitive code and data. SGX enables the execution of processes within a protected address space called ``enclave''. Enclaves safeguard confidentiality and integrity by protecting it from specific forms of hardware attacks as well as other software on the same host, including the operating system. The protected memory region is called the Processor Reserved Memory range (PRM). Enclave Page Cache (EPC) paging enables the mapping of trusted pages in PRM to the untrusted memory when memory usage exceeds its limitations. It serves to enhance overall system performance. EPC paging takes more time than common paging because it takes cryptographic operations to protect trusted pages. Most versions of SGXs have 128 MB or 256 MB of PRM \cite{EnclavePageCache}. 

Although data is effectively protected within the isolated space created by SGX, it may still be vulnerable during transmission. The encryption of communication channels can provide data protection, but it cannot guarantee the authenticity of communication parties. Intel provides an advanced capability known as Remote Attestation (RA) \cite{Anati_Gueron_Johnson_Scarlata, gunn2018circumventing}, which is designed to offer enhanced assurance in the integrity and authenticity of an entity to a remote service provider. RA verifies three items: the application's identity, its integrity (that it has not been tampered with), and whether it is running safely within an enclave on an Intel SGX-enabled platform. It also shares the session key between the two parties, thus encrypting the transmitted data to ensure confidentiality and significantly improve trust.

\subsection{Blockchain and Smart Contracts}
Blockchain \cite{nakamoto2008bitcoin, wood2014ethereum} is a technique as a ledger maintained by distributed nodes. The ledger takes the form of a chain data structure consisting of chronologically ordered blocks containing transactions sent by users and other information that guarantees security. Due to its distributed consensus and cryptography-based data structures, such as Merkle Tree of Bitcoin \cite{nakamoto2008bitcoin}, data stored on the blockchain can be trusted and tamper-proof. Smart contracts are a secure and decentralized computing paradigm provided by the blockchain. It is a program executed by a network of participators who agree on the states of the program. The contract developer defines the code logic of the contract, and the contract user invokes the different interfaces provided by the contract through the specified input to obtain the output of the contract execution. 

Existing smart contract-enabled blockchain systems take the form of replicating data and computation across all nodes in the system so that a single node can verify the correct execution of the contract. Full replicated execution on all nodes provides high robustness and availability. However, double counting on all blockchain nodes leads to a severe waste of computational power. To prop up computational power consumption, the inherent blockchain design renders users to pay for this overhead, which brings a huge economic burden for users. Therefore, \cite{RaymondCheng2018EkidenAP, fang2021high} proposed solutions that integrate TEE into blockchains to enable confidentiality and improve computational power. Inspired by them, we propose a TEE-enabled blockchain framework oriented for FL scenarios. The blockchain can be leveraged to record and synchronize TEE's results to maintain integrity and consistency.

\subsection{Desired Properties}
We conclude Voltran's challenges and desired properties here. The main target of Voltran is to back up the general execution of FL tasks with the following properties:

\subsubsection{Security}
The security is two-fold in our design, including the authenticity and confidentiality:

\begin{itemize}
	\item \textit{Authenticity:} Intuitively, authenticity means that an adversary (including a corrupt client, execution node or other situations like collusion) cannot forge the participant's identity and convince a receiver to accept data which is not the expected content. Due to data transmission across multiple entities, it is necessary to guarantee that all the recipients can trust the data senders during the whole process.
	\item \textit{Confidentiality:} Existing research (Melis et al., 2019) suggests that inference attacks can be used to steal training data from clients through model updates. In a vanilla FL framework, the centralized server has access to the model data passed by all clients, which enables inference attacks. In Voltran, we replace the centralized server mode with a distributed cluster of SGX nodes to perform aggregation tasks. This design allows the clients to send their local models to the execution nodes, which load models into SGX enclaves. Nevertheless, the curious adversary $\mathcal{N}$ has the possibility of accessing the client's data. Voltran must ensure that the confidentiality of client-supplied models participating in the aggregation process is protected. In the absence of TEE violations, Voltran guarantees that FL inputs and outputs are kept secret from all parties except the clients themselves. Further privacy leakage by clients or task owners is not considered in this paper.
\end{itemize}

\subsubsection{Correctness} 
The FL task is executed correctly, which means that the contract code is executed correctly, SGX performs the correct computation, the results are verified correctly on the chain, and clients get the correct global model to start the next round of training.

\section{Overview of Voltran}\label{Overview}
\subsection{Overview}
\begin{figure}[t]
	\centering
	\includegraphics[width=1.02\linewidth]{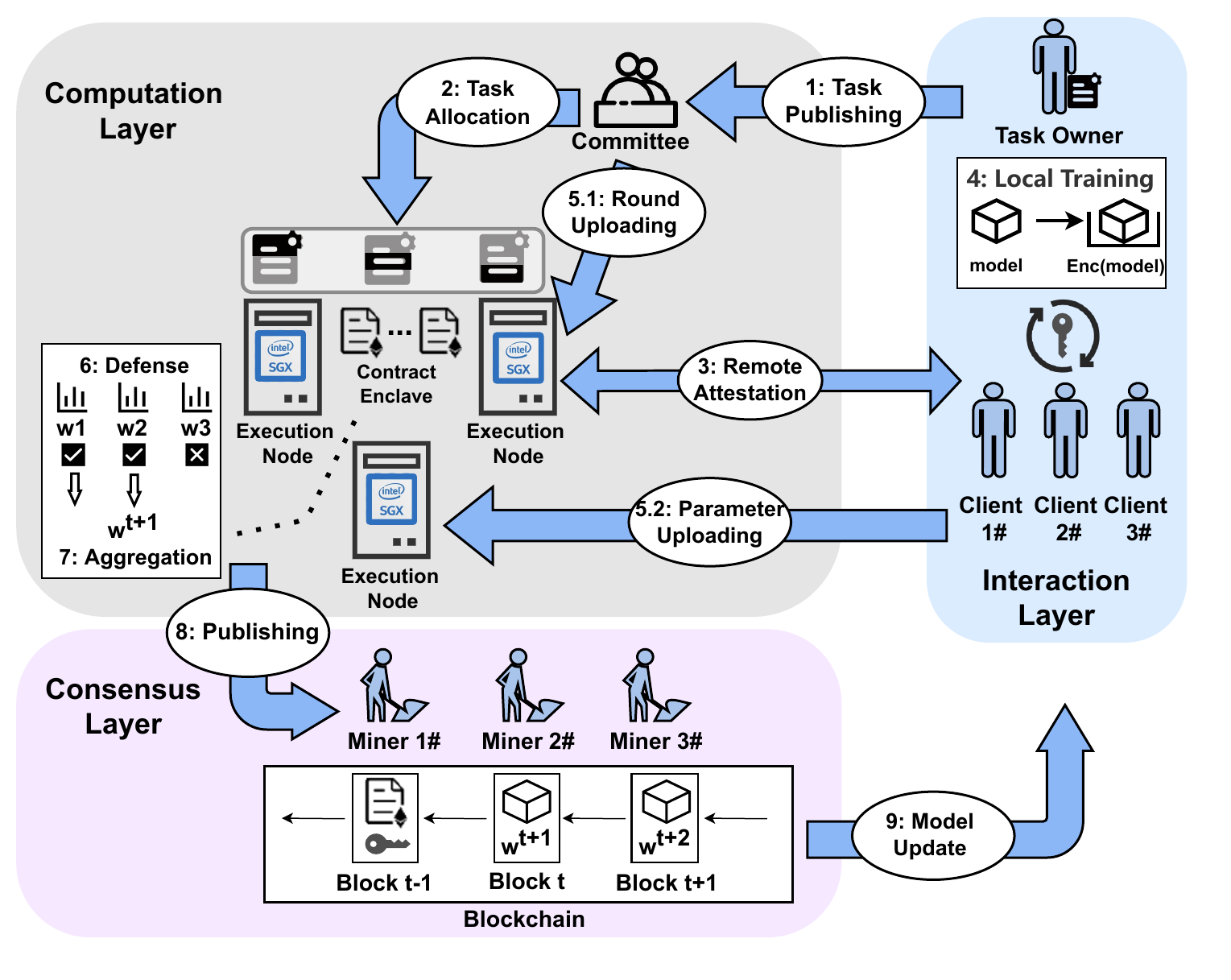}
	\caption{System workflow of Voltran.}
	\vspace{-0.5cm}
	\label{overview}
\end{figure}

Essentially, Voltran provides a decentralized, trusted and secure execution environment for FL aggregation by the distributed blockchain nodes with SGXs. Fig. \ref{overview} presents an overview of Voltran. We logically separate the FL computation away from the blockchain consensus process. Therefore, the system is divided into three layers according to different functionalities:

\textbf{Interaction Layer}. The Interaction Layer provides an interface for users to perform and participate in FL tasks using Voltran. In Voltran, users are categorized into distinct roles. They can be the Task Owners $\mathcal{\brown{O}}$ (detailed Notation is defined in Appendix A), responsible for owning and dispatching the original models to designated clients for local training. The local model results are sent to SGXs for aggregation. The underlying logic for this process is implemented through a smart contract named Contract Enclave {\fontfamily{cmss}\selectfont Con}$_{\textnormal{encl}}$. $\mathcal{\brown{O}}$ can implement the secure aggregation algorithm alongside advanced functionalities of attack detection in $\texttt{Con}_{\textnormal{encl}}$. Users can also act as clients $\mathcal{\brown{C}}$, who retrieve the global model from Voltran, train it with their private dataset, and subsequently send the training results to $\texttt{Con}_{\textnormal{encl}}$. Details on {\fontfamily{cmss}\selectfont Con}$_{\textnormal{encl}}$ are described in Section IV-D.

To record and verify the authenticity of results after each aggregation round, there is also a Contract Storage denoted as $\texttt{Con}_{\textnormal{stor}}$. This contract is deployed on the blockchain and provides interfaces for uploading and storing aggregation results, as well as some additional functionalities, such as judging and executing rewards/penalties. These two contracts are designed and implemented by $\mathcal{\brown{O}}$ and sent to Voltran for installation.

\textbf{Computation Layer}. The Computation Layer comprises a swarm of execution nodes $\mathcal{\brown{N}}$ equipped with Intel SGX. In our decentralized FL framework, the Computation Layer essentially acts as an aggregator of the centralized server in the vanilla FL. By running {\fontfamily{cmss}\selectfont Con}$_{\textnormal{encl}}$ given by $\mathcal{\brown{O}}$ in SGX, the aggregated global model and authenticity proof are generated and sent to the blockchain on Consensus Layer. We guarantee that every task running in SGX enclaves will not exceed the memory size of SGX by two proposed execution strategies to schedule and split the aggregation work into several pieces. Each piece is executed in one SGX to guarantee the capacity is enough.  Moreover, a committee \textit{Comm} is accompanied to be set up to schedule and assign tasks to the swarm of execution nodes. We will elaborate on the design details in Section \ref{committee}.

\textbf{Consensus Layer}. The Consensus Layer is basically the same as the architecture of the general blockchain, in which distributed miners include the received transactions into blocks and maintain the consistency of the state of the distributed ledger according to the consensus protocol. After executing the computation, $\mathcal{\brown{N}}$ sends the aggregated results and proof to the chain for public audit. We also consider scenarios where clients may lack motivation to participate in training. Therefore, we also set incentives by virtue of the economic properties of blockchain on the Consensus level, which can be implemented on {\fontfamily{cmss}\selectfont Con}$_{\textnormal{stor}}$.

\begin{figure}[t]
	\centering
	\includegraphics[width=3.2in]{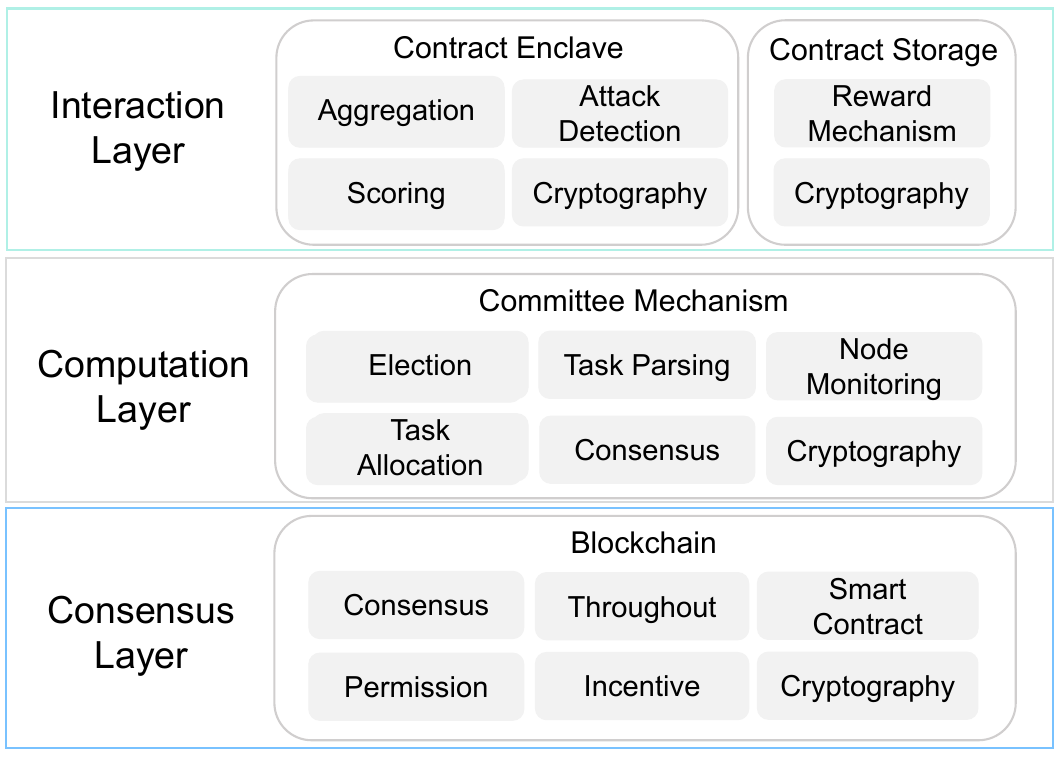}
	\caption{Architecture of Voltran. Voltran is designed and modularized in three layers. Each module can be implemented according to the requirements of different tasks.}
	\vspace{-0.2cm}
	\label{architecture}
\end{figure}

One of the major advantages of Voltran is its high composability and scalability. Each sub-function in each module shown in Fig. \ref{architecture} can be substituted and combined arbitrarily without any restrictions, which provides convenience for supporting various FL scenarios. Implementing a scaleable, pluggable FL framework that removes the restriction of blockchain type helps to better adapt to FL scenarios. We isolate the computation layer from the consensus layer logically. In fact, SGX nodes can also be blockchain nodes concurrently. For contracts, Voltran provides the concept of \textit{Composite Smart Contracts}, where contracts can call each other to extend larger functionality. {\fontfamily{cmss}\selectfont Con}$_{\textnormal{encl}}$ and {\fontfamily{cmss}\selectfont Con}$_{\textnormal{stor}}$ are not single specific contract but the abstract concepts. They can be composed of sub-contracts that implement diverse functions. Moreover, contracts can be adopted and reused for the following tasks with the same requirements.

A series of initializations, including node registration and committee election, needs to be set up when a Voltran prototype is instantiated. SGX nodes need to register to become an execution node $\mathcal{\brown{N}}$. The Committee \textit{Comm} is an autonomous internal management system. Given that Voltran may perform multiple tasks simultaneously, $\mathcal{\brown{N}}$ needs to be managed and prioritized. \textit{Comm} facilitates $\mathcal{\brown{N}}$ discovery and load balancing by maintaining a coordinator that produces a real-time optimal selection strategy.

\subsection{Threat Model and Assumptions}

\subsubsection{SGX} 
Assuming that SGX is well manufactured and its security protocol are secure, adversaries cannot break them to forge the identity of SGX or the identity of clients interacting with SGX. The key used for SGX communication is secure and will not be cracked within a valid time. Moreover, SGX may face various side-channel attacks, which may target the SGX units used in Voltran to compromise security and privacy. Although Voltran itself is not designed to withstand these attacks, it may be possible to defend against them by integrating existing studies aimed at these attacks, such as ShuffleFL \cite{zhang2021shufflefl}, HybCache \cite{dessouky2020hybcache}, DR.SGX \cite{brasser2019dr1} and other schemes \cite{crone2021towards, hu2021sok, hosseinzadeh2018mitigating}. Voltran’s data transmission can be based on secure communication protocols such as Transport Layer Security (TLS) \cite{oppliger2006ssl, esfahani2019efficient, knauth2018integrating} to resist man-in-the-middle attacks.

\subsubsection{Committee} 
The committee is composed of multiple executive nodes $\mathcal{\brown{N}}$ with SGX elected by a determinate strategy. The resulting decisions are generated through internal consensus. The security of the committee system is analogous to the security of the blockchain system, which depends on the number of selected nodes and the security of the consensus algorithm. To simplify Voltran's security considerations, we assume that the committee is secure and that the decisions and data it produces are trusted. Some common vulnerabilities in distributed systems, such as 51\% attacks or Sybil attacks, are not considered in this paper.

\subsubsection{Blockchain} 

Voltran is independent of the consensus layer. We take minimal requirements for blockchain. Voltran can be deployed on any blockchain implementation as long as it satisfies the smart contract functionality. We assume the blockchain architecture is secure and trustworthy, will perform the specific computations correctly, and will always be available (i.e., be of liveness). Data on the chain cannot be tampered with. Miners are rational and will not deviate from the intent of maintaining the consistency of the system. We do not consider attacks on the blockchain level.

\subsubsection{Threat Model}

Although the SGX hardware is assumed secure, its host is not trusted and has the possibility of misbehaving. It will honestly execute the protocol to finish the aggregation but may try to infer the privacy information from the incoming models. Adversaries may attack on the SGX nodes to arbitrarily determine the execution of the process and the message flow. They can create and cancel processes at will, delay or reject incoming messages, and try to forge the outgoing messages from SGX. Clients act as external system users, participating in training tasks and gaining rewards. They are rational and will perform the task in compliance with the protocol. They need to perform an internal key negotiation to generate {\fontfamily{cmss}\selectfont msk}. Each client has a {\fontfamily{cmss}\selectfont uid} that uniquely identifies them (e.g. an address in the blockchain) and cannot be impersonated. They do not intentionally interrupt the execution of a task, e.g. by deliberately not sending a local model or deliberately not obtaining the new model from the chain. However, differences in individual configurations, network environments and local data volumes can lead to different uploading speeds and even timeouts for clients. Clients can also be malicious and may execute attacks to disrupt model convergence. Therefore, our aim is to help FL aggregation execute correctly and protect FL models' confidentiality and authenticity.

\subsection{Workflow}
As shown in Fig. 1, we divide the process into three phases: Task creation, FL execution, and On-chain operations.

\subsubsection{Task Creation}
As steps 1 to 3 of Fig. \ref{overview} show, the task owner $\mathcal{\brown{O}}$ publishes the training task and designs the smart contract {\fontfamily{cmss}\selectfont Con}$_{\textnormal{encl}}$, which is an executable program for SGX. Also, $\mathcal{\brown{O}}$ needs to decide clients $\mathcal{\brown{C}}$ who participate in this task. Then, $\mathcal{\brown{O}}$ sends $\texttt{Con}_{\mathrm{Encl}}$ with initial model and client information {\fontfamily{cmss}\selectfont Con}$_{\textnormal{encl}}$, {\fontfamily{cmss}\selectfont mod}, {\fontfamily{cmss}\selectfont cli}) to the execution node committee \textit{Comm}. {\fontfamily{cmss}\selectfont Con}$_{\textnormal{stor}}$ also derives from it and is deployed on blockchain. \textit{Comm} evaluates the task and allocates one or several execution nodes $\mathcal{N}$ to participate in this task according to the model size and client number. A configuration file \textit{conf} containing $\mathcal{\brown{C}}$ and $\mathcal{\brown{N}}$ participating in each round of the task is generated and distributed to all parties. The files are loaded into the chosen SGXs and initialized. Afterwards, every pair of $\mathcal{\brown{C}}$ and SGX make RA to verify the enclaves' authenticity. A shared session key {\fontfamily{cmss}\selectfont ssk} is generated to transfer the encrypted model weights between $\mathcal{\brown{C}}$ and SGXs. Moreover, $\mathcal{\brown{O}}$ needs to perform a key agreement algorithm with $\mathcal{\brown{C}}$ to generate a master secret key {\fontfamily{cmss}\selectfont msk} for SGX to encrypt the computed global models. After these pre-operations, the task was delivered to designated clients and SGXs. Every $\mathcal{\brown{C}}$ and SGX have established a connection and a secure private channel.

\subsubsection{FL Execution}
After the contract is deployed and secure channels are built, the FL task execution process begins, corresponding to steps 4 to 7 in Fig. \ref{overview}. $\mathcal{\brown{C}}$ execute the local training with their private data to generate their local models. They use {\fontfamily{cmss}\selectfont ssk} to encrypt their local model and {\fontfamily{cmss}\selectfont msk} and send the ciphertexts to $\mathcal{\brown{N}}$s. $\mathcal{\brown{N}}$ loads data into the enclaves as the input of the contract {\fontfamily{cmss}\selectfont Con}$_{\textnormal{encl}}$. {\fontfamily{cmss}\selectfont Con}$_{\textnormal{encl}}$ contains the encryption/decryption function and a secure aggregation algorithm. It decrypts the ciphertext with {\fontfamily{cmss}\selectfont ssk}, obtains the plaintext local models of different clients and {\fontfamily{cmss}\selectfont msk} and executes the aggregation. The results are encrypted again by {\fontfamily{cmss}\selectfont msk} for client access. Moreover, enclaves also generate a digital signature on the ciphertext using the verification key {\fontfamily{cmss}\selectfont vk} to guarantee the integrity and authenticity of the computation results. In total, enclaves send the ciphertext and signature to {\fontfamily{cmss}\selectfont Con}$_{\textnormal{stor}}$. We will elaborate on the integrated secret key flow in Section \ref{secretkeyflow}. 

\subsubsection{On-chain Operations}
Another smart contract {\fontfamily{cmss}\selectfont Con}$_{\textnormal{stor}}$, runs on the blockchain to record model updates, verify authenticity, and perform incentive/punishment mechanisms. First, \textit{Comm} generates the verification key pair {\fontfamily{cmss}\selectfont vk} and sends the public key of it {\fontfamily{cmss}\selectfont pk}$_{\mathrm{vk}}$ to {\fontfamily{cmss}\selectfont Con}$_{\textnormal{stor}}$, which will play the role of authenticity verification throughout the task. Then, {\fontfamily{cmss}\selectfont Con}$_{\textnormal{stor}}$ provides $\mathcal{\brown{N}}$ with an interface to send model updates onto the chain and $\mathcal{\brown{N}}$ sends the encrypted model to the contract along with the signature mentioned above. Blockchain nodes use {\fontfamily{cmss}\selectfont pk}$_{\mathrm{vk}}$ to check the signature, and if passed, confirm the transaction and include it into a block. It also provides an interface for $\mathcal{\brown{C}}$ to get the model update of each round to perform the next round of local training. In addition, reward/punishment mechanisms can be written in $\texttt{Con}_{\mathrm{stor}}$ to incentivize $\mathcal{\brown{C}}$ and SGXs to behave positively and honestly.

\section{Building Blocks}\label{blocks}

\subsection{Remote Attestation and Secret Key Flow}\label{secretkeyflow}

\begin{figure}[t]
	\centering
	\includegraphics[width=3.55in]{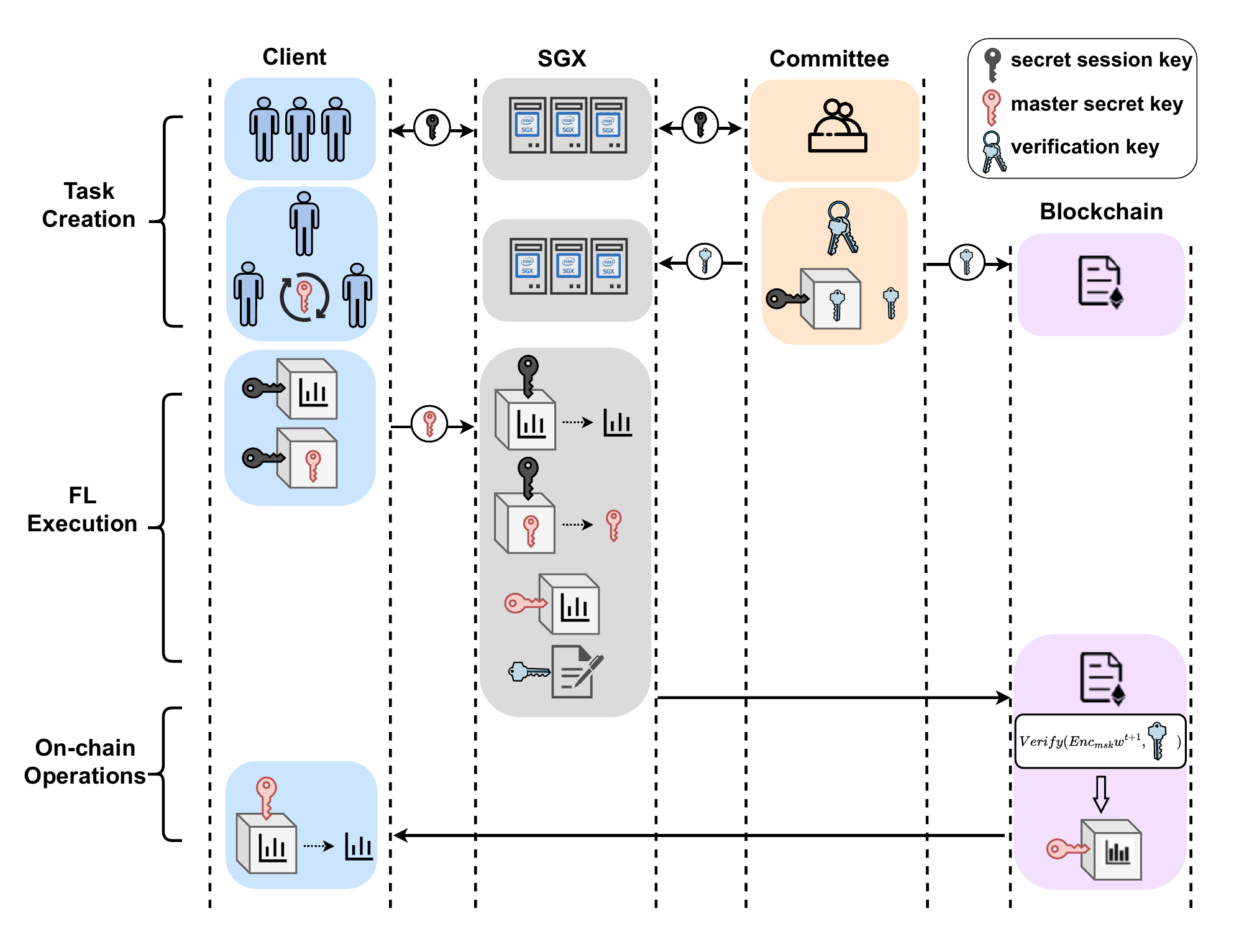}
	\caption{The data transmission mechanism depicted in a secret key flow in our system.}
	\label{keyflow}
\end{figure}

To realize the confidentiality and authenticity of Voltran, we propose a secure data transmission mechanism by combining the Intel Remote Attestation and cryptographic primitives. We depict this mechanism in Voltran in a form of a secret key flow in Fig. \ref{keyflow}. 

Firstly, in the Task Creation phase, each of the selected SGXs needs to perform multiple RA due to the requirement of authenticity verification and creating secure channels, which occurs in: a). between it and the client $\mathcal{\brown{C}}$ involved in training; b). between it and the committee \texttt{Comm}. Each RA results in a symmetric session key {\fontfamily{cmss}\selectfont ssk}, indicated by the black key in the figure. Then, $\mathcal{\brown{C}}$ performs a key agreement algorithm internally to generate a symmetric master key {\fontfamily{cmss}\selectfont msk} to encrypt the global model, indicated by the red key. Meanwhile, after the scheduling, \texttt{Comm} generates a pair of asymmetric verification key pair ({\fontfamily{cmss}\selectfont vk}$_{\mathrm{pk}}$, {\fontfamily{cmss}\selectfont vk}$_{\mathrm{sk}}$) for each SGX participating in the task, and sends the private key $\texttt{sk}_{vk}$ to SGX (encrypted by {\fontfamily{cmss}\selectfont ssk}). The public key is sent to $\texttt{Con}_{\mathrm{Stor}}$ on the chain.

In the FL Execution phase, in each round, $\mathcal{\brown{C}}$ participating in the training encrypts the local model and {\fontfamily{cmss}\selectfont msk} with  {\fontfamily{cmss}\selectfont ssk} negotiated with the appointed SGX and sends {\fontfamily{cmss}\selectfont ct}$_{\mathrm{in}}$ =  (\texttt{ENC}({\fontfamily{cmss}\selectfont ssk}, {\fontfamily{cmss}\selectfont m}$_{\mathrm{cid}}$), \texttt{ENC}({\fontfamily{cmss}\selectfont ssk}, {\fontfamily{cmss}\selectfont msk})) to $\texttt{Con}_{\mathrm{Encl}}$. The data has to go through $\mathcal{\brown{N}}$ before being loaded into the enclave. However, because of the encryption, $\mathcal{\brown{N}}$ can not obtain any information or do any malicious behaviour without being detected and can only load it into the enclave. SGX computes  ({\fontfamily{cmss}\selectfont mod}, {\fontfamily{cmss}\selectfont msk}) = \texttt{Dec}({\fontfamily{cmss}\selectfont ssk}, {\fontfamily{cmss}\selectfont ct}$_{\mathrm{in}}$) by using {\fontfamily{cmss}\selectfont ssk} for decryption and restoring the original model and {\fontfamily{cmss}\selectfont msk} for the further execution of $\texttt{Con}_{\mathrm{Encl}}$. When $\texttt{Con}_{\mathrm{Encl}}$ completes the aggregation, the resulting global model is encrypted by {\fontfamily{cmss}\selectfont msk} and gets {\fontfamily{cmss}\selectfont ct}$_{\mathrm{out}}$ = \texttt{ENC}({\fontfamily{cmss}\selectfont msk}, {\fontfamily{cmss}\selectfont m}$_{\mathrm{glob}}$). Moreover, to guarantee the authenticity, it also computes $\tau_{\mathrm{sgx}}$ = \texttt{Sig}({\fontfamily{cmss}\selectfont vk}$_{\mathrm{sk}}$, {\fontfamily{cmss}\selectfont ct}$_{\mathrm{out}}$) to generate a signature as the proof. Hence, the output {\fontfamily{cmss}\selectfont out}$_{\mathrm{Encl}}$ = ({\fontfamily{cmss}\selectfont ct}$_{\mathrm{out}}$, $\tau_{\mathrm{sgx}})$. $\mathcal{\brown{N}}$ sends transactions as the input of {\fontfamily{cmss}\selectfont out}$_{\mathrm{Encl}}$ to $\texttt{Con}_{\mathrm{Stor}}$. 

Finally, in the On-chain Operations phase, upon receiving the transactions, blockchain miners receive and verify the transactions. They verify the proof $\tau_{\mathrm{sgx}}$ by checking whether \texttt{Verify}({\fontfamily{cmss}\selectfont vk}$_{\mathrm{pk}}$, {\fontfamily{cmss}\selectfont ct}$_{\mathrm{out}}$, $\tau_{\mathrm{sgx}}$) = \texttt{TRUE}. If passing, they include the transactions into a block and publish it. $\mathcal{\brown{C}}$ can retrieve {\fontfamily{cmss}\selectfont ct}$_{\mathrm{out}}$ from $\texttt{Con}_{\mathrm{Stor}}$ and decrypt it with {\fontfamily{cmss}\selectfont msk} to get {\fontfamily{cmss}\selectfont m}$_{\mathrm{glob}}$ = \texttt{Dec}({\fontfamily{cmss}\selectfont msk}, {\fontfamily{cmss}\selectfont ct}$_{\mathrm{out}}$) for the local training of the next round. 

Note that we additionally bring in a novel verification mechanism for SGX computing results to ensure the authenticity. We add this mechanism because the native SGX verification mechanism additionally requires access to Intel Attestation Service (IAS). The absence of a substantive ecosystem of trustworthy out-of-band network access for smart contracts poses a challenge for deploying this mechanism on them \cite{zhang2016town}. Our design offers the benefit of avoiding access dependence on IAS (e.g., bringing in any relay or server).

\subsection{Committee}\label{committee}

Voltran is an FL-oriented service platform. Its purpose is to support various FL tasks with different models and aggregation algorithms. It requires Voltran to have the ability to parse different tasks and allocate computational resources reasonably. Since in Voltran, the computational layer with $\mathcal{\brown{N}}$ is logically separated from the blockchain, it is challenging to manage and schedule a batch of execution nodes. Moreover, the on-chain result verification needs a pair of keys to perform the signature. The generation and management of verification keys also require a trusted party.

Therefore, based on the above considerations, we bring in the committee mechanism in the computational layer to perform FL task reception, parsing and distribution. Essentially, the committee mechanism is a trusted management system. The Task Owner $\mathcal{\brown{O}}$ sends the task to the committee \textit{Comm}, which includes $\texttt{Con}_{\mathrm{Encl}}$, $\texttt{Con}_{\mathrm{Stor}}$, the unique identifications of the client group, the initial model, the basic training information (including the number of rounds, number of clients in each round), and other extra information. Based on the information and the network condition, \textit{Comm} generates a configuration file \textit{conf}. \textit{conf} records the one-to-one correspondence between $\mathcal{\brown{C}}$ and $\mathcal{\brown{N}}$ in each round of tasks, i.e., which clients need to send their local models to which SGX for execution. Moreover, in order to prevent tasks from the single point of failure problem, \textit{Comm} also needs a monitoring mechanism to detect the delay and replace the faulty nodes timely. In addition to scheduling, the committee also needs to generate the verification key ({\fontfamily{cmss}\selectfont vk}$_{\mathrm{pk}}$, {\fontfamily{cmss}\selectfont vk}$_{\mathrm{sk}}$). Due to its trustworthiness, the possibility of using {\fontfamily{cmss}\selectfont vk} for the forgery of signatures is not considered.

We sketch a kind of implementation of the committee inspired by \cite{che2022decentralized}. It is composed of multiple execution nodes as the committee members. In other words, the committee is a distributed organization in which execution nodes manage themselves autonomously. They vote for node scheduling and maintain a mechanism to reach a consensus. Committee nodes also play the role of a sentinel \cite{Redis}, receiving heartbeats pulsed from $\mathcal{\brown{N}}$ that participate in the execution to determine whether the node is offline. Once a disconnection is found, automatic fault migration is performed, and a new candidate $\mathcal{\brown{N}}$ is selected to replace the dropped $\mathcal{\brown{N}}$ to prevent task stagnation. The final decision on the disconnection is made by the joint decision of multiple committee nodes. Note that the committee also has to consider its own fault tolerance. Therefore, it can take the Byzantine Fault Tolerance (BFT) consensus algorithms as the election strategy of the committee to ensure security, decentralization and high robustness. The agreement of the verification key ({\fontfamily{cmss}\selectfont vk}$_{\mathrm{pk}}$, {\fontfamily{cmss}\selectfont vk}$_{\mathrm{sk}}$) can be realized by a threshold secret sharing scheme.

\subsection{Task Scheduling} \label{scheduling}

Maintaining a multi-entity framework like Voltran requires considering the stability of each end-to-end connection. To ensure the smooth execution of tasks, several issues need to be considered: a). the network conditions of clients, as clients with limited bandwidth for model uploading can slow down the aggregation progress; b). the size of the trusted area that SGX can create is limited, and larger aggregation tasks cannot be performed in a single SGX; c). the capacity of transactions allowed by the blockchain is limited, and the aggregated results cannot be uploaded in a single transaction. Therefore, in Voltran, we propose a new execution strategy that splits the computation task into several subtasks and puts them into multiple SGXs to execute an efficient and well-organized schedule for resource-intensive tasks.

\begin{figure}
	\centering
	\includegraphics[width=3.6in]{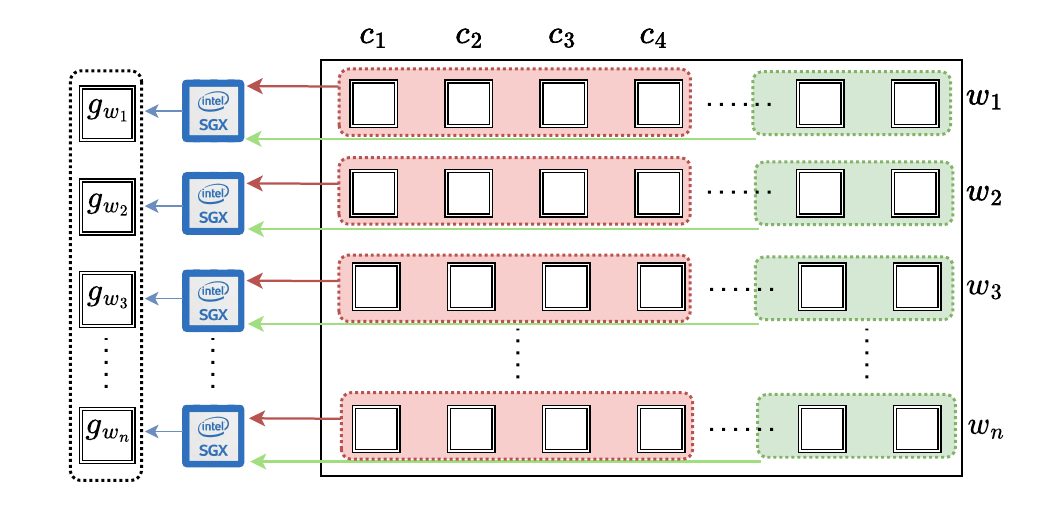}
	\caption{Spilt by clients. This strategy contains metadata from most different clients into one single SGX. In this figure's case, assuming one SGX can contain four metadata, the strategy puts metadata of $w_1$ from $c_1$ to $c_4$ into SGX 1. SGXs carrying data from $w_1$ are divided into a partition to compute and generate $g_{w_1}$ of the global model.}
	\vspace{-0.5cm}
	\label{Spilt by clients}
	
\end{figure}

Let us take a holistic view of the data that requires SGX computation. It is the machine learning model sent by multiple clients. The models of different clients have the same network structure but different weights. All the data can be seen as a matrix, with each layer as a row and each client as a column. There are two priority strategies for splitting this data: \textit{ClientMax} and \textit{LayerMax}. First, in Fig. \ref{Spilt by clients}, \textit{ClientMax} is a multi-SGX parallel strategy, which means that more client blocks are accommodated in priority in each subpartition. The advantage of this partition is that it adheres to the general aggregation logic where weights of the same layer should be placed in the same SGX as much as possible. This allows for quick aggregation of layer weights with multiple SGXs. Each SGX is responsible for aggregation on weights of one layer and sends results onto {\fontfamily{cmss}\selectfont Con}$_{\textnormal{stor}}$ respectively. It takes the pressure off the bandwidth but brings more communication. The second strategy conforms to the client's common upload logic. In this mode, all the clients send all the parameters of their local model to one single SGX. Data are stored in the virtual memory outside the SGX and wait for the \textit{ecall} instruction. It needs to perform EPC paging and change the data accessing context. Aggregation results are cumulated and form the global model. This mode needs less SGX but requires serial execution. Both strategies have to consider the transaction capacity limitation and may need to divide the outputs into several pieces to be sent on the blockchain when the model size is large. In summary, these two data-splitting strategies are generic for computations on FL. Choosing a suitable splitting scheme and the right scheduling strategy can help significantly improve efficiency.

Our framework can also provide high \textit{robustness}. For the computational layer, distributed execution nodes allow the possibility of performing repeated computations similar to the execution mechanism of blockchain. In other words, to prevent the single point of failure issue of a single execution node, tasks can be concurrently assigned to multiple nodes (or node groups, depending on whether the \textit{ClientMax} or \textit{LayerMax} mode is adopted) for execution. Replication-based computation can improve the reliability of the computation results and enhance fault tolerance. Besides, when configuring \textit{conf}, the Task Owner can also add standby nodes to restore task execution to avoid stagnation quickly.

\section{Protocol and Security Analysis}\label{protocol}
In this section, we define the protocol of Voltran $\textbf{Prot}_{\texttt{Volt}}$ and security properties based on assumptions. We consider the threat model and give the security proof of $\textbf{Prot}_{\texttt{Volt}}$.

\subsection{Protocol}

The protocol of Voltran $\textbf{Prot}_{\texttt{Volt}}$ is formally specified. $\textbf{Prot}_{\texttt{Volt}}$ relies on $\mathcal{F}_{\text {sgx}}$ and $\mathcal{F}_{\text {blockchain }}$, ideal functionality for SGX operations and the blockchain. $\textbf{Prot}_{\texttt{Volt}}$ also utilizes a digital signature scheme $\Sigma(\mathrm{KGen}, \mathrm{Sig}, \mathrm{Verify})$ and two symmetric encryption schemes $\mathcal{SE}_1, \mathcal{SE}_2$(KGen, Enc, Dec). 

\begin{figure}[!t]
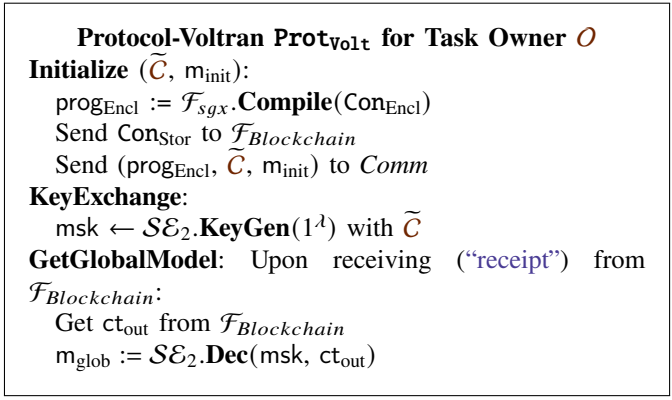

	
	\begin{framed}
		
		\centerline{\textbf{Protocol-Voltran $\texttt{Prot}_{\texttt{Volt}}$ for Task Owner $\mathcal{\brown{O}}$}}
		
		\noindent \textbf{Initialize} $(\widetilde{\mathcal{\brown{C}}},$  {\fontfamily{cmss}\selectfont m}$_{\mathrm{init}})$:  
		
		\quad {\fontfamily{cmss}\selectfont prog}$_{\mathrm{Encl}}$ := $\mathcal{F}_{sgx}.\textbf{Compile}(\texttt{Con}_{\mathrm{Encl}})$
		
		\quad Send $\texttt{Con}_{\mathrm{Stor}}$ to $\mathcal{F}_{Blockchain}$
		
		
		\quad Send ({\fontfamily{cmss}\selectfont prog}$_{\mathrm{Encl}}$, $\widetilde{\mathcal{\brown{C}}}$, {\fontfamily{cmss}\selectfont m}$_{\mathrm{init}})$ to \textit{Comm}
		
		
		
		\noindent \textbf{KeyExchange}:
		
		\quad {\fontfamily{cmss}\selectfont msk} $\leftarrow\mathcal{SE}_2.\mathrm{\textbf{KeyGen}}(1^{\lambda})$ with $\widetilde{\mathcal{\brown{C}}}$

		\noindent \textbf{GetGlobalModel}: 
		Upon receiving (\purple{``receipt''}) from $\mathcal{F}_{Blockchain}$: 
		
		\quad Get {\fontfamily{cmss}\selectfont ct}$_{\mathrm{out}}$ from $\mathcal{F}_{Blockchain}$

		\quad {\fontfamily{cmss}\selectfont m}$_{\mathrm{glob}} := \mathcal{SE}_2.\mathrm{\textbf{Dec}}(${\fontfamily{cmss}\selectfont msk}, {\fontfamily{cmss}\selectfont ct}$_{\mathrm{out}})$
	\end{framed}
	
	\caption{Protocol of Voltran for the Task Owner.}
	\label{TaskOwner}
	\vspace{-0.4cm}
\end{figure}

\begin{figure}[!t]
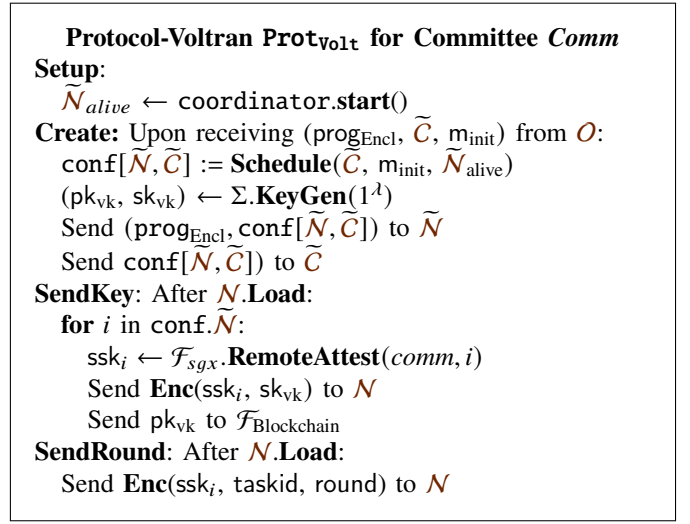

	\begin{framed}
		\centerline{\textbf{Protocol-Voltran $\texttt{Prot}_{\texttt{Volt}}$ for Committee \textit{Comm}}}
		
		\noindent \textbf{Setup}:

		\quad $\widetilde{\mathcal{\brown{N}}}_{alive}$ $\leftarrow$ \texttt{coordinator}.\textbf{start}()
		
		\noindent \textbf{Create:} Upon receiving ({\fontfamily{cmss}\selectfont prog}$_{\mathrm{Encl}}$, $\widetilde{\mathcal{\brown{C}}}$, {\fontfamily{cmss}\selectfont m}$_{\mathrm{init}})$ from $\mathcal{\brown{O}}$:  
		
		\quad $\texttt{conf}[\widetilde{\mathcal{\brown{N}}}, \widetilde{\mathcal{\brown{C}}}] := \textbf{Schedule} (\widetilde{\mathcal{\brown{C}}},$  {\fontfamily{cmss}\selectfont m}$_{\mathrm{init}}$, $\widetilde{\mathcal{\brown{N}}}_{\mathrm{alive}})$
		
		\quad ({\fontfamily{cmss}\selectfont pk}$_{\mathrm{vk}}$, {\fontfamily{cmss}\selectfont sk}$_{\mathrm{vk}})$ $\leftarrow \Sigma.\mathrm{\textbf{KeyGen}}(1^{\lambda})$

		\quad Send $(\texttt{prog}_{\mathrm{Encl}}, \texttt{conf}[\widetilde{\mathcal{\brown{N}}}, \widetilde{\mathcal{\brown{C}}}])$ to 
		$\widetilde{\mathcal{\brown{N}}}$
		\vspace{0.03cm}
		
		\quad Send $\texttt{conf}[\widetilde{\mathcal{\brown{N}}}, \widetilde{\mathcal{\brown{C}}}])$ to $\widetilde{\mathcal{\brown{C}}}$
		
		\noindent \textbf{SendKey}: After  $\mathcal{\brown{N}}$.\textbf{Load}:  
		
		\quad \textbf{for} $i$ in $\texttt{conf}.\widetilde{\mathcal{\brown{N}}}$:
		
		\quad \quad {\fontfamily{cmss}\selectfont ssk}$_i \leftarrow \mathcal{F}_{sgx}.\textbf{RemoteAttest}(\textit{comm}, i)$
		
		\quad \quad Send $\textbf{Enc}$({\fontfamily{cmss}\selectfont ssk}$_i$, {\fontfamily{cmss}\selectfont sk}$_{\mathrm{vk}})$ to $\mathcal{\brown{N}}$
		
		\quad \quad Send {\fontfamily{cmss}\selectfont pk}$_{\mathrm{vk}}$ to $\mathcal{F}_{\mathrm{Blockchain}}$

		\noindent \textbf{SendRound}: After $\mathcal{\brown{N}}$.\textbf{Load}: 
		
		\quad Send $\textbf{Enc}$({\fontfamily{cmss}\selectfont ssk}$_i$, {\fontfamily{cmss}\selectfont taskid}, {\fontfamily{cmss}\selectfont round}) to $\mathcal{\brown{N}}$

	\end{framed}
	
	\caption{Protocol of Voltran for the Execution Node Committee.}
	\label{comm}
\end{figure}

\begin{figure}[!t]
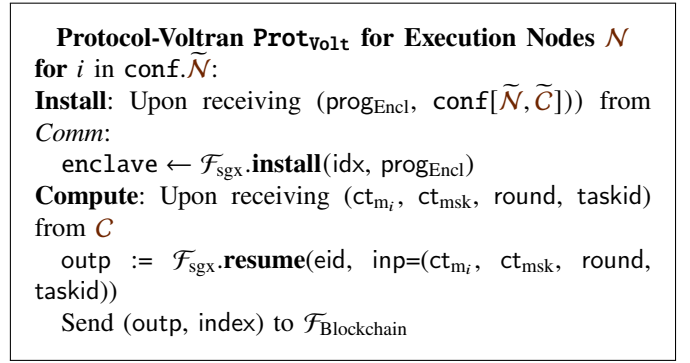

	\begin{framed}
		\centerline{\textbf{Protocol-Voltran $\texttt{Prot}_{\texttt{Volt}}$ for Execution Nodes $\mathcal{\brown{N}}$}}
		
		\textbf{for} $i$ in $\texttt{conf}.\widetilde{\mathcal{\brown{N}}}$:
		
		\noindent \textbf{Install}: Upon receiving ({\fontfamily{cmss}\selectfont prog}$_{\mathrm{Encl}}$, $\texttt{conf}[\widetilde{\mathcal{\brown{N}}}, \widetilde{\mathcal{\brown{C}}}]))$ from \textit{Comm}:  
		
		\quad $\texttt{enclave} \leftarrow         \mathcal{F}_{\mathrm{sgx}}.\textbf{install}(${\fontfamily{cmss}\selectfont idx}, {\fontfamily{cmss}\selectfont prog}$_{\mathrm{Encl}}$)

		\noindent \textbf{Compute}: Upon receiving ({\fontfamily{cmss}\selectfont ct}$_{\mathrm{m}_i}$, {\fontfamily{cmss}\selectfont ct}$_{\mathrm{msk}}$, {\fontfamily{cmss}\selectfont round}, {\fontfamily{cmss}\selectfont taskid}) from $\mathcal{\brown{C}}$
		
		\quad {\fontfamily{cmss}\selectfont outp} := $\mathcal{F}_{\mathrm{sgx}}$.\textbf{resume}({\fontfamily{cmss}\selectfont eid}, \textbf{{\fontfamily{cmss}\selectfont inp}}=({\fontfamily{cmss}\selectfont ct}$_{\mathrm{m}_i}$, {\fontfamily{cmss}\selectfont ct}$_{\mathrm{msk}}$, {\fontfamily{cmss}\selectfont round}, {\fontfamily{cmss}\selectfont taskid}))

		\quad Send ({\fontfamily{cmss}\selectfont outp}, {\fontfamily{cmss}\selectfont index}) to $\mathcal{F}_{\mathrm{Blockchain}}$ 

	\end{framed}
	
	\caption{Protocol of Voltran for the execution nodes.}
	\label{EN}
	\vspace{-0.2cm}
\end{figure}

\begin{figure}[!t]
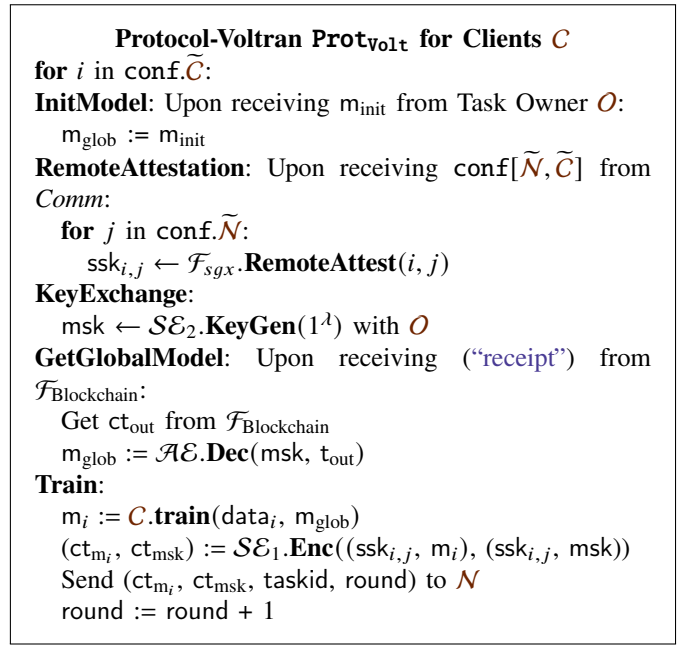

	\begin{framed}
		\centerline{\textbf{Protocol-Voltran $\texttt{Prot}_{\texttt{Volt}}$ for Clients {$\mathcal{\brown{C}}$}}}
		
		\textbf{for} $i$ in $\texttt{conf}$.$\widetilde{{\mathcal{\brown{C}}}}$:
		
		\noindent \textbf{InitModel}: Upon receiving {\fontfamily{cmss}\selectfont m}$_{\mathrm{init}}$ from Task Owner \brown{$\mathcal{O}$}: 
		
		\quad {\fontfamily{cmss}\selectfont m}$_{\mathrm{glob}}$ := {\fontfamily{cmss}\selectfont m}$_{\mathrm{init}}$
		
		\noindent \textbf{RemoteAttestation}: Upon receiving $\texttt{conf}[\widetilde{\mathcal{\brown{N}}}, \widetilde{\mathcal{\brown{C}}}]$ from \textit{Comm}:
		
		\quad \textbf{for} $j$ in $\texttt{conf}$.$\widetilde{\mathcal{\brown{N}}}$:
		
		\quad \quad {\fontfamily{cmss}\selectfont ssk}$_{i,j} \leftarrow \mathcal{F}_{sgx}.\textbf{RemoteAttest}(i,j)$

		\noindent \textbf{KeyExchange}:

		\quad {\fontfamily{cmss}\selectfont msk} $\leftarrow\mathcal{SE}_2.\mathrm{\textbf{KeyGen}}(1^{\lambda})$ with $\mathcal{\brown{O}}$
		
		\noindent \textbf{GetGlobalModel}: Upon receiving (\purple{``receipt''}) from $\mathcal{F}_{\mathrm{Blockchain}}$: 
		
		\quad Get {\fontfamily{cmss}\selectfont ct}$_{\mathrm{out}}$ from $\mathcal{F}_{\mathrm{Blockchain}}$
		
		\quad {\fontfamily{cmss}\selectfont m}$_{\mathrm{glob}} := \mathcal{AE}.\mathrm{\textbf{Dec}}(${\fontfamily{cmss}\selectfont msk}, {\fontfamily{cmss}\selectfont t}$_{\mathrm{out}})$
		
		\noindent \textbf{Train}: 
		
		\quad {\fontfamily{cmss}\selectfont m}$_{i} := \mathcal{\brown{C}}.\textbf{train}(${\fontfamily{cmss}\selectfont data}$_i$, {\fontfamily{cmss}\selectfont m}$_{\mathrm{glob}})$
		
		\quad ({\fontfamily{cmss}\selectfont ct}$_{\mathrm{m}_i}$, {\fontfamily{cmss}\selectfont ct}$_{\mathrm{msk}}) := \mathcal{SE}_1.\mathrm{\textbf{Enc}}((${\fontfamily{cmss}\selectfont ssk}$_{i,j},$ {\fontfamily{cmss}\selectfont m}$_i),$ ({\fontfamily{cmss}\selectfont ssk}$_{i,j}$, {\fontfamily{cmss}\selectfont msk}))
		
		\quad Send ({\fontfamily{cmss}\selectfont ct}$_{\mathrm{m}_i}$, {\fontfamily{cmss}\selectfont ct}$_{\mathrm{msk}}$, {\fontfamily{cmss}\selectfont taskid}, {\fontfamily{cmss}\selectfont round}) to $\mathcal{\brown{N}}$
		
		\quad {\fontfamily{cmss}\selectfont round} $:=$ {\fontfamily{cmss}\selectfont round} + 1

	\end{framed}
	
	\caption{Protocol of Voltran for the clients.}
	\label{Client}
	\vspace{-0.2cm}
\end{figure}

To clearly depict our multi-role system, we divide the protocol into four parts based on four entities. Fig. \ref{TaskOwner} stipulates the behaviours of the Task Owner. $\mathcal{\brown{O}}$ first executes the \textbf{Initialize} functionality to create a new FL task with training details. $\mathcal{\brown{O}}$ compiles {\fontfamily{cmss}\selectfont Con}$_{\textnormal{encl}}$ to generate an executable file {\fontfamily{cmss}\selectfont prog}$_{\mathrm{Encl}}$. Then, {\fontfamily{cmss}\selectfont Con}$_{\textnormal{encl}}$, the list of clients $\widetilde{\mathcal{\brown{C}}}$ and the initial model {\fontfamily{cmss}\selectfont m}$_{\mathrm{init}}$ are sent to \textit{Comm}. {\fontfamily{cmss}\selectfont m}$_{\mathrm{init}}$ is also sent to $\widetilde{\mathcal{\brown{C}}}$. {\fontfamily{cmss}\selectfont Con}$_{\textnormal{stor}}$ is sent onto the blockchain and initialized. Also, $\mathcal{\brown{O}}$ executes the key generation algorithm to get a pair of Verification Key ({\fontfamily{cmss}\selectfont pk}$_{\mathrm{vk}}$, {\fontfamily{cmss}\selectfont sk}$_{\mathrm{vk}}$). Also, after {\fontfamily{cmss}\selectfont prog}$_{\mathrm{encl}}$ is loaded into SGX, $\mathcal{\brown{O}}$ performs RA to generate {\fontfamily{cmss}\selectfont ssk} with each participating execution node and sends the ciphertext of {\fontfamily{cmss}\selectfont sk}$_{\mathrm{vk}}$ by {\fontfamily{cmss}\selectfont ssk} and {\fontfamily{cmss}\selectfont pk}$_{\mathrm{vk}}$ to {\fontfamily{cmss}\selectfont Con}$_{\textnormal{stor}}$. Fig. \ref{comm} presents the operations of \textit{Comm}. It executes the \textbf{Setup} function to conduct the election of committee members and the \textbf{Create} function to take the proper execution nodes to generate \texttt{conf}. Then, \textit{Comm} sends {\fontfamily{cmss}\selectfont prog}$_{\mathrm{Encl}}$ and $\texttt{conf}$ to each participating execution node $\mathcal{\brown{N}}$. Fig. \ref{EN} and Fig. \ref{Client} depict the behaviours of execution nodes and clients. After receiving \textit{conf}, \textbf{InitModel} functionality makes each $\mathcal{\brown{C}}$ set {\fontfamily{cmss}\selectfont m}$_{\mathrm{init}}$ as the global model of the first round. When \textit{Comm} has finished the scheduling, $\mathcal{\brown{N}}$ performs \textbf{Install} to generate the enclave by {\fontfamily{cmss}\selectfont prog}$_{\mathrm{Encl}}$, while $\mathcal{\brown{C}}$ executes \textbf{KeyExchange} with $\mathcal{\brown{O}}$ to generate {\fontfamily{cmss}\selectfont msk} and build connections with $\mathcal{\brown{N}}$. Then, the training computation begins. \textbf{Train} and \textbf{Compute} functionalities are executed in sequence by $\mathcal{\brown{C}}$ and $\mathcal{\brown{N}}$. Whenever $\mathcal{\brown{N}}$ uploads the computing output onto the blockchain, $\mathcal{\brown{C}}$ executes \textbf{GetGlobalModel} to acquire it and decrypt it.

\subsection{Blockchain Design}

Our framework is a hybrid system of blockchain and TEE. The main effect of blockchain is reflected in the decentralized verification of TEE calculation results. Although SGX provides RA services to verify the authenticity of its identity, once an unexpected error occurs, non-blockchain architecture cannot automatically verdict the correctness so as to solve the dispute. Hence, it has to bring in a trusted third party for arbitration. We believe that blockchain with smart contracts provides a perfect arbitration platform. The contract can automatically review and verify the computation results by executing the pre-written verification code, thus avoiding disputes. Moreover, blockchain is also a secure and trusted distributed database. Storing data on the blockchain achieves traceability and immutability. In addition, the incentive mechanism carried by the blockchain can provide rewards to clients and execution nodes, which motivate $\mathcal{\brown{C}}$ to contribute their data to participate in training and motivate $\mathcal{\brown{N}}$ to contribute SGXs to help computation respectively.

\subsubsection{Contract Design}

Other than current blockchain-only systems, Voltran is not limited to a specified form of contract implementation. We define all the programs executed in our system as smart contracts. Hence, both the enclave executable files in SGXs and the traditional contract deployed on the blockchain are seen as smart contracts. We bring in the concept of \textit{Composite Smart Contract}, i.e., through the cross combination of contracts to achieve the dynamic collocation of different functions. In general, we encapsulate the specifications that a contract needs to meet in Voltran. The user needs to follow the provided wrapper to implement the service logic in the FL task, which we abstract into two main contracts: Contract Enclave {\fontfamily{cmss}\selectfont Con}$_{\textnormal{encl}}$ and Contract Storage {\fontfamily{cmss}\selectfont Con}$_{\textnormal{stor}}$.

\begin{figure}[t]
	
	\begin{framed}
		
		\centerline{\textbf{Contract Enclave Wrapper {\fontfamily{cmss}\selectfont Con}$_{\textnormal{encl}}$}}
		1. \textbf{\blue{On input}} (\purple{``install"}, {\fontfamily{cmss}\selectfont idx}, {\fontfamily{cmss}\selectfont prog}):
		
		\quad\quad \textbf{Return} {\fontfamily{cmss}\selectfont eid}.

		2. \textbf{\blue{On input}} (\purple{``getsk"}, {\fontfamily{cmss}\selectfont vk}$_{\mathrm{sk}}$):
		
		\quad\quad {\fontfamily{cmss}\selectfont vk}$_{\mathrm{sk}}$ := {\fontfamily{cmss}\selectfont vk}$_{\mathrm{sk}}$;
		
		3. \textbf{\blue{On input}} (\purple{``getRound"}, {\fontfamily{cmss}\selectfont round}, {\fontfamily{cmss}\selectfont round}$_{\mathrm{c}}$):
		
		\quad\quad Assert {\fontfamily{cmss}\selectfont round}$_{\mathrm{c}}$ = {\fontfamily{cmss}\selectfont round}, abort if \texttt{FALSE}
		
		\quad\quad {\fontfamily{cmss}\selectfont round} := {\fontfamily{cmss}\selectfont round}
		
		4. \textbf{\blue{On input}} (\purple{``decrypt"}, {\fontfamily{cmss}\selectfont ct}$_{\mathrm{msk}}$, {\fontfamily{cmss}\selectfont ct}$_{\mathrm{m}}$):
		
		\quad\quad {\fontfamily{cmss}\selectfont m} := Dec({\fontfamily{cmss}\selectfont sk}, {\fontfamily{cmss}\selectfont ct}$_{\mathrm{m}}$);
		
		\quad\quad {\fontfamily{cmss}\selectfont msk} := Dec({\fontfamily{cmss}\selectfont sk}, {\fontfamily{cmss}\selectfont ct}$_{\mathrm{msk}}$);
		
		\quad\quad \textbf{Return} ({\fontfamily{cmss}\selectfont m}, {\fontfamily{cmss}\selectfont msk}).
		
		\textit{\# can add attack detection or scoring functionality here}\\
		5. \textbf{\blue{On input}} (\purple{``aggregation"}, $\widetilde{\text{\fontfamily{cmss}\selectfont m}}$):

		\quad\quad {\fontfamily{cmss}\selectfont m}$_{\mathrm{glob}}$ := Aggregate($\widetilde{\text{\fontfamily{cmss}\selectfont m}}$));
		
		\quad\quad \textbf{Return} {\fontfamily{cmss}\selectfont m}$_{\mathrm{glob}}$.
		
		6. \textbf{\blue{On input}} (\purple{``encrypt"}, {\fontfamily{cmss}\selectfont m}$_{\mathrm{glob}}$, {\fontfamily{cmss}\selectfont msk}):
		
		\quad\quad {\fontfamily{cmss}\selectfont ct}$_{\mathrm{out}}$ := Enc({\fontfamily{cmss}\selectfont msk}, {\fontfamily{cmss}\selectfont m}$_{\mathrm{glob}}$);
		
		\quad\quad \textbf{Return} {\fontfamily{cmss}\selectfont ct}$_{\mathrm{out}}$.
		
		7. \textbf{\blue{On input}} (\purple{``sign"}, {\fontfamily{cmss}\selectfont m}$_{\mathrm{glob}}$, {\fontfamily{cmss}\selectfont round}, {\fontfamily{cmss}\selectfont vk}$_{\mathrm{sk}}$):
		
		\quad\quad {\fontfamily{cmss}\selectfont $\sigma$}$_{\mathrm{sgx}}$ := Sign({\fontfamily{cmss}\selectfont vk}$_{\mathrm{sk}}$, ({\fontfamily{cmss}\selectfont ct}$_{\mathrm{out}}$, {\fontfamily{cmss}\selectfont round}));
		
		\quad\quad \textbf{Return} {\fontfamily{cmss}\selectfont $\sigma$}$_{\mathrm{sgx}}$.
	\end{framed}
	
	\caption{Contract Enclave Wrapper}
	\label{Enclave}
	\vspace{-0.455cm}
\end{figure}

{\fontfamily{cmss}\selectfont Con}$_{\textnormal{encl}}$ is the code running on SGX in $\mathcal{\brown{N}}$. It is the core content to perform the FL aggregation. We give the wrapper of the contract in Fig. \ref{Enclave}. Users need to provide the following specific functions according to the logic specified by the wrapper to meet the running requirements of Voltran: 1. Encryption and decryption; 2. Attack detection/scoring (optional); 3. Aggregation; 4. Digital signature. In essence, Voltran does not provide a service of secure aggregation but an environment capable of performing secure, trusted and privacy-preserving computations. Our framework does not restrict any data, model or algorithm. Users can arbitrarily choose any secure aggregation algorithm to design their contracts according to their personalized demands.

\begin{figure}[t]
	\label{contract Storage}
	\begin{framed}
		\centerline{\textbf{Contract Storage Wrapper {\fontfamily{cmss}\selectfont Con}$_{\textnormal{stor}}$}}
		1:  \textbf{\blue{On input}} (\purple{``create'', {\fontfamily{cmss}\selectfont eid}, {\fontfamily{cmss}\selectfont taskid}}):
		
		\quad\quad Set cid := {\fontfamily{cmss}\selectfont eid}  
		
		\quad\quad Set {\fontfamily{cmss}\selectfont round} := 0;
		
		\quad\quad Set {\fontfamily{cmss}\selectfont taskid} := {\fontfamily{cmss}\selectfont taskid};

		2:  \textbf{\blue{On input}} (\purple{``uploadPK''}, {\fontfamily{cmss}\selectfont vk}$_{\mathrm{pk}}$):
		
		\quad\quad Assert msg.sender = \textit{Comm};
		
		\quad\quad {\fontfamily{cmss}\selectfont vk}$_{\mathrm{pk}}$ := {\fontfamily{cmss}\selectfont vk}$_{\mathrm{pk}}$.
		
		\quad\quad \textbf{Return} \texttt{TRUE}.

		4:  \textbf{\blue{On input}} (\purple{``uploadGlobalModel''}, {\fontfamily{cmss}\selectfont ct}$_{\mathrm{out}}$, {\fontfamily{cmss}\selectfont $\sigma$}$_{\mathrm{sgx}}$, {\fontfamily{cmss}\selectfont round'}, 
		
		\quad {\fontfamily{cmss}\selectfont taskid'}, {\fontfamily{cmss}\selectfont index}):
		
		\quad\quad Assert {\fontfamily{cmss}\selectfont taskid'} = {\fontfamily{cmss}\selectfont taskid};
		{\fontfamily{cmss}\selectfont round'} = {\fontfamily{cmss}\selectfont round};

		\quad\quad Verify ({\fontfamily{cmss}\selectfont vk}$_{\mathrm{pk}}$, {\fontfamily{cmss}\selectfont ct}$_{\mathrm{out}}$, {\fontfamily{cmss}\selectfont $\sigma$}$_{\mathrm{sgx}}$, {\fontfamily{cmss}\selectfont round'}) = \texttt{TRUE};
		
		\quad\quad {\fontfamily{cmss}\selectfont Stroage[round][index]} $:=$ {\fontfamily{cmss}\selectfont ct}$_{\mathrm{out}}$.
		
		\quad\quad \textbf{Return} \texttt{TRUE}.
		
		\textit{\# can add the reward/punishment functionality here.}

	\end{framed}
	\caption{Contract Storage Wrapper}
	\label{Storage}
	\vspace{-0.2cm}
\end{figure}

{\fontfamily{cmss}\selectfont Con}$_{\textnormal{stor}}$ is the common smart contract deployed on the blockchain. It is built to store the ciphertext of an aggregated global model of each round and verify its authentication. The $\texttt{Con}_{\mathrm{Stor}}$ wrapper is proposed in Fig. \ref{Storage}. {\fontfamily{cmss}\selectfont Con}$_{\textnormal{stor}}$ is also capable of implementing the incentive/punishment mechanism for $\mathcal{\brown{C}}$ and $\mathcal{\brown{N}}$ to raise enthusiasm and guarantee honesty due to the requirements of $\mathcal{O}$. 

Note that our proposed wrappers for these two contracts are the minimum requirements for implementing the contract functions. Users can extend more functions, such as attack detection or reward and punishment mechanisms, by implementing more composite sub-contracts.

\subsection{Security Analysis}

Voltran can realize the security goals of \textit{Authenticity} and \textit{Confidentiality}. We give simple formal definitions of them in Definition 1-2 and Theorem 1-2 characterize how $\textbf{Prot}_{\texttt{Volt}}$ capture these properties. We give an abbreviated proof here. The complete formal definitions and security proof sketch is in Appendix C.

\textbf{Definition 1} (\textit{Authenticity}). We say that $\texttt{Prot}_{\mathrm{Volt}}$ satisfies \textit{authenticity} if, for any polynomial-time adversary \brown{$\mathcal{A}$} that can interact arbitrarily with $\texttt{Prot}_{\mathrm{Volt}}$, \brown{$\mathcal{A}$} cannot cause an honest verifier to accept the following three situations:
\begin{enumerate}
\item \brown{$\mathcal{A}$} forges $\mathcal{C}$ to send a ``sendModel'' message with a dummy input ({\fontfamily{cmss}\selectfont ct}$^{\prime}$ = ({\fontfamily{cmss}\selectfont ct}$^{\prime}_{\mathrm{m}_i}$, {\fontfamily{cmss}\selectfont ct}$^{\prime}_{\mathrm{msk}}$), {\fontfamily{cmss}\selectfont id} = ({\fontfamily{cmss}\selectfont taskid}, {\fontfamily{cmss}\selectfont round}));
\item \brown{$\mathcal{A}$} forges $\mathcal{N}$ to install a dummy {\fontfamily{cmss}\selectfont prog}$^{\prime}_{\mathrm{Encl}}$ on SGX;
\item \brown{$\mathcal{A}$} forges $\mathcal{F}_{\mathrm{sgx}}$ to send a ``uploadGlobalModel'' message with a dummy input ({\fontfamily{cmss}\selectfont outp}$^{\prime}$ = ({\fontfamily{cmss}\selectfont ct}$^{\prime}_{\mathrm{out}}$, {\fontfamily{cmss}\selectfont $\sigma$}$^{\prime}_{\mathrm{sgx}}$), {\fontfamily{cmss}\selectfont param} = ({\fontfamily{cmss}\selectfont round}, {\fontfamily{cmss}\selectfont taskid}, {\fontfamily{cmss}\selectfont index})).
\end{enumerate}

\textbf{Theorem 1} (\textit{Authenticity}). Assume that the RA mechanism of Intel SGX is secure and the signature algorithm is existentially unforgeable under chosen message attacks (EU-CMA), then $\texttt{Prot}_{\mathrm{Volt}}$ achieves \textit{authenticity} under Definition 2.

\textit{Proof.} Considering the three cases in Definition 2, the adversary $\mathcal{A}$ needs to violate at least one of the security of $\mathcal{F}_{\mathrm{sgx}}$ and the EU-CMA security of the signature $\Sigma$ to forge a dummp input. Hence, the \textit{authenticity} is proved.

\textbf{Definition 2} (\textit{Confidentiality}). We say that $\texttt{Prot}_{\mathrm{Volt}}$ satisfies \textit{confidentiality} if, for any polynomial-time adversary \brown{$\mathcal{A}$} that can interact arbitrarily with $\texttt{Prot}_{\mathrm{Volt}}$, \brown{$\mathcal{A}$} cannot obtain information about the plaintexts from ciphertexts during the protocol execution under the chosen plaintext attack (CPA) security. It requires an attacker cannot reveal the encapsulated key from the ciphertexts.

\textbf{Theorem 2} (\textit{Confidentiality}). Assume that the encryption algorithms of $\mathcal{F}_{sgx}$ and $\mathcal{SE}$ are IND-CPA secure, then the protocol achieves \textit{confidentiality} under Definition 3.

\textit{Proof.} According to Definition 3, the adversary $\mathcal{A}$ needs to break the IND-CPA security of $\mathcal{SE}$ to obtain information. Hence, the \textit{confidentiality} is proved.

	\subsection{Correctness Analysis}
	
	We also analyze the desired properties of \textit{Correctness}. Similarly, the complete definitions and proof sketch in Appendix C. 
	
	\textbf{Definition 3} (\textit{Correctness}). We say that $\texttt{Prot}_{\mathrm{Volt}}$ satisfies \textit{correctness} if, for any polynomial-time adversary \brown{$\mathcal{A}$} that can interact arbitrarily with $\texttt{Prot}_{\mathrm{Volt}}$, \brown{$\mathcal{A}$} cannot cause an honest party to return a wrong result the following two situations:
	\begin{enumerate}
		\item \brown{$\mathcal{A}$} forces an enclave to return a dummy output \text{\fontfamily{cmss}\selectfont outp}$^{\prime}$ with a specific input ({\fontfamily{cmss}\selectfont ct} =({\fontfamily{cmss}\selectfont ct}$_{\mathrm{m}_i}$, {\fontfamily{cmss}\selectfont ct}$_{\mathrm{msk}}$), {\fontfamily{cmss}\selectfont id} = ({\fontfamily{cmss}\selectfont round}, {\fontfamily{cmss}\selectfont taskid})) from \brown{$\mathcal{C}$};
		\item \brown{$\mathcal{A}$} forces $\mathcal{F}_{\mathrm{blockchain}}$ to store a dummy input (\text{\fontfamily{cmss}\selectfont outp}$^{\prime}$ = ({\fontfamily{cmss}\selectfont ct}$^{\prime}_{\mathrm{out}}$, {\fontfamily{cmss}\selectfont $\sigma$}$^{\prime}_{\mathrm{sgx}}$), \text{\fontfamily{cmss}\selectfont id} = ({\fontfamily{cmss}\selectfont round}, {\fontfamily{cmss}\selectfont taskid}, {\fontfamily{cmss}\selectfont index})) with (\text{\fontfamily{cmss}\selectfont outp} = ({\fontfamily{cmss}\selectfont ct}$_{\mathrm{out}}$, {\fontfamily{cmss}\selectfont $\sigma$}$_{\mathrm{sgx}}$), \text{\fontfamily{cmss}\selectfont id} = ({\fontfamily{cmss}\selectfont round},  {\fontfamily{cmss}\selectfont taskid}, {\fontfamily{cmss}\selectfont index})) from \brown{$\mathcal{N}$}.
	\end{enumerate}

	\textbf{Theorem 3} (\textit{Correctness}). Assume $\mathcal{F}_{\mathrm{sgx}}$ and $\mathcal{F}_{\mathrm{blockchain}}$ are secure, then $\texttt{Prot}_{\mathrm{Volt}}$ achieves \textit{correctness} under Definition 1.
	
	\textit{Proof.} Considering the two cases in Definition 1, the adversary $\mathcal{A}$ needs to violate at least one of the security of $\mathcal{F}_{\mathrm{sgx}}$ and $\mathcal{F}_{\mathrm{blockchain}}$ to generate the wrong output. Hence, the \textit{correctness} is proved.

\section{Implementation}\label{implement}
\subsection{Setup}

We build an end-to-end instantiation of Voltran Voltran-Fabric by choosing Fabric Hyperledger V2.4.6 as the blockchain with Fabric Java SDK. We build up the communication module in JAVA to invoke APIs to deploy contracts and send transactions. We employ a server equipped with an Intel$^\circledR$ Xeon$^\circledR$ Gold 6330 CPU @ 2.00 GHz, which supports Intel SGX as execution nodes. This server runs on a Linux OS, Ubuntu 20.04.3. The training process of clients is simulated through multi-thread programming on a server equipped with a GPU with seven cores and 16 GB. We use Python to implement local training and recover a new round of the global model from the blockchain. Our TEE module is indeed implemented in C++, including the enclaves, remote attestation, and APIs. {\fontfamily{cmss}\selectfont Con}$_{\textnormal{stor}}$ is implemented in Golang, particularly including a signature verification implementation. We leverage AES-GCM as the encryption algorithm with a 128-bit key length \cite{kasper2009faster}. We utilize ECDSA and implement it using the NIST p-256 curve \cite{adalier2015efficient} for our digital signature algorithm. The interaction between Python codes (i.e., client side) and C++ codes (i.e., SGX) is realized through a socket-based remote procedure call (RPC) implementation on each side. All experiments are done with 10\% of the total number of clients participating in each round. We conduct our experiments on our prototype to measure system performance. Each FL task has been executed five times to reduce errors due to chance events. The implementation is open-source in the github\footnote{https://github.com/W-ScorPioN/Voltran.}.

\subsection{Model and Dataset}
We use five datasets with six models including various tasks, such as image classification and natural language processing (NLP), to conduct experimental results in Table I. The aggregation algorithm is the classic algorithm \textit{FedAvg} \cite{mcmahan2017communication} except for some evaluations using specific secure aggregation algorithms.

\begin{table}

\vspace{-0.15cm}
\caption{Detailed information for FL tasks evaluated on Voltran.}
\label{dataset}
\scalebox{1.01}{
	\begin{tabular}{cccccc}
		\toprule
		No. & Model    & Dataset  & Parameters & Size     & Rounds \\ \midrule
		1   & MLP      & Adult    & 10,901     & 42.58 KB & 50     \\
		2   & CNN      & MNIST    & 22,340     & 85.31 KB & 50     \\
		3   & ResNet18 & CIFAR-10 & 11.18M & 42.64 MB & 50     \\
		4   & ResNet50 & CelebA   & 21.29M & 81.20 MB & 30     \\ 
		5   & AlexNet  & CIFAR-10 & 1.25M & 4.76 MB & 50 \\
		6   & Bert & THUCNews   & 97.54M
		& 390.16MB & 30     \\ 
		\bottomrule
\end{tabular}}
\vspace{-0.3cm}
\end{table}

\section{Evaluation and Discussion}\label{evaluation}
In this section, we present the experimental evaluation of Voltran by answering a set of key questions. First, we propose questions A-C to demonstrate Voltran's feasibility. Then, we evaluate Voltran's additional overhead by questions D and E. Finally, we present our scalability by questions F-H.

\subsection{What are the advantages and disadvantages of Voltran compared to other solutions?}
To help readers better understand the comparative advantages of Voltran, we present a literature comparison to highlight Voltran's superiority In Table \ref{compare}.

\begin{table*}[]
\caption{Literature Comparison between Voltran and related work.}
\label{compare}
\centering
\setlength{\tabcolsep}{1.8mm}{
	\begin{tabular}{cccccccc}
		\toprule
		Scheme          & Confidential?                          & Decentralized? & Practical? & Integrity                                         & Scalable?    & Parallel? & Hardware-based?\\
		\midrule

		\multirow{2}{*}{Voltran }                  & \multirow{2}{*}{\checkmark}                                       & \multirow{2}{*}{\checkmark} & \multirow{2}{*}{\checkmark}              & \multirow{2}{*}{\checkmark}                                                     & \multirow{2}{*}{\checkmark}            & \multirow{2}{*}{\checkmark}    & \multirow{2}{*}{\checkmark}            \\
		&&&&&&&
		\\\midrule
		
		\multirow{2}{*}{Scheme \cite{ChuanMa2020WhenFL}}           & Performance loss by & \multirow{2}{*}{\checkmark}                & Impractical on-                      & \multirow{2}{*}{\checkmark}& \multirow{2}{*}{\checkmark}             & \multirow{2}{*}{×}       & \multirow{2}{*}{×}          \\
		&  differential privacy &  & chain computation &&&&
		
		\\\midrule
		
		\multirow{2}{*}{Scheme \cite{LingchenZhao2021SEARSA}}           & \multirow{2}{*}{\checkmark}                                         & \multirow{2}{*}{×}    & \multirow{2}{*}{\checkmark}                                                    & \multirow{2}{*}{\checkmark} & \multirow{2}{*}{Not mentioned }  & \multirow{2}{*}{×}     & \multirow{2}{*}{\checkmark}              \\
		
		&&&&&&
		\\\midrule
		
		
		\multirow{2}{*}{Scheme \cite{bao2019flchain}}         & \multirow{2}{*}{×}                                        & \multirow{2}{*}{\checkmark}                 & \multirow{2}{*}{\checkmark}                      &\multirow{2}{*}{\checkmark} & \multirow{2}{*}{Not mentioned } & \multirow{2}{*}{×}            & \multirow{2}{*}{×}      \\
		&&&&&&&
		
		\\\midrule

		\multirow{2}{*}{Scheme \cite{wei2020federated}} & Performance loss by      & \multirow{2}{*}{×}               & \multirow{2}{*}{\checkmark}    &\multirow{2}{*}{×}     &   \multirow{2}{*}{Not mentioned }       & \multirow{2}{*}{×}     & \multirow{2}{*}{×}            \\
		
		&differential privacy&&&&&
		\\\midrule
		
          & ×       & ×           \\
		
		\multirow{2}{*}{Scheme \cite{miao2022privacy}}  & Based on strong       &  \multirow{2}{*}{×}               & Limited usage of block-  & \multirow{2}{*}{\checkmark} &\multirow{2}{*}{Not mentioned }               & \multirow{2}{*}{×}       & \multirow{2}{*}{×}           \\
		
		&trust assumptions&&chain for only storage&&&
		\\\midrule
		
		\multirow{2}{*}{Scheme \cite{kalapaaking2022blockchain}}  & Data leakage on       & \multirow{2}{*}{\checkmark}                & Impractical TEE    & \multirow{2}{*}{\checkmark}              & \multirow{2}{*}{\checkmark}     &\multirow{2}{*}{×}& \multirow{2}{*}{\checkmark}             \\
		
		& blockchain nodes &&verification and demands &&&&
		\\\midrule

		\multirow{2}{*}{Scheme \cite{jeon2021privacy}}   & Trade-off between privacy  & \multirow{2}{*}{\checkmark}                & \multirow{2}{*}{\checkmark}  & \multirow{2}{*}{×} & \multirow{2}{*}{Not mentioned }              & \multirow{2}{*}{×}       & \multirow{2}{*}{×}          \\
		
		&  and time overhead &&&&&&
		\\\midrule
		
		\multirow{2}{*}{Scheme \cite{bonawitz2017practical}}   &\multirow{2}{*}{\checkmark}& \multirow{2}{*}{×}               & \multirow{2}{*}{\checkmark}  & \multirow{2}{*}{\checkmark} & \multirow{2}{*}{×}              & \multirow{2}{*}{×}       & \multirow{2}{*}{×}          \\
		
		& &&&&&&
		\\\midrule
		
		\multirow{2}{*}{Scheme \cite{chen2022feddual}}   &\multirow{2}{*}{\checkmark}& \multirow{2}{*}{\checkmark}              & \multirow{2}{*}{\checkmark}  & \multirow{2}{*}{×}& \multirow{2}{*}{×}              & \multirow{2}{*}{×}       & \multirow{2}{*}{×}          \\
		
		& &&&&&&
		\\
		
		\midrule
		
		\multirow{2}{*}{Scheme \cite{zhou2022privacy}}   &\multirow{2}{*}{\checkmark}& \multirow{2}{*}{×}             & \multirow{2}{*}{\checkmark}  & \multirow{2}{*}{\checkmark}& \multirow{2}{*}{×}              & \multirow{2}{*}{×}       & \multirow{2}{*}{×}          \\
		
		& &&&&&&
		\\

		\bottomrule              
	\end{tabular}
}
\end{table*}

Similar to Voltran, schemes \cite{ChuanMa2020WhenFL, bao2019flchain, kalapaaking2022blockchain} pay attention to decentralized FL. Scheme \cite{ChuanMa2020WhenFL} utilize differential privacy to achieve confidentiality, which may lead to the trade-off between privacy and model performance. Also, they put the aggregation computation on the chain directly, which may be an impractical design because the blockchain makes it hard to process large-scale computation. Scheme \cite{bao2019flchain} does not consider the confidentiality of model updates. Scheme \cite{kalapaaking2022blockchain} also leverages the blockchain and TEE to realize DFL. However, their design leads to data leakage on the blockchain nodes, and they do not separate TEE and the blockchain, which means each blockchain node has to be equipped with a TEE, making their work impractical. Furthermore, their TEE result verification mechanism is based on the IAS, which needs a node to access the off-chain environment. This operation requires the node's own subjective judgment of correctness rather than automatic execution by smart contracts, which makes this design not feasible from a practical point of view. Scheme \cite{LingchenZhao2021SEARSA} proposes a TEE-based secure aggregation scheme. Unlike us, they perform aggregation based on a single TEE node, resulting in a single point of failure problem. Scheme \cite{wei2020federated} presents a privacy-preserving FL scheme based on differential privacy, which has the same problem with \cite{ChuanMa2020WhenFL}. Also, they do not mention the decentralization. Scheme \cite{miao2022privacy} proposes a homomorphic encryption-based FL scheme. Their confidentiality relies on the strong trust assumptions of the Verifier and Solver in their design. Also, they leverage the blockchain only to store procedure values, which does not improve their reliability. Scheme \cite{jeon2021privacy} applies the Alternating Direction Method of Multiplier algorithm for DFL without introducing extra components. Similar to differential privacy, their privacy-preserving strength also involves a trade-off with a ``gap'', meaning that the greater the privacy protection strength, the larger the communication overhead required. In addition, Voltran presents a multi-SGX parallel execution mode, which is unique from other schemes. However, our scheme is based on hardware support due to the usage of TEE.

\subsection{What is the performance gap between Voltran and the vanilla FL training?}
To compare the performance gap between Voltran and the vanilla FL, we conduct experiments to measure the model performance and total aggregation time on tasks 1-4 in Table \ref{dataset}. We vary the number of clients to 10, 50, 100, and 500. The experimental results are shown in Table \ref{tab:performance}. The table presents the accuracy of the models as denoted by $Acc$ and the total aggregation time denoted by $T_{agg}$. 

\begin{table*}  
\caption{Comparison between Voltran and the vanilla FL on model accuracy and aggregation time of one round. $T_{agg}$ is measured in milliseconds.}  
\label{tab:performance}  
\resizebox{\textwidth}{22mm}{  
	\begin{tabular}{cccccccccc}  
		\toprule  
		\multirow{2}{*}{Model} & \multirow{2}{*}{Paradigm} & \multicolumn{2}{c}{10 Clients} & \multicolumn{2}{c}{50 Clients} & \multicolumn{2}{c}{100 Clients} & \multicolumn{2}{c}{500 Clients}\\  
		&  & $Acc$ & $T_{agg}$ & $Acc$ & $T_{agg}$ & $Acc$ & $T_{agg}$ & $Acc$ & $T_{agg}$ \\  
		\midrule  
		
		\multirow{2}{*}{MLP} & \makecell[c]{FL} & 85.64 ± 0.01 & 0.27 ± 0.03 & 85.82 ± 0.06 & 0.52 ± 0.09 & 85.75 ± 17.61 & 0.83 ± 0.14 & 85.04 ± 0.06 & 3.31 ± 0.41 \\  
		& Voltran & 85.66 ± 0.05 & 4.62 ± 0.37 & 85.72 ± 0.02 & 15.01 ± 0.71 & 85.53 ± 0.04 & 31.33 ± 1.06 & 85.15 ± 0.07 & 123.62 ± 3.58	 \\  
		\midrule   
		
		\multirow{2}{*}{CNN} & FL &  99.09 ± 0.04 & 0.60 ± 0.02 & 98.87 ± 0.02 & 0.94 ± 0.08 & 98.53 ± 0.02 & 1.61 ± 0.41 & 96.71 ± 0.13 & 6.66 ± 0.94 \\  
		& Voltran &  98.93 ± 0.03 & 6.91 ± 0.15 & 98.87 ± 0.05 & 26.13 ± 0.53 & 98.55 ± 0.05 & 53.32 ± 0.94 & 96.42 ± 0.21 & 265.92 ± 2.79 \\  
		\midrule  
		
		\multirow{2}{*}{ResNet18} & FL & 75.43 ± 0.16 & 64.69 ± 1.94 & 72.97 ± 0.27 & 145.87 ± 3.08 & 70.45 ± 0.38 & 266.70 ± 4.27 & 60.17 ± 0.28 & 1201.90 ± 11.69 \\  
		& Voltran & 75.06 ± 0.43 & 258.84 ± 8.74 & 72.68 ± 0.33 & 1069.68 ± 13.48 & 70.32 ± 0.34 & 2198.43  ± 31.55 & 60.05 ± 0.13  & 10207.73 ± 97.20  \\  
		\midrule   
		
		\multirow{2}{*}{ResNet50} & FL & 75.43 ± 0.16 & 108.83 ± 6.41 & 72.97 ± 0.27 & 289.49 ± 9.98 & 70.45 ± 0.38 & 536.04 ± 20.04 & 60.17 ± 0.28 & 2108.59 ± 76.96 \\  
		& Voltran & 75.06 ± 0.43 & 289.24 ± 15.45 & 72.68 ± 0.33 & 1501.05 ± 91.01 & 70.32 ± 0.34 & 3054.16  ± 167.77 & 60.05 ± 0.13  & 14471.57 ± 405.47  \\  
		\midrule   
		
		\multirow{2}{*}{Bert} & FL & 97.31 ± 1.22 &  409.32 ± 62.31 & 97.50 ± 0.98 & 1421.07 ± 28.68  & 97.08 ± 1.02 &  1941.95 ± 216.03 & 96.11 ± 1.29 &   8672.21 ± 412.98 \\  
		& Voltran & 97.30 ± 1.63 & 1661.62 ± 47.92 & 97.49 ± 1.16 &  5912.43 ± 341.98 & 97.12 ± 0.76 & 8647.67 ± 496.98   & 95.97 ± 1.44  &  26534.12 ± 906.98  \\  
		\bottomrule  
		
	\end{tabular}  
	\vspace{-0.5cm}
}  
\end{table*}  

\subsubsection{Model performance}
Table \ref{tab:performance} illustrates that the performance disparity between models executed on Voltran and vanilla FL is marginal, with an accuracy difference of less than 1\%. This observation indicates that the Voltran framework does not introduce a significant deviation in model performance. Therefore, Voltran demonstrates its feasibility without compromising the accuracy of the model.

\subsubsection{Time performance}
We also illustrate the time cost of the aggregation gap between Voltran and vanilla FL in Table \ref{tab:performance}. Due to cryptographic operations and additional time overhead brought by SGX, Voltran's aggregation time is longer than vanilla FL. Deeply, the gap becomes larger when the model size gets larger. This is because the larger data amount brings more cryptographic operations. Meanwhile, if the data amount exceeds the maximum enclave capacity, it needs EPC paging, which takes more time. However, we claim our platform meets the feasibility criteria because Voltran achieves confidentiality-preserving aggregation, and compared to other state-of-the-art privacy-preserving aggregation schemes, Voltran presents higher efficiency. We display the experimental results and analysis in the following question D.

\subsection{How do two scheduling modes affect the FL performance?}

\begin{figure*}[h]
\centering
\begin{subfigure}[b]{0.244\textwidth}
	\includegraphics[width=\textwidth]{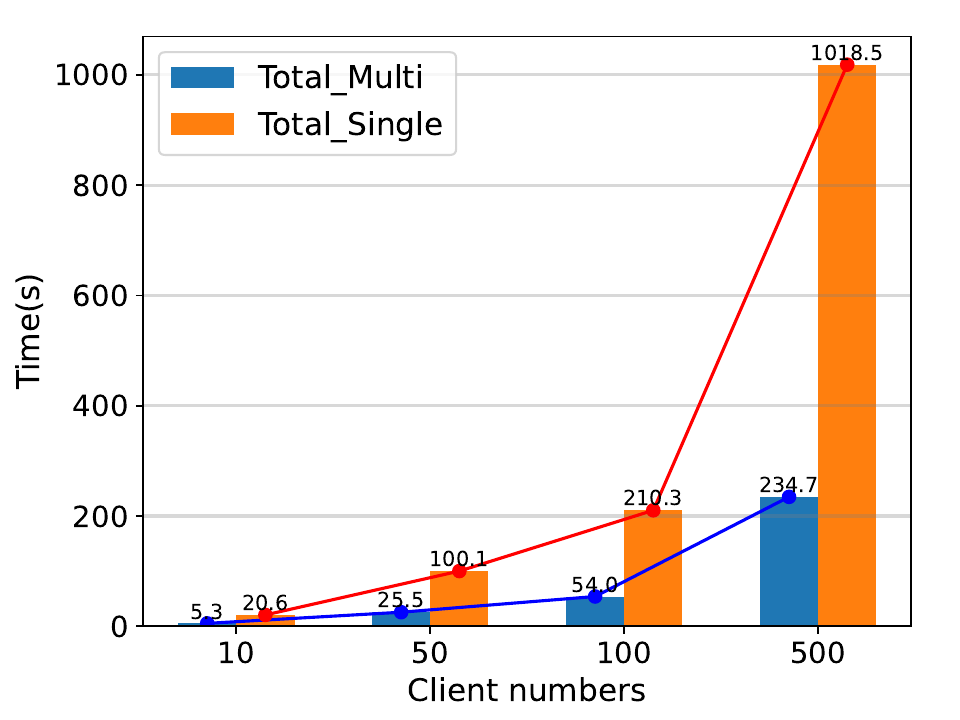}
	\caption{SendModeltoSGX}
	\label{SendModeltoSGX}
\end{subfigure}
\hfill
\begin{subfigure}[b]{0.244\textwidth}
	\includegraphics[width=\textwidth]{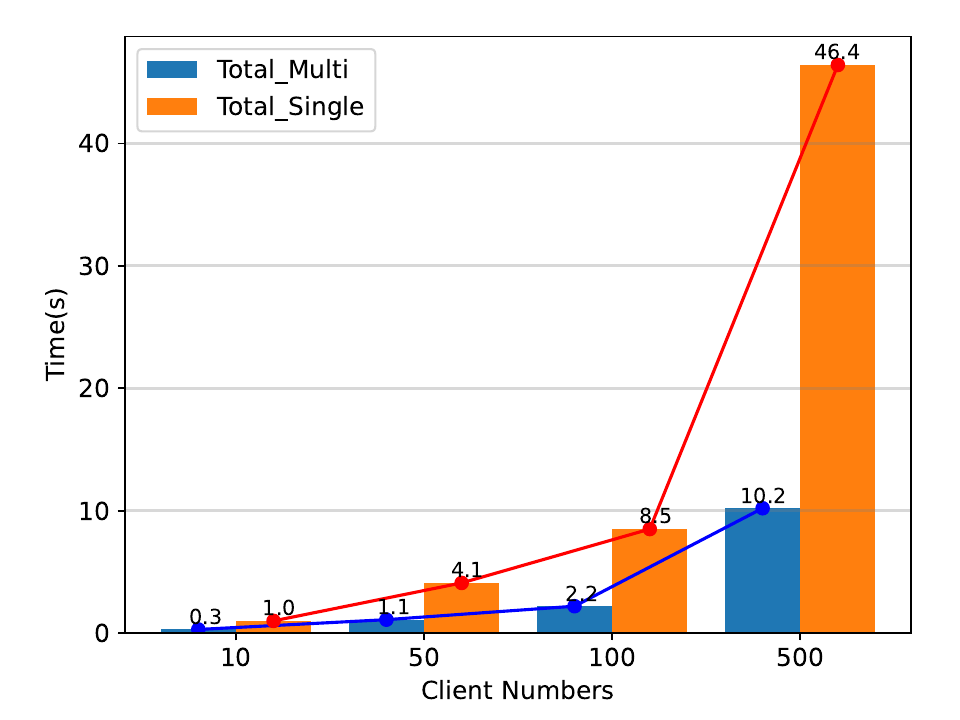}
	\caption{Aggregate}
	\label{Aggregate}
\end{subfigure}
\begin{subfigure}[b]{0.244\textwidth}
	\includegraphics[width=\textwidth]{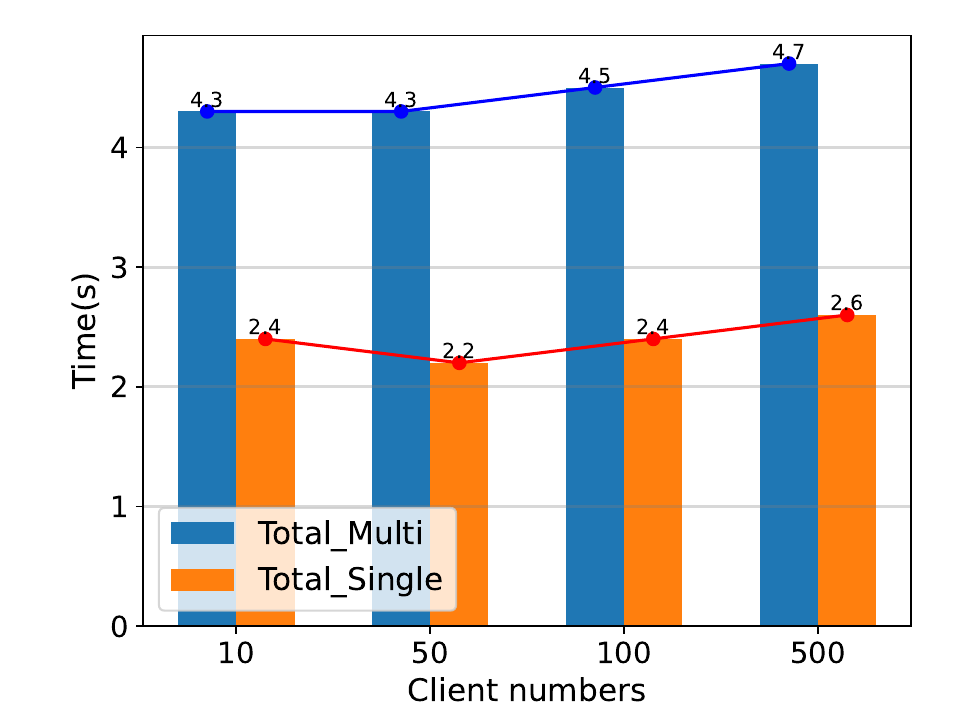}
	\caption{SendResulttoChain}
	\label{SendResulttoChain}
\end{subfigure}
\hfill
\begin{subfigure}[b]{0.244\textwidth}
	\includegraphics[width=\textwidth]{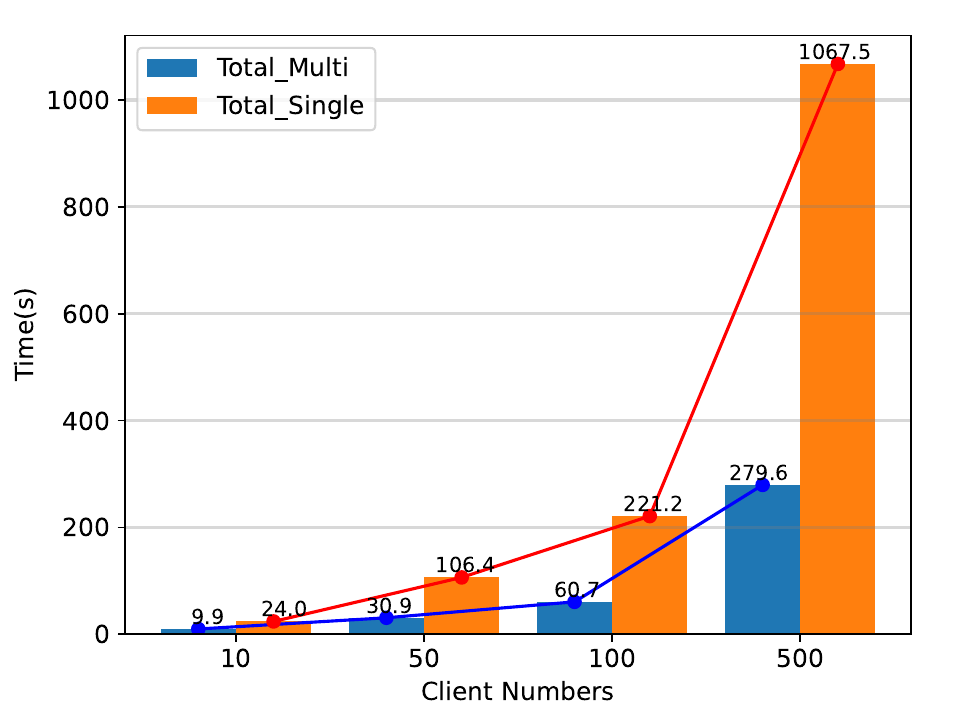}
	\caption{Total}
	\label{Total}
\end{subfigure}
\caption{Time cost of each step between single-SGX and multi-SGX on Resnet18.}
\label{multisingle}
\end{figure*}

Voltran provides 1 to \textit{n} execution nodes with SGX to provide secure computation service for FL aggregation, which brings two execution modes: parallel processing of multiple SGXs and sequential execution of one single SGX. The mode choice depends on the task's size and the node's liveness. One-SGX execution can undertake tasks based on models with fewer parameters because network bandwidth and SGX EPC paging bring little influence to efficiency. When large models with more layers and computational costs are encountered, the multi-SGX parallel strategy will take effect. We test the performance of the two scheduling strategies for different numbers of clients under ResNet18. Fig. \ref{multisingle} shows the experimental results. The parallel strategy becomes more and more advantageous than individual execution as the client number gets larger. The reason is that parallel execution reduces the bandwidth limitation and the number of times SGX performs EPC paging, significantly improving execution efficiency. 

In summary, Voltran can take the single SGX execution strategy for simple tasks, which simplifies our scheme to single TEE-based FL schemes such as \cite{LingchenZhao2021SEARSA}. For large workloads, the multi-SGX execution strategy can greatly improve efficiency. It is a \textbf{unique} solution for Voltran compared to TEE-based aggregation schemes \cite{LingchenZhao2021SEARSA,  kalapaaking2022blockchain}. In addition, with the presence of the committee mechanism, when a single SGX generates a single point of failure, Voltran also provides the replacement to guarantee that tasks continue to execute. 

\subsection{Is the Voltran performance comparable to state-of-the-art privacy-preserving aggregation schemes?}

As Voltran can achieve secure aggregation, we compare its efficiency with state-of-the-art privacy-preserving aggregation schemes \cite{ChuanMa2020WhenFL,liu2022efficient,LingchenZhao2021SEARSA, zhang2020batchcrypt} to evaluate its performance. BatchCrypt \cite{zhang2020batchcrypt} is a homomorphic encryption (HE) secure aggregation scheme using Paillier encryption based on cross-silo FL. Liu \textit{et al.} \cite{liu2022efficient} proposed a privacy-preserving aggregation scheme based on the additive homomorphic property of Shamir's secret sharing scheme, which is one of secure multi-party computation (MPC) techniques. Ma \textit{et al.} \cite{ChuanMa2020WhenFL} utilizes differential privacy (DP) on models for privacy protection. Zhao \textit{et al.} \cite{LingchenZhao2021SEARSA} sends the ciphertext models into TEE by encryption. In contrast, Voltran achieves aggregation on plaintexts in SGXs and provides parallel multi-SGX execution, which theoretically can be faster than methods based on ciphertext or complex protocols without any performance loss. 

\textbf{Voltran vs. HE, MPC and TEE }
To ensure comparability across schemes, we adjust the key length bits to maintain the same level of security strength and take the same models and datasets as the original experiments in their papers. Based on these preparations, we perform and derive the experimental results shown in Table \ref{sota}. We see that Voltran significantly speeds up the aggregation runtime: \textbf{6.194$\times$} for AlexNet with 50 clients against BatchCrypt and \textbf{199.78$\times$} for ResNet18 with 500 clients against \cite{liu2022efficient}. Moreover, the total runtime in one FL round is also accelerated. It implies that as the increase in the number of clients and model size brings a corresponding rise in the volume of data, the advantage of Voltran’s plaintext-based aggregation becomes even more pronounced. Compared to schemes that leverage TEE for aggregation, such as SEAR \cite{LingchenZhao2021SEARSA}, our proposed multi-SGX parallel execution strategy significantly reduces computation and communication time.

\begin{table}
\caption{Comparison of Voltran and schemes based on HE, MPC and TEE on time overhead and accuracy.}
\label{sota}
\scalebox{0.995}{
	\begin{tabular}{cccccc}
		\toprule
		Scheme              & Clients     & Model   & $T_{agg}$(s)   & $T_{total}$(s) & \textit{Acc}              \\ 
		\midrule
		BatchCrypt \cite{zhang2020batchcrypt}   &\multirow{2}{*}{50}  & \multirow{2}{*}{AlexNet}   & 1.1       &27.453 &73.97\\
		
		Voltran &     &   & \textbf{0.178}      &  \textbf{15.579}      &74.08            \\ 
		\midrule
		
		Liu \textit{et al.} \cite{liu2022efficient}  &\multirow{3}{*}{500}        & \multirow{3}{*}{ResNet18}  & 2039.3  & 3057.8  & 60.07\\
		
		SEAR \cite{LingchenZhao2021SEARSA} &   & & 16.44 & 1032.8 & 60.01\\ 		
		
		Voltran&     &  & \textbf{10.21}   &    \textbf{279.6}           &   60.05       \\ 
		
		\bottomrule
\end{tabular}}
\vspace{-0.3cm}
\end{table}

\begin{figure}[t]
\centering
\begin{subfigure}[b]{0.24 \textwidth}
	\includegraphics[width=\textwidth]{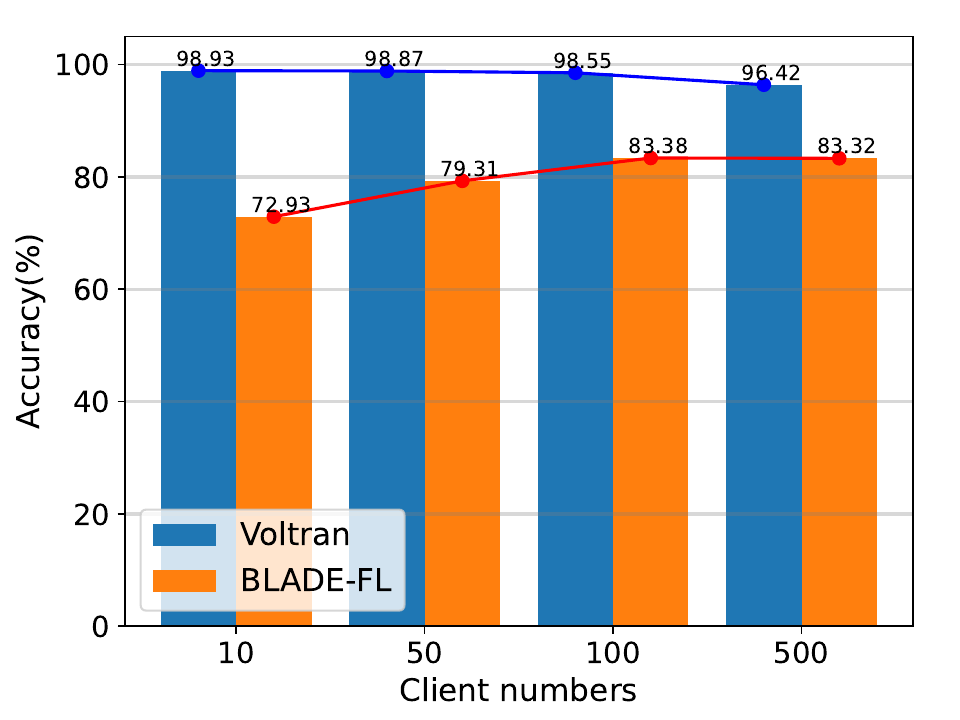}
	\caption{AccComparisonWithDP}
	\label{AccComparisonWithDP}
\end{subfigure}
\hfill
\begin{subfigure}[b]{0.24 \textwidth}
	\includegraphics[width=\textwidth]{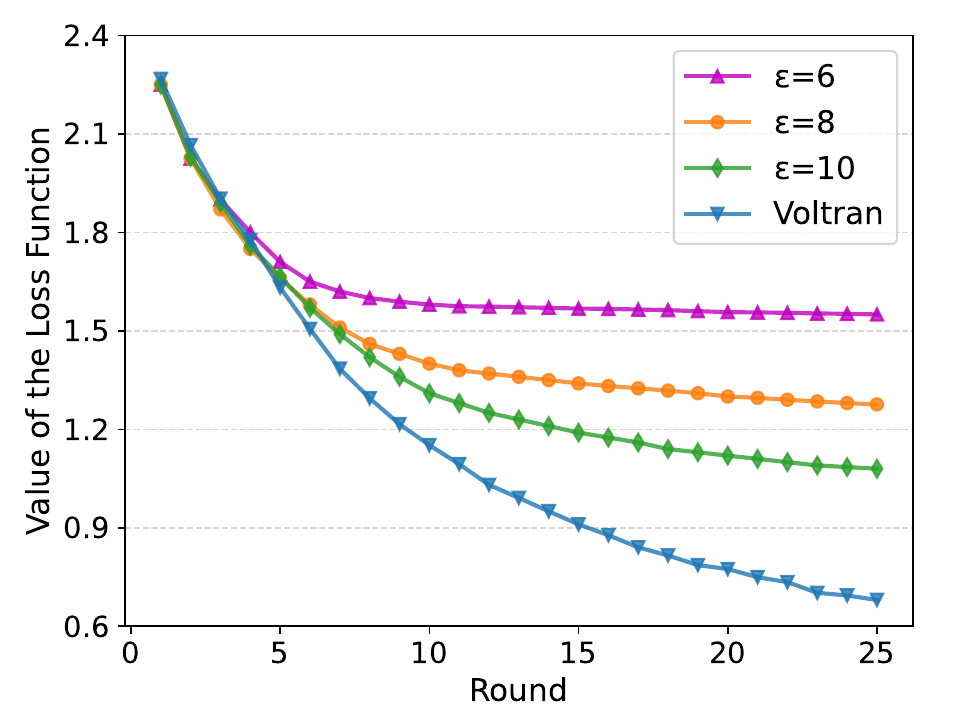}
	\caption{LossComparionWithDP}
	\label{TrainingLossComparionWithDP}
\end{subfigure}
\hfill
\caption{The comparison of accuracy and training loss between Voltran and BLADE-FL.}
\label{CompWithDP}
\end{figure}

\textbf{Voltran vs. DP} Although DP can be seen as a lightweight privacy-preserving method compared to encryption, due to its impact on the original data, it may degrade the model performance and become difficult to converge. Therefore, we compare Voltran with \cite{ChuanMa2020WhenFL} on the metrics of model accuracy and iteration rounds. Experimental results are shown in Fig. \ref{CompWithDP}. First, Fig. \ref{AccComparisonWithDP} depicts the comparison of model accuracy between Voltran and DP-based BLADE-FL \cite{ChuanMa2020WhenFL} on the CNN task with various client numbers. It can be seen that Voltran takes an obvious advantage over \cite{ChuanMa2020WhenFL} because the noise added to the data affects the performance of the aggregated model. Fig. \ref{TrainingLossComparionWithDP} presents the conditions of loss function based on different DP levels $\epsilon$ = 6, 8, 10 compared to Voltran with 50 clients. As the aggregation proceeds, the loss function value decreases. Furthermore, As $\epsilon$ increases (indicating lower privacy protection), the loss function value decreases. This is because a higher privacy protection level of DP increases the standard deviation of additive noise terms and decreases the model quality. Overall, although DP may have good efficiency, the inevitable trade-off between its privacy level and accuracy degrades the model performance, while Voltran still maintains the same model performance as non-privacy protection models due to its support for plaintext aggregation.

\textbf{Discussion.} The sensitive data held by the user locally is first trained on their local system. Subsequently, the trained model is encrypted using the session key {\fontfamily{cmss}\selectfont ssk} generated through RA, and securely transmitted into SGX. After undergoing aggregation processing within SGX, the global model is obtained. It is then encrypted using the master key {\fontfamily{cmss}\selectfont msk} and placed on the chain. Upon retrieval from the blockchain, the user decrypts the model locally, restoring it to plaintext, and then proceeds with the next round of training. The entire process ensures that no information is leaked, guaranteeing the privacy protection of sensitive data.

End-to-end encryption can provide strong security and privacy for data, but this is limited to data transmission. When data needs to be used and analyzed, it has to be decrypted into plaintext, thus leading to privacy leakage. Therefore, trivial encryption schemes generally cannot provide privacy protection on data usage. Differential privacy (DP) can ensure data privacy protection for transmission and usage, but the noise added to the data can lead to distortion. Our design brings a higher level of privacy protection brought by end-to-end encryption and TEE. We guarantee data privacy throughout the entire process of data transmission and usage while providing higher security.

	\subsection{How is the time \& traffic overhead of each step in Voltran?}
	
\begin{figure*}[h]
\centering
\begin{subfigure}[b]{0.2441\textwidth}
	\includegraphics[width=\textwidth]{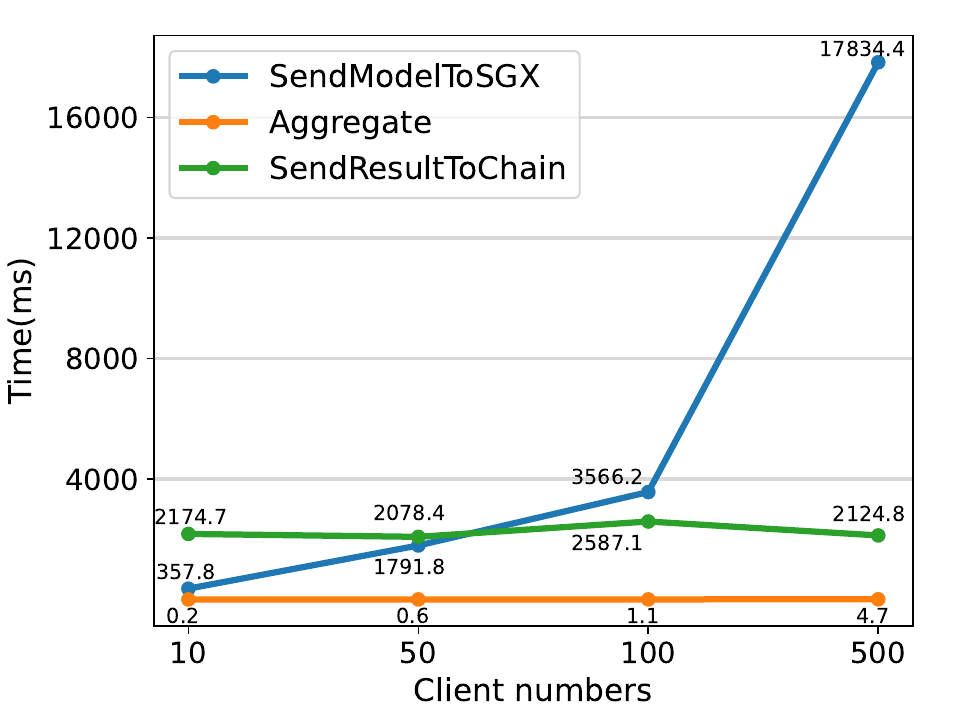}
	\caption{MLP}
	\label{Time of each step on MLP}
\end{subfigure}
\hfill
\begin{subfigure}[b]{0.2441\textwidth}
	\includegraphics[width=\textwidth]{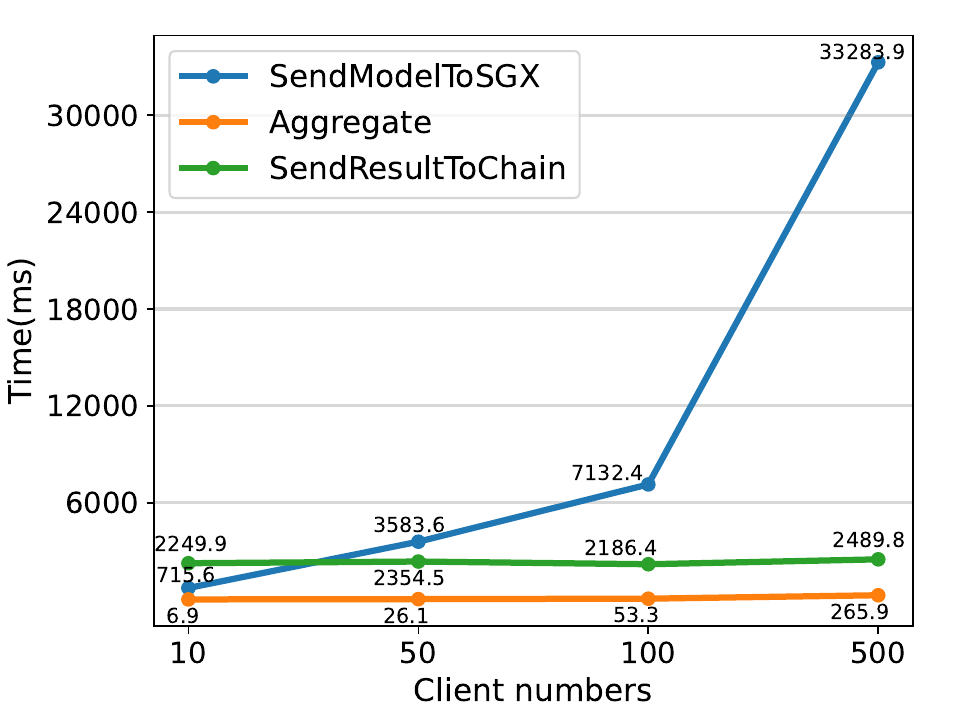}
	\caption{CNN}
	\label{Time of each step on CNN}
\end{subfigure}
\begin{subfigure}[b]{0.2441\textwidth}
	\includegraphics[width=\textwidth]{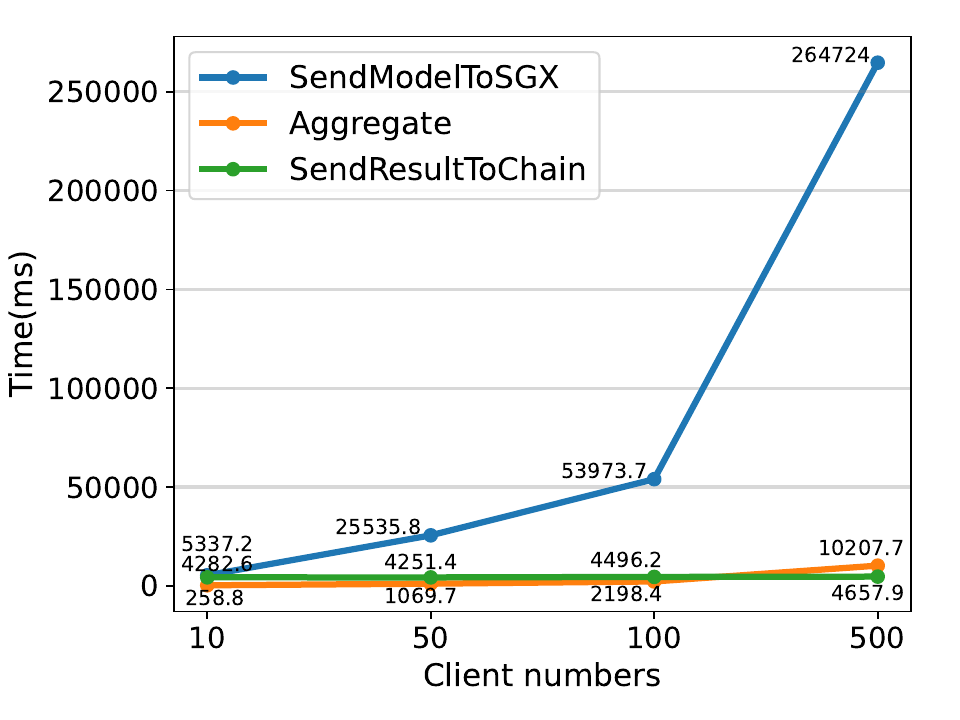}
	\caption{ResNet18}
	\label{Time of each step on ResNet18}
\end{subfigure}
\hfill
\begin{subfigure}[b]{0.2441\textwidth}
	\includegraphics[width=\textwidth]{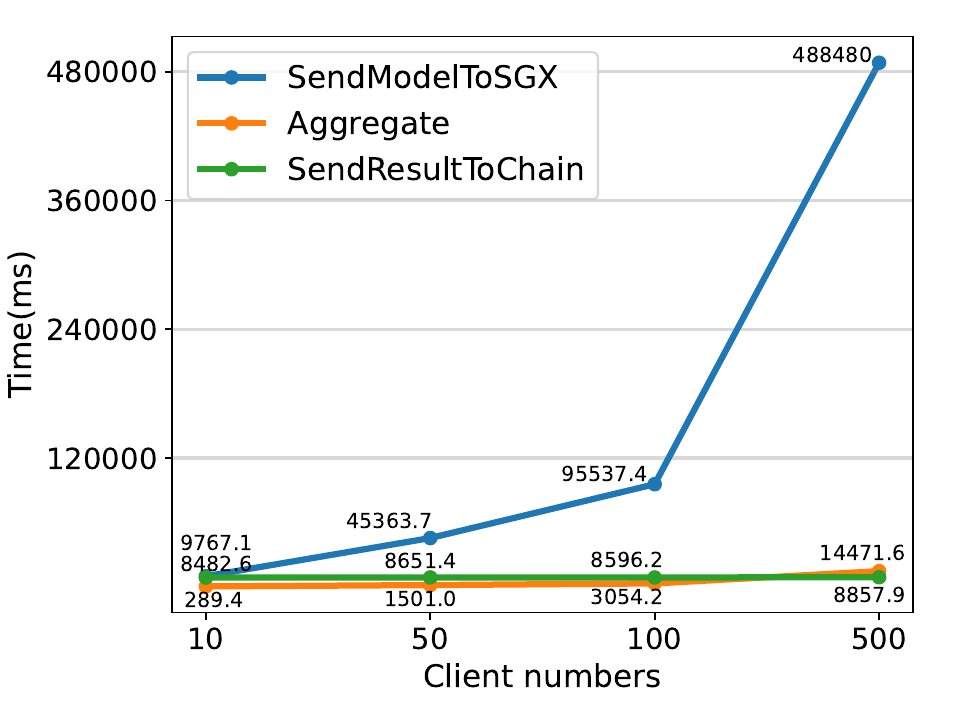}
	\caption{ResNet50}
	\label{Time of each step on ResNet50}
\end{subfigure}
\caption{Time cost of each step in four FL tasks.}
\label{Time of each step}
\vspace{-0.2cm}
\end{figure*}

	We evaluate the time cost of each step in Voltran with the different number of clients on the four tasks shown in Fig. \ref{Time of each step}. Tasks on MLP and CNN choose the single-SGX mode, and ResNet18 and ResNet50 choose the multi-SGX mode. We set that when the first round starts, the client is trained locally and encrypted, and then the ciphertext is sent to SGX $\mathcal{\brown{N}}$ before the FL task starts, and the steps such as contract creation or RA are seen as pre-processing. As we can see, the model transmission phase \textit{SendResultToChain} occupies the vast majority of the time, while the computation phase \textit{Aggregate} takes a small fraction. In the MLP and CNN task (Fig. \ref{Time of each step on MLP} and \ref{Time of each step on CNN}), the step \textit{SendResultToChain} takes a notable portion of time, which is led by the blockchain latency. In the ResNet tasks (Fig. \ref{Time of each step on ResNet18} and \ref{Time of each step on ResNet50}), the overhead of \textit{Aggregate} and \textit{SendResultToChain} is negligible. The results indicate that with regard to larger models, the additional overhead of two steps, \textit{Aggregate} and \textit{SendResultToChain}, have less impact on the performance, which means Voltran is more suitable for executing large-scale tasks with large numbers of clients.  
	
Table \ref{Communication cost of each step} presents the Communication cost of each step on the four FL tasks. Voltran's additional communication overhead lies in the on-chain operation. Hence, the deviation between the vanilla FL and Voltran is a model size. As the number of clients grows, the disparity between the two diminishes and eventually becomes negligible.

\begin{table}
\centering
\caption{Communication cost of each step on four FL tasks between Voltran and vanilla FL.}
\label{Communication cost of each step}
\scalebox{1.05}{
	\begin{tabular}{cccccc}
		\toprule
		\multirow{2}{*}{Model} &\multirow{2}{*}{Paradigm}& \multicolumn{4}{c}{Client Number}  \\ \cmidrule{3-6}
		&&10&50&100&500
		\\
		
		\midrule
		\multirow{2}{*}{MLP} & FL & 42 & 210 & 420 & 2100  \\
		& Voltran & 44.1 & 212.1 & 422.1 & 2102.1   \\
		\midrule
		\multirow{2}{*}{CNN} &FL & 85 & 425 & 850 & 4250 \\
		& Voltran & 89.25 & 429.25 & 854.25 & 4254.25 \\
		\midrule
		\multirow{2}{*}{ResNet18} & FL & 42640   & 213200 &   426400 &   2132000  
		\\
		& Voltran & 44772 & 215332 & 428532 & 2134132
		\\
		\midrule
		\multirow{2}{*}{ResNet50} & FL & 81200 &  406000 &  812000 &  4060000  
		\\
		& Voltran & 85260 & 410060 & 816060 & 4064060
		\\ 
		
		\bottomrule
\end{tabular}}
\vspace{-0.5cm}
\end{table}

\subsection{How well does Voltran defend against various attacks?}
Due to the programmability, Voltran can effectively employ off-the-shelf schemes for targeted defence against all types of attacks, such as Backdoor attacks and Byzantine attacks. We will describe how we conduct experiments on defending against these two attacks and evaluate our performance.

\textbf{Backdoor attack.} We choose CRFL \cite{xie2021crfl} as the backdoor defense method. We apply CRFL to both Voltran and the vanilla FL and compare their effectiveness in defending backdoor attacks on two scenarios on the datasets MNIST and FMNIST. Experimental results are presented in Table \ref{tab: backdoor}, which displays the Clean Data Accuracy (CDA) and Attack Success Rate (ASR). As indicated in the results presented in Table \ref{tab: backdoor}, the defense performance of implementing CRFL in Voltran is comparable to implementing it in vanilla FL. 

\begin{figure*}[h]
\centering
\begin{subfigure}[b]{0.32\textwidth}
	\includegraphics[width=\textwidth]{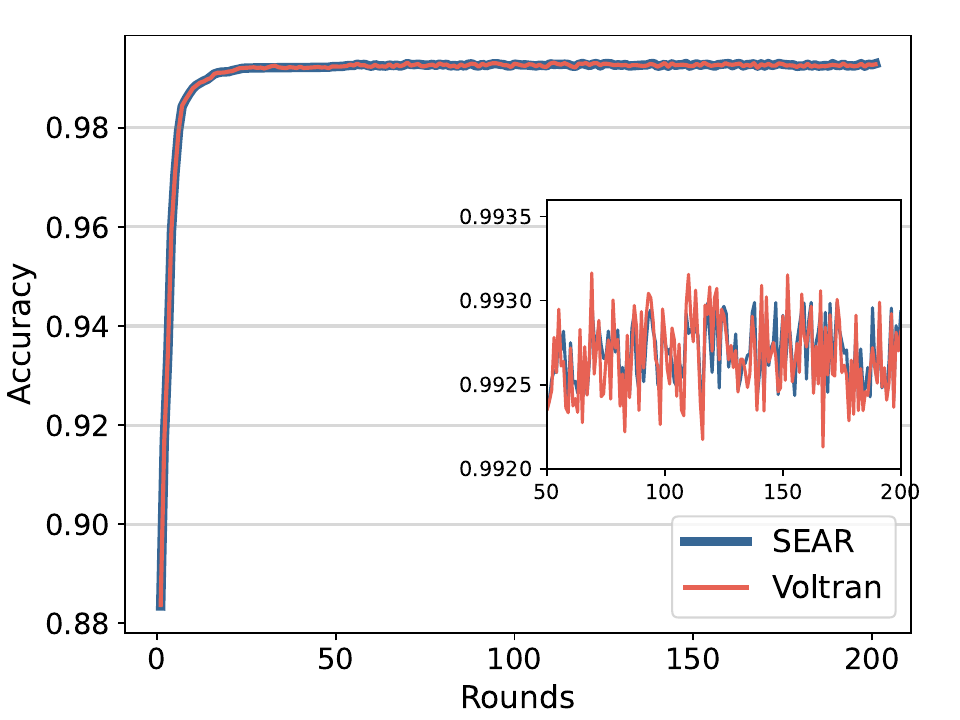}
	\caption{Without adversaries}
	\label{Without adversaries}
\end{subfigure}
\hfill
\begin{subfigure}[b]{0.32\textwidth}
	\includegraphics[width=\textwidth]{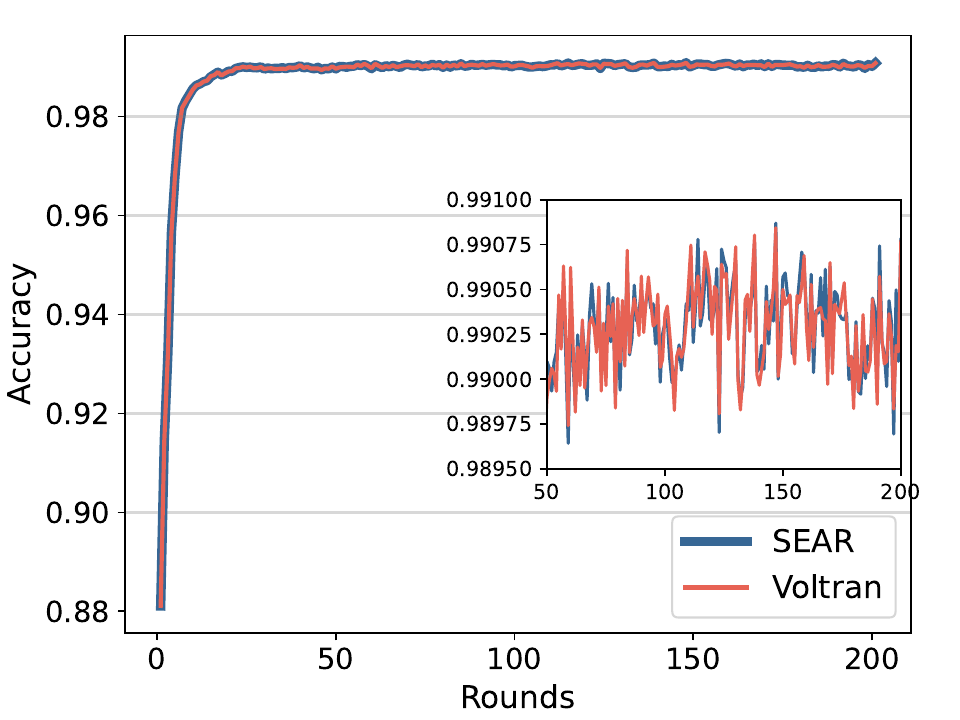}
	\caption{Don't collude}
	\label{don't collude}
\end{subfigure}
\begin{subfigure}[b]{0.32\textwidth}
	\includegraphics[width=\textwidth]{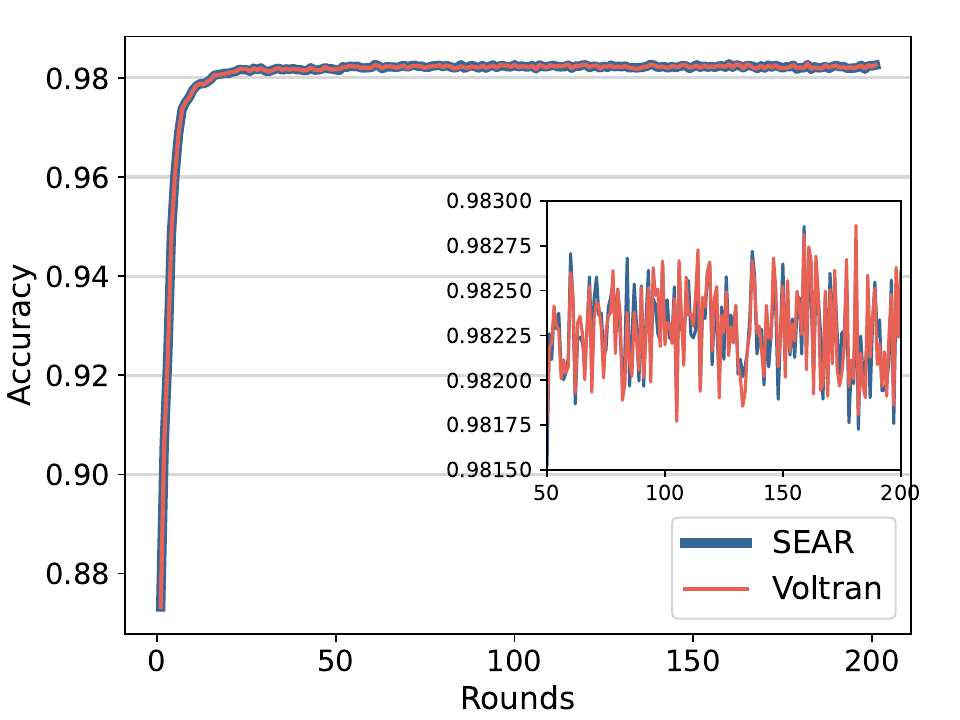}
	\caption{Collude}
	\label{collude}
\end{subfigure}
\hfill
\caption{Performance comparison on Byzantine attacks between the defense scheme SEAR and its evaluation on Voltran.}
\label{Performance comparison between SEAR and its implementation on Voltran}
\vspace{-0.3cm}
\end{figure*}

\textbf{Byzantine attack.} We choose SEAR \cite{LingchenZhao2021SEARSA} as the Byzantine attack defense method. Specifically, we conduct the Byzantine attack on the MNIST-1-7 dataset and integrate the attack defense algorithm from SEAR into Voltran to measure the defense performance. We follow the experiment settings of SEAR to perform evaluations on the MNIST dataset using a CNN model and set the batch size $B$ = 32, the learning rate $\eta$ = 0.01, the client number $n$ = 100 and the adversary number $f$ = 20. Similar to SEAR, we also evaluate the three situations in which there are no Byzantine adversaries; the Byzantine adversaries do not collude, and they collude. Experimental results are shown in Fig. \ref{Performance comparison between SEAR and its implementation on Voltran}. The difference between the two curves of SEAR and Voltran comes from randomness and is negligible. We can see that Voltran can perform equally well compared to the native algorithm in SEAR. 

\begin{table}[tp]
\caption{Comparison of the performance of implementing CRFL on vanilla FL and Voltran. ``CDA'' represents the model accuracy tested with clean target data. ``ASR'' represents the attack success rate using target data with triggers. }
\label{tab: backdoor}
\centering
\resizebox{\columnwidth}{!}
{
	\begin{tabular}{cccccc}
		\toprule
		\multirow{2}{*}{Rate} & \multirow{2}{*}{Paradigm} &  \multicolumn{2}{c}{MNIST} & \multicolumn{2}{c}{FMNIST} \\ \cmidrule{3-4} \cmidrule{5-6}
		&  &CDA(\%) & ASR(\%) & CDA(\%) & ASR(\%) \\
		
		\midrule
		\multirow{2}{*}{10\%} &FL & 97.36 ± 0.07 & 0.32 ± 0.05 & 85.38 ± 0.45 & 2.93 ± 0.66 \\
		& Voltran &  97.26 ± 0.05 & 0.30 ± 0.02 & 85.29 ± 0.34 & 3.08 ± 0.77 \\
		
		\midrule
		\multirow{2}{*}{20\%} & FL&  96.43 ± 0.17 & 0.47 ± 0.04 & 85.09 ± 0.51 & 3.05 ± 0.53\\
		& Voltran & 97.02 ± 0.05 & 0.46 ± 0.02 & 85.15 ± 0.33 & 3.21 ± 0.44 \\
		
		\midrule 
		\multirow{2}{*}{30\%} & FL & 95.84 ± 0.24 & 0.59 ± 0.02 & 84.98 ± 0.63 & 2.96 ± 0.55 \\
		
		& Voltran & 95.73 ± 0.19 & 0.60 ± 0.01 & 85.01 ± 0.54 & 2.87 ± 0.41 \\
		
		\midrule 
		\multirow{2}{*}{40\%} & FL & 94.09 ± 0.49 & 0.74 ± 0.18 & 84.61 ± 0.63 & 3.04 ± 0.39 \\
		
		& Voltran & 94.37 ± 0.51 & 0.72 ± 0.15 & 84.63 ± 0.71 & 3.05 ± 0.54  \\

		\bottomrule
	\end{tabular}
}

\end{table}

\subsection{How do blockchain settings affect the FL performance?}
\begin{figure}
\centering
\vspace{-0.2cm}
\includegraphics[width=2.6in]{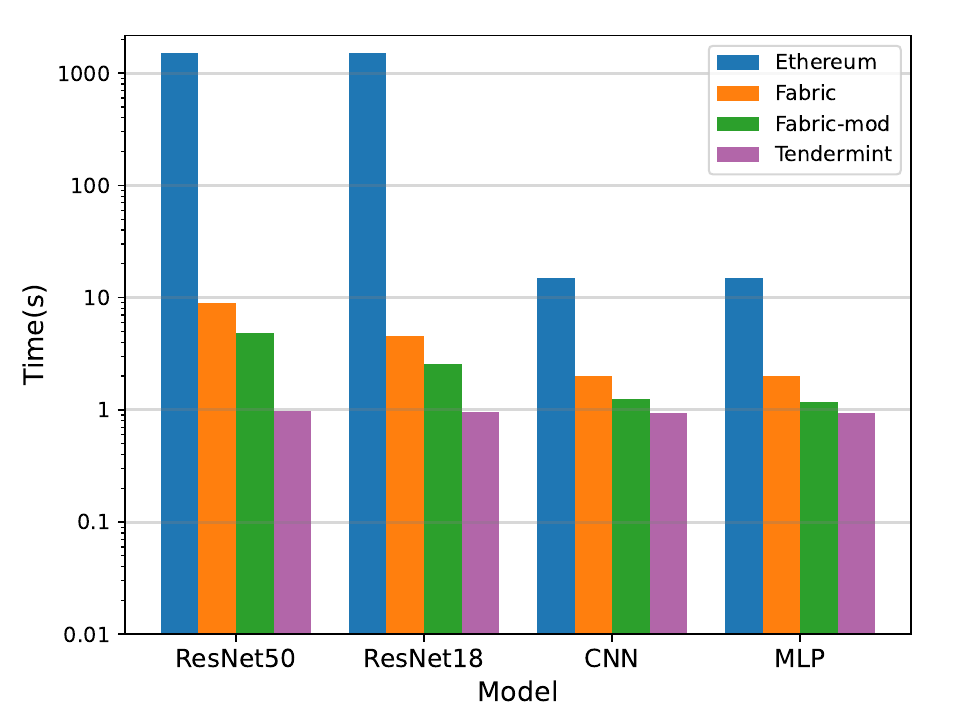}
\caption{Time cost comparison between different blockchains.}
\label{blockchain}
\vspace{-0.5cm}
\end{figure}

We attempt to minimize the negative impact of the blockchain. We count the end-to-end latency of different blockchains to measure their impact on FL performance, including Ethereum, Fabric and Tendermint. For Ethereum, we create an Ethereum private chain without any modifications to the official main chain. Our prototype Voltran-Fabric is used to measure latency by Fabric. Furthermore, to show the high scalability of Voltran, we provide another Fabric instantiation by decreasing the block interval from the default value of 2 seconds to 1 seconds and expand the block size up to 60 MB to support more extensive model storage. In Tendermint \cite{buchman2016tendermint}, the block size is not fixed and can be dynamically adjusted as needed. Therefore, it can handle larger blocks and effectively process a significant amount of transaction data. We run FL tasks 1-4 in Table I on each blockchain setting. We plot a bar chart to display the distinction of end-to-end latency of FL tasks on different blockchain settings. Results are shown in Fig. \ref{blockchain}. Because of Ethereum's slow block generation speed and small data capacity, when faced with ResNet models over 40 MB, it takes THOUSANDS of blocks to contain. Thus, we consider the native Ethereum setting unsuitable for large models. Fabric and Fabric-mod are able to take on this amount of model data with a low time overhead. In addition, Tendermint shows better performance by controlling the blockchain's overhead below one second per round. These results demonstrate the excellent scalability of our framework to dynamically adapt the configuration to different tasks for better and more efficient completion. 

\subsection{How is the scalability of Voltran on large-scale environments?}

We evaluate the performance of Voltran when the number of nodes increases, or the FL network expands to highlight the scalability of our system by concluding the experimental results above. We will discuss the following three dimensions: the number of clients, model size, and execution nodes. First, we consider the increase in the number of FL client nodes. In our experiments, we set the maximum client number as 500. At this size, we first compare Voltran with vanilla FL, as shown in Table III, and find that in terms of aggregation time and accuracy, the performance does not experience an additional decrease due to an increase in the number of clients. We also make comparisons with other privacy-preserving schemes, as shown in Table IV, and outperform the performance of \cite{liu2022efficient} and \cite{LingchenZhao2021SEARSA}. Second, we consider the larger model size. In Table I, we conduct experiments on ResNet18 (42.64MB), ResNet50 (81.20MB) and Bert (390.16MB), which can be as large models. Experimental results are shown in Table III, indicating that these large models can perform well within Voltran. Third, we consider the increase of execution nodes in our system. Due to our support of multi-SGX parallel execution, increasing execution nodes can provide more efficient computation. We present the significant performance improvement of multi-SGX execution compared to single-SGX execution in Fig. \ref{multisingle}.

Moreover, additional evaluations and discussions on Voltran are presented in Appendix D.

\section{Conclusion}

In this paper, we propose Voltran, a novel platform specifically designed to enable confidentiality-preserving and trustful FL aggregation based on a hybrid architecture combined with Intel SGX and the blockchain. Benefiting from our secure data transmission protocol, Voltran can guarantee the correctness, authenticity and confidentiality of the clients' local model data. Further, we consider several challenges when implementing Voltran, including the deficiency of the throughput and capacity, and propose a multi-SGX parallel mechanism to address them. A prototype of Voltran is implemented and six diverse tasks are extensively evaluated on it. Experimental results demonstrate the feasibility and efficiency of Voltran.

In future work, we plan to consider the applicability of Voltran on larger-scale tasks, such as GPT \cite{eloundou2023gpts}. Also, we plan to exploring more efficient model compression techniques such as Knowledge Distillation \cite{polino2018model} to further optimize computational costs and keep communication overhead low. Furthermore, in read-world applications, TEE can take on more forms. We will consider the potential threats on TEE and integrating more types of TEE into Voltran. We also consider further optimization of resource scheduling for multiple SGX in our future work, with potential optimization solutions including Ring Allreduce \cite{RingAllreduce}.


\bibliographystyle{IEEEtran}
\bibliography{ref}
\clearpage


\appendices
\section{Notation}
Table V summarizes the notations used in this paper. 
\begin{table}[htbp]
	\caption{Notations}
	\label{table_notation}
	\centering
	\setlength{\tabcolsep}{8mm}{
		\begin{tabular}{cc}
			\toprule
			Description                     & Notations                      \\ \midrule
			Task Owner                      & $\mathcal{\brown{O}}$                  \\
			Execution Node                  & $\mathcal{\brown{N}}$                  \\
			Client                          & $\mathcal{\brown{C}}$                  \\
			Contract Enclave                & {\fontfamily{cmss}\selectfont Con}$_{\textnormal{encl}}$                    \\
			Contract Storage                & {\fontfamily{cmss}\selectfont Con}$_{\textnormal{stor}}$                    \\
			Execution Node Committee        & \textit{Comm}\\
			Shared Session Key              & {\fontfamily{cmss}\selectfont ssk}   \\
			Master Secret Key               & {\fontfamily{cmss}\selectfont msk}  \\
			Verification Key                & ${\fontfamily{cmss}\selectfont vk}$\\
			Verification Key -- Public Key  & ${\fontfamily{cmss}\selectfont vk}_{\textit{pk}}$                      \\
			Verification Key -- Private Key & ${\fontfamily{cmss}\selectfont vk}_{\textit{sk}}$   \\
			\bottomrule
	\end{tabular}}
\end{table}

\section{Related Work}
\subsection{Federated Learning on decentralization and privacy-preserving}

The current FL paradigm is mostly based on the assumption that a single centralized server is trustworthy and will perform fair and correct aggregation computations. However, this assumption is not always appropriate. In real-world scenarios, the central server often exhibits dishonest behavior, showing bias towards selected clients, thus affecting the actual aggregation results. Additionally, the stability of the aggregation process depends on the central server orchestrating the application of the centralized aggregator. Therefore, a single point of failure can lead to the collapse of the entire task. Hence, there is a strong motivation to develop a decentralized FL framework. Moreover, although FL is designed to protect clients' local data, information can still be inferred by analyzing the shared gradients. Therefore, existing studies focus on the privacy-preserving FL solutions.

Kim et al. \cite{HyesungKim2018BlockchainedOF} proposed a blockchain on-device FL framework to exchange and verify the uploaded model updates. They evaluated their performance and gave the optimal block generation rate. Lu et al. \cite{YunlongLu2020BlockchainAF} focused on the industrial Internet of Things paradigm and designed a blockchain-enabled secure FL architecture to utilize the training process as the computation workload of the blockchain consensus. Bao et al. \cite{XianglinBao2019FLChainAB} considered the incentive to the trainers and introduced the FLChain to build a public auditable and incentive FL system. Wang \cite{ShufenWang2019BlockFedMLBF} gave a blockchain-enabled solution to address the two challenges of gradient leakage and integrity attacks for FL. Ma et al. \cite{ChuanMa2020WhenFL} proposed a blockchain-enabled FL framework BLADE-FL to create an autonomous and self-motivated FL system for clients. In their design, there is no server, and the aggregation is fully executed by the clients themselves decentralized. They also consider the issues of privacy and lazy clients and give their countermeasures. \cite{liao2023adaptive} introduces FedHP, an efficient decentralized federated learning method that addresses system and statistical heterogeneity in edge computing by adaptively controlling local update frequency and network topology to enhance training convergence and model accuracy. FedDual \cite{chen2022feddual} is a privacy-preserving and efficient gradient aggregation algorithm for federated learning in large decentralized networks, enhancing communication efficiency and model performance through local differential privacy, pair-wise gossip communication, and a noise reduction trick based on Private Set Intersection. \cite{zhou2022privacy} introduces the PVFL framework, which integrates technologies such as Differential Privacy, Homomorphic Hashing, Symmetric Encryption, and Digital Signatures to achieve privacy protection, data integrity verification, and efficient aggregation in federated learning for edge computing environments.

These efforts give a variety of blockchain-enabled FL frameworks and focus on different priorities, but their privacy protection is not good enough. Although \cite{ChuanMa2020WhenFL, ShufenWang2019BlockFedMLBF, chen2022feddual, zhou2022privacy} utilize differential privacy to protect the model's privacy. However, the problem of differential privacy is that the effect of privacy protection is contradictory to the accuracy of the model. More noise will sacrifice accuracy. \cite{zhou2022privacy} presents the solution to eliminate the noise, but they do not consider the centralized problem. Therefore, in our scenario, we use TEE to isolate the aggregation process into a closed space and encrypt the input and output, thus protecting privacy without sacrificing accuracy. We combined TEE with blockchain to provide stronger integrity, robustness and availability guarantees.


\subsection{Secure Aggregation with Trusted Execution Environment}
Secure aggregation is a computation paradigm in FL that enables a number of clients to send their local values (usually trained models) to a server and generate an aggregated result \cite{mcmahan2017communication}. Trusted hardware, particularly Intel SGX, has seen a wide spectrum of applications in FL to achieve secure aggregation and parameter preserving. However, due to the upper limitation of TEE memory size, previous studies run only part of the model (e.g., sensitive layers) inside the TEE, such as \cite{FanMo2020DarkneTZTM, FlorianTramr2018SlalomFV}. PPFL \cite{mo2021ppfl} deployed TrustZone on clients for local training and SGX on the server for aggregation, respectively. With regard to memory size, they took greedy layer-wise training to get around it. SEAR \cite{LingchenZhao2021SEARSA} utilized Intel SGX to execute secure and Byzantine-robust aggregation and protect clients' private models. Moreover, they considered the limitation of trusted memory size and provided data storage modes to enhance efficiency. PPFL do not consider Byzantine resilience and poisoning attacks; these two schemes ignore the single point of failure on the SGX server \cite{mo2019efficient}.

\section{Proof of our Protocol}
\label{appproof}
Here we give the complete proof of our protocol, which is given in Section IV.

\begin{figure}[!t]
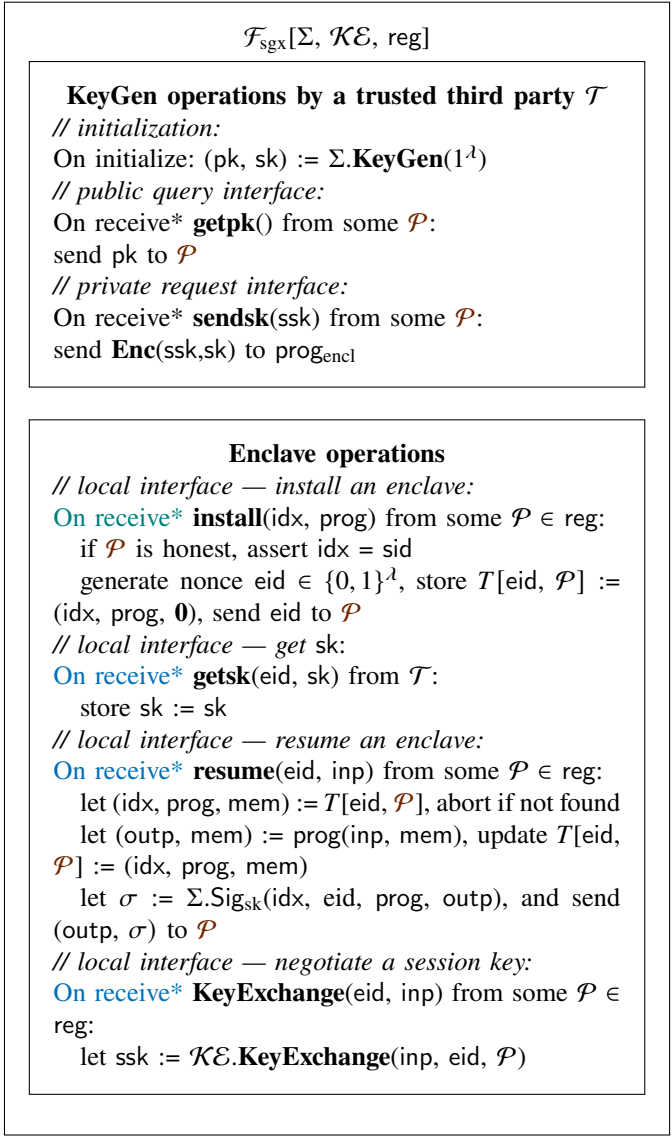

	\begin{framed}
		\centerline{$\mathcal{F}_{\mathrm{sgx}}$[$\Sigma$, $\mathcal{KE}$, {\fontfamily{cmss}\selectfont reg}]}
		
		\begin{framed}
			\centerline{\textbf{KeyGen operations by a trusted third party $\mathcal{T}$}}
			\textit{// initialization:}\\
			On initialize: ({\fontfamily{cmss}\selectfont pk}, {\fontfamily{cmss}\selectfont sk}) := $\Sigma$.\textbf{KeyGen}($1^{\mathrm{\lambda}}$)
			
			\textit{// public query interface:}\\
			On receive* \textbf{getpk}() from some \brown{$\mathcal{P}$}: \\
			send {\fontfamily{cmss}\selectfont pk} to \brown{$\mathcal{P}$}   
			
			\textit{// private request interface:}\\
			On receive* \textbf{sendsk}({\fontfamily{cmss}\selectfont ssk}) from some \brown{$\mathcal{P}$}: \\
			send \textbf{Enc}({\fontfamily{cmss}\selectfont ssk},{\fontfamily{cmss}\selectfont sk}) to {\fontfamily{cmss}\selectfont prog}$_{\mathrm{encl}}$  
			
		\end{framed}

		\begin{framed}
			\centerline{\textbf{Enclave operations}}

			\textit{// local interface --- install an enclave:}
			
			{\color{teal}{On receive*}} \textbf{install}({\fontfamily{cmss}\selectfont idx}, {\fontfamily{cmss}\selectfont prog}) from some $\mathcal{P} \in$ {\fontfamily{cmss}\selectfont reg}:
			
			\quad if \brown{$\mathcal{P}$} is honest, assert {\fontfamily{cmss}\selectfont idx} = {\fontfamily{cmss}\selectfont sid}
			
			\quad generate nonce {\fontfamily{cmss}\selectfont eid} $\in\{0,1\}^\lambda$, store $T[${\fontfamily{cmss}\selectfont eid}, $\mathcal{P}]$ := ({\fontfamily{cmss}\selectfont idx}, {\fontfamily{cmss}\selectfont prog}, $\mathbf{0}$), send {\fontfamily{cmss}\selectfont eid} to \brown{$\mathcal{P}$}
			
			\textit{// local interface --- get} {\fontfamily{cmss}\selectfont sk}:
			
			\blue{On receive*} \textbf{getsk}({\fontfamily{cmss}\selectfont eid}, {\fontfamily{cmss}\selectfont sk}) from $\mathcal{T}$:
			
			\quad store {\fontfamily{cmss}\selectfont sk} := {\fontfamily{cmss}\selectfont sk}
			
			
			
			
			\textit{// local interface --- resume an enclave:}
			
			\blue{On receive*} \textbf{resume}({\fontfamily{cmss}\selectfont eid}, {\fontfamily{cmss}\selectfont inp}) from some $\mathcal{P} \in$ {\fontfamily{cmss}\selectfont reg}:
			
			\quad let ({\fontfamily{cmss}\selectfont idx}, {\fontfamily{cmss}\selectfont prog}, {\fontfamily{cmss}\selectfont mem}) := $T$[{\fontfamily{cmss}\selectfont eid}, \brown{$\mathcal{P}$}], abort if not found
			
			\quad let ({\fontfamily{cmss}\selectfont outp}, {\fontfamily{cmss}\selectfont mem}) := {\fontfamily{cmss}\selectfont prog}({\fontfamily{cmss}\selectfont inp}, {\fontfamily{cmss}\selectfont mem}), update $T$[{\fontfamily{cmss}\selectfont eid}, \brown{$\mathcal{P}$}] := ({\fontfamily{cmss}\selectfont idx}, {\fontfamily{cmss}\selectfont prog}, {\fontfamily{cmss}\selectfont mem})
			
			\quad let $\sigma$ := $\Sigma$.{\fontfamily{cmss}\selectfont Sig}$_{\text {sk}}$({\fontfamily{cmss}\selectfont idx}, eid, {\fontfamily{cmss}\selectfont prog}, {\fontfamily{cmss}\selectfont outp}), and send ({\fontfamily{cmss}\selectfont outp}, $\sigma)$ to \brown{$\mathcal{P}$}

			\textit{// local interface --- negotiate a session key:}\\
			\blue{On receive*} \textbf{KeyExchange}({\fontfamily{cmss}\selectfont eid}, {\fontfamily{cmss}\selectfont inp}) from some $\mathcal{P} \in$ {\fontfamily{cmss}\selectfont reg}: 
			
			\quad let {\fontfamily{cmss}\selectfont ssk} := $\mathcal{KE}$.\textbf{KeyExchange}({\fontfamily{cmss}\selectfont inp}, {\fontfamily{cmss}\selectfont eid}, $\mathcal{P}$)
			
		\end{framed}
	\end{framed}
	\caption{The ideal functionality of SGX $\mathcal{F}_{\text {sgx}}$. Functionalities in blue (and starred) denote reentrant activation points. Functionalities in green are executed at most once. The proof of enclave outputs is included in an anonymous attestation $\sigma$.}
	\label{F_sgx}
\end{figure}


\subsection{Formal Modelling}
\subsubsection{SGX Formal Modelling}
With respect to the ideal functionality of SGX, we adopt the formal modelling from \cite{pass2017formal}. We refer the reader to \cite{pass2017formal} for a more detailed overview of $\mathcal{F}_{\mathrm{sgx}}$. The main idea behind this SGX modelling of \cite{pass2017formal} is to regard SGX as a trusted third party defined by a global functionality $\mathcal{F}_{\mathrm{sgx}}$. We mainly use two functionalities \texttt{install} and \texttt{resume}. The \texttt{install} functionality makes an executable program {\fontfamily{cmss}\selectfont prog} loaded into a trusted hardware. Users call \texttt{resume} to get an output {\fontfamily{cmss}\selectfont outp} with an attestation $\sigma_{\text{TEE}}$ = $\Sigma_{\text{TEE}}$.{\fontfamily{cmss}\selectfont Sig}$(\textsf{sk}_{\text{TEE}}$,({\fontfamily{cmss}\selectfont prog, outp})) on a given input {\fontfamily{cmss}\selectfont inp}. To comply with our new verification mechanism, we modify $\sigma_{\mathrm{sgx}}$ on $\mathcal{F}_{\mathrm{sgx}}$ and replace it with our signature design. As our mechanism is similar to the native design of $\sigma_{\mathrm{sgx}}$, we think of it as an equivalent substitute. 

In addition, in our protocol, \textit{remote attestation} is required for clients to verify the identity of enclaves and negotiate a session key to securely transmit privacy model data into the enclaves for computation. In particular, we decouple the functionality of \textit{remote authentication} from $\mathcal{F}_{\mathrm{sgx}}$ to highlight its function in generating secure session keys in a \textbf{KeyExchange} interface extension. We assert that remote attestation holds a high degree of security, effectively safeguarding the datagram transmitted via the secret channel it establishes against interception or unauthorized modification. The whole abstraction is shown in Fig. \ref{F_sgx}.




\subsubsection{Blockchain Formal Modelling}

For the blockchain, we follow the ideal functionality $\mathcal{F}_{\mathrm{blockchain}}$ in Fig. 16 proposed by \cite{RaymondCheng2018EkidenAP}. $\mathcal{F}_{\mathrm{blockchain}}$ specify a simplified generic blockchain protocol run as a distributed append-only ledger. It depicts intuitively the basic interfaces of blockchain criteria, including \textit{init}, \textit{read} and \textit{write}. 

\begin{figure}[!t]
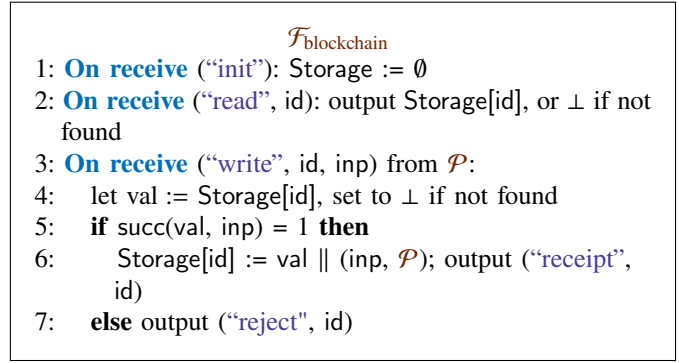

	\begin{framed}
		\centerline{\brown{$\mathcal{F}_{\text {blockchain}}$} }
		
		1: \textbf{\blue{On receive}} (\purple{``init''}): {\fontfamily{cmss}\selectfont Storage} := $\emptyset $
		
		2: \textbf{\blue{On receive}} (\purple{``read''}, {\fontfamily{cmss}\selectfont id}): output {\fontfamily{cmss}\selectfont Storage[id]}, or $\perp$ if not 
		
		\quad found
		
		3: \textbf{\blue{On receive}} (\purple{``write''}, {\fontfamily{cmss}\selectfont id}, {\fontfamily{cmss}\selectfont inp}) from \brown{$\mathcal{P}$}:
		
		4:  \quad let $\text{val} :=$ {\fontfamily{cmss}\selectfont Storage[id]}, set to $\perp$ if not found 
		
		5: \quad \textbf{if} {\fontfamily{cmss}\selectfont succ}({\fontfamily{cmss}\selectfont val}, {\fontfamily{cmss}\selectfont inp}) = 1 \textbf{then} 
		
		6: \quad \quad {\fontfamily{cmss}\selectfont Storage[id]} := {\fontfamily{cmss}\selectfont val} $\|$ ({\fontfamily{cmss}\selectfont inp}, \brown{$\mathcal{P}$}); output (\purple{``receipt''}, 
		
		\quad \quad \quad  {\fontfamily{cmss}\selectfont id})
		
		7: \quad \textbf{else} output (\purple{``reject"}, {\fontfamily{cmss}\selectfont id})
	\end{framed}
	\label{Fbc}
	\caption{The ideal functionality of blockchain $\mathcal{F}_{\mathrm{blockchain}}$.}
\end{figure}

\subsection{Proof of Security}

\textbf{Definition 1} (\textit{Authenticity}). We say that $\texttt{Prot}_{\mathrm{Volt}}$ satisfies \textit{authenticity} if, for any polynomial-time adversary \brown{$\mathcal{A}$} that can interact arbitrarily with $\texttt{Prot}_{\mathrm{Volt}}$, \brown{$\mathcal{A}$} cannot cause an honest verifier to accept the following two situations:
\begin{enumerate}
	\item \brown{$\mathcal{A}$} forges $\mathcal{N}$ to install a dummy {\fontfamily{cmss}\selectfont prog}$^{\prime}_{\mathrm{Encl}}$ on SGX;
	\item \brown{$\mathcal{A}$} forges $\mathcal{F}_{\mathrm{sgx}}$ to send a ``uploadGlobalModel'' message with a dummy input ({\fontfamily{cmss}\selectfont outp}$^{\prime}$ = ({\fontfamily{cmss}\selectfont ct}$^{\prime}_{\mathrm{out}}$, {\fontfamily{cmss}\selectfont $\sigma$}$^{\prime}_{\mathrm{sgx}}$), {\fontfamily{cmss}\selectfont param} = ({\fontfamily{cmss}\selectfont round}, {\fontfamily{cmss}\selectfont taskid}, {\fontfamily{cmss}\selectfont index})).
\end{enumerate}   


Formally, for any polynomial-time adversary \brown{$\mathcal{A}$}, \\

$\Pr \left[  
\begin{array}{ll}
	
	
	(\Sigma_{sgx}.\textbf{Verify}(\text{\fontfamily{cmss}\selectfont pk}, \sigma,\text{\fontfamily{cmss}\selectfont prog}_{\mathrm{Encl}}= \texttt{True})) \vee\\
	
	(\Sigma.\textbf{Verify}(\text{{\fontfamily{cmss}\selectfont pk}}_{\mathrm{vk}},\sigma^*, m^*) = \texttt{True}):\\

	(\text{\fontfamily{cmss}\selectfont pk},\text{{\fontfamily{cmss}\selectfont prog}}^{\prime}_{\mathrm{Encl}}) \leftarrow \mathcal{A}^{\mathcal{F}_{\text{sgx}}}(1^{\lambda});\\

	(\text{{\fontfamily{cmss}\selectfont pk}}_{\mathrm{vk}},\sigma,m,\mathrm{st}) \leftarrow \mathcal{A}_{\text{query}}^{\mathcal{O}_{\sigma}}(1^{\lambda});\\


	
	
	
\end{array}       
\right]\\
$

$\le negl(\lambda)$, for security parameter $\lambda$.

\textbf{Theorem 1} (\textit{Authenticity}). Assume that the remote attestation mechanism of Intel SGX is secure and the signature algorithm is existentially unforgeable under chosen message attacks (EU-CMA), then $\texttt{Prot}_{\mathrm{Volt}}$ achieves \textit{authenticity} under Definition 2.

\textit{Proof.} The property of \textit{Authenticity} can be categorized into two cases:

Case 1: The input ({\fontfamily{cmss}\selectfont ct}, {\fontfamily{cmss}\selectfont id}) of a client and the enclave code \text{\fontfamily{cmss}\selectfont prog}$_{\mathrm{Encl}}$ can be authenticated by the SGX. If there exists an adversary $\mathcal{A}$ that can forge a dummy input ({\fontfamily{cmss}\selectfont ct'}, {\fontfamily{cmss}\selectfont id}) or a dummy enclave code \text{\fontfamily{cmss}\selectfont prog}$'_{\mathrm{Encl}}$, which can be authenticated by SGX, this violates the property of $\mathcal{F}_{sgx}$ and the EU-CMA security of $\Sigma$. 

Case 2: The output {\fontfamily{cmss}\selectfont outp} of SGX can be publicly authenticated. Suppose there exists a probability polynomial-time (PPT) adversary $\mathcal{A}$ that can break the EU-CMA security of the proposed $\texttt{Prot}_{\mathrm{Volt}}$, then we can build a simulator $\mathcal{B}$ that can break the EU-CMA security of $\Sigma$.

\textbf{Query Phase 1.} The adversary $\mathcal{A}$ makes the signature $\textbf{Sig}(m)$ query. Whenever $\mathcal{A}$ issues a signature query on a message $m$, the simulator $\mathcal{B}$ first generates a random key $msk$ and then encrypts {\fontfamily{cmss}\selectfont m} using the symmetric encryption algorithm $C=\mathcal{SE}.\textbf{Enc}(msk,m)$. Then $\cal B$ passes $C$ to the challenger $\mathcal{C}$ and gets the signature $\sigma$ from the challenger $\cal C$. Then $\cal B$ returns $C$ to the adversary $\cal A$.

\textbf{Challenge.} $\mathcal{B}$ randomly generates a message $m^*$ and encrypts $m^*$ using a random key $msk^*$ to get a ciphertext $CT^*=\mathcal{SE}.\textbf{Enc}(msk^*,m^*)$. Then, $\mathcal{B}$ passes $CT^*$ to $\mathcal{A}$. $\mathcal{A}$ returns a signature $\sigma^*$ , where $\sigma^*$ is a valid signature on $CT^*$. Then, $\mathcal{B}$ returns ($CT^*$, $\sigma^*$) to the challenger $\mathcal{C}$ as its response. This completes the simulation of $\mathcal{B}$. 

\textbf{Definition 2} (\textit{Confidentiality}). We say that $\texttt{Prot}_{\mathrm{Volt}}$ satisfies \textit{confidentiality} if, for any polynomial-time adversary \brown{$\mathcal{A}$} that can interact arbitrarily with $\texttt{Prot}_{\mathrm{Volt}}$, \brown{$\mathcal{A}$} cannot obtain information about the plaintexts from ciphertexts during the protocol execution under the chosen plaintext attack (CPA) security. It requires an attacker cannot reveal the encapsulated key from the ciphertexts.


Formally, for any polynomial-time adversary \brown{$\mathcal{A}$}, \\

$\bigg | \Pr \left[  
\begin{array}{ll}
	
	b^{\prime}=b \vee c^{\prime}=c:\\
	
	(\text{ssk},(m_0, m_1), \mathrm{st}) \leftarrow \mathcal{A}_{\text{query}}^{\mathcal{O}_{\text{ssk}}}(1^{\lambda}) ;\\
	(\text{msk},(m'_0, m'_1), \mathrm{st}) \leftarrow \mathcal{A}_{\text{query}}^{\mathcal{O}_{\text{msk}}}(1^{\lambda}) ;\\
	\{b,c\} \stackrel{\$}{\leftarrow}\{0,1\}; \\\mathrm{CT}_1^* \leftarrow {\cal SE}_1.\operatorname{Enc}(\text{ssk}^*, m_b) ;  \\

	
	\mathrm{CT}_2^* \leftarrow {\cal SE}_2.\operatorname{Enc}(\text{msk}^*, m'_c) ;

\end{array}       
\right]
- \frac{1}{2} \bigg | \\$

$\le negl(\lambda)$, for security parameter $\lambda$.


\textbf{Theorem 2} (\textit{Confidentiality}). Assume that the encryption algorithms of $\mathcal{F}_{sgx}$ and $\mathcal{SE}$ are IND-CPA secure, then the protocol achieves \textit{confidentiality} under Definition 3.

\textit{Proof.} Suppose there exists a probability polynomial-time (PPT) adversary $\mathcal{A}$ that can break the IND-CPA security of the proposed $Prot_{Volt}$, then we can build a simulator $\mathcal{B}$ that can break the IND-CPA security of $\mathcal{SE}$.

\textbf{Query Phase 1.} The adversary $\mathcal{A}$ makes the two queries: the KeyGen({\fontfamily{cmss}\selectfont ssk}) query and the KeyGen({\fontfamily{cmss}\selectfont msk}) query. Whenever $\mathcal{A}$ issues the two queries, the simulator $\mathcal{B}$ passes it to the challenger $\mathcal{C}$ and returns the result answered by $\mathcal{C}$ to $\mathcal{A}$.

\textbf{Challenge.} $\mathcal{B}$ calls $\mathcal{A}$ to get two tuple of equal length messages $(m_0, m_1)$ and $(m'_0, m'_1)$ and sends to the challenger $\mathcal{C}$. The challenger $\mathcal{C}$ generates the challenge ciphertext $CT_b^*$, $CT_c^*$, where $b,c \in\{0,1\}$ and $C T_b^*$ and $CT_c^*$ are an $\mathcal{SE}$ ciphertext of $m_b$ and $m'_c$, respectively. Then $\mathcal{C}$ sends $CT_b^*$ and $CT_c^*$ to $\mathcal{B}$.


\textbf{Guess.} $\mathcal{B}$ passes $C T_b^*$ and $CT_c^*$ to $\mathcal{A}$ and gets the response $b'$ and $c'$. $\mathcal{B}$ outputs $b'$ and $c'$ as its guess.

This completes the simulation. We here analyze the probability of the simulator $\mathcal{B}$ to break the IND-CPA security of $\mathcal{SE}$.

Case 1: If $\cal A$ breaks the IND-CPA of $\mathcal{SE}_1$, which means $\cal A$ can win the above game with a non-negligible advantage $\epsilon$. Then we can get
$Pr[b' = b]$ = 1/2 + $\epsilon$. 

Case 2:  If $\cal A$ breaks the IND-CPA of $\mathcal{SE}_2$, which means $\cal A$ can win the above game with a non-negligible advantage $\epsilon$. Then we can get
$Pr[c' = c]$ = 1/2 + $\epsilon$. 

In both cases, $\mathcal{B}$ breaks the IND-CPA security of $\mathcal{SE}$. This completes the proof of Theorem 3.

\subsection{Proof of Correctness}

\textbf{Definition 3} (\textit{Correctness}). We say that $\texttt{Prot}_{\mathrm{Volt}}$ satisfies \textit{correctness} if, for any polynomial-time adversary \brown{$\mathcal{A}$} that can interact arbitrarily with $\texttt{Prot}_{\mathrm{Volt}}$, \brown{$\mathcal{A}$} cannot cause an honest party to return a wrong result the following two situations:
\begin{enumerate}
	\item \brown{$\mathcal{A}$} forces an enclave to return a dummy output \text{\fontfamily{cmss}\selectfont outp}$^{\prime}$ with a specific input ({\fontfamily{cmss}\selectfont ct} =({\fontfamily{cmss}\selectfont ct}$_{\mathrm{m}_i}$, {\fontfamily{cmss}\selectfont ct}$_{\mathrm{msk}}$), {\fontfamily{cmss}\selectfont id} = ({\fontfamily{cmss}\selectfont round}, {\fontfamily{cmss}\selectfont taskid})) from \brown{$\mathcal{C}$};
	\item \brown{$\mathcal{A}$} forces $\mathcal{F}_{\mathrm{blockchain}}$ to store a dummy input (\text{\fontfamily{cmss}\selectfont outp}$^{\prime}$ = ({\fontfamily{cmss}\selectfont ct}$^{\prime}_{\mathrm{out}}$, {\fontfamily{cmss}\selectfont $\sigma$}$^{\prime}_{\mathrm{sgx}}$), \text{\fontfamily{cmss}\selectfont id} = ({\fontfamily{cmss}\selectfont round}, {\fontfamily{cmss}\selectfont taskid}, {\fontfamily{cmss}\selectfont index})) with a specific output (\text{\fontfamily{cmss}\selectfont outp} = ({\fontfamily{cmss}\selectfont ct}$_{\mathrm{out}}$, {\fontfamily{cmss}\selectfont $\sigma$}$_{\mathrm{sgx}}$), \text{\fontfamily{cmss}\selectfont id} = ({\fontfamily{cmss}\selectfont round},  {\fontfamily{cmss}\selectfont taskid}, {\fontfamily{cmss}\selectfont index})) from \brown{$\mathcal{N}$}.
\end{enumerate}  

Formally, for any polynomial-time adversary \brown{$\mathcal{A}$}, \\

$ \Pr \left[     
\begin{array}{l}    
(\text{\fontfamily{cmss}\selectfont prog}_{\text{\fontfamily{cmss}\selectfont Encl}}.\text{\fontfamily{cmss}\selectfont Resume}(\text{\fontfamily{cmss}\selectfont id}, \text{\fontfamily{cmss}\selectfont ct'}) = \text{\fontfamily{cmss}\selectfont prog}_{\text{\fontfamily{cmss}\selectfont Encl}}.\text{\fontfamily{cmss}\selectfont Resume}(\text{\fontfamily{cmss}\selectfont id}, \text{\fontfamily{cmss}\selectfont ct})) \vee \\    
(\text{\fontfamily{cmss}\selectfont Storage}[\text{\fontfamily{cmss}\selectfont id}] =  \text{\fontfamily{cmss}\selectfont outp'} \wedge \mathcal{F}_{\text{\fontfamily{cmss}\selectfont blockchain}}.\text{\fontfamily{cmss}\selectfont write}(\text{\fontfamily{cmss}\selectfont id}, \text{\fontfamily{cmss}\selectfont outp})):\\  
(\text{\fontfamily{cmss}\selectfont ct'},\text{\fontfamily{cmss}\selectfont outp'}, \text{\fontfamily{cmss}\selectfont id})\gets \mathcal{A}^{\mathcal{F}_{\text{\fontfamily{cmss}\selectfont blockchain}}}(1^{\lambda})\\  
\end{array}           
\right]  $

$\le negl(\lambda)$, for security parameter $\lambda$.
\textbf{Theorem 3} (\textit{Correctness}). Assume $\mathcal{F}_{\mathrm{sgx}}$ and $\mathcal{F}_{\mathrm{blockchain}}$ are secure, then $\texttt{Prot}_{\mathrm{Volt}}$ achieves \textit{correctness} under Definition 1.

\textit{Proof.} The property of \textit{Correctness} is twofold. On one side,  If there exists an adversary $\cal A$ can force an enclave to return a dummy output {\fontfamily{cmss}\selectfont outp} that is not the correct execution result of a specific input ({\fontfamily{cmss}\selectfont id}, {\fontfamily{cmss}\selectfont ct}), this violates the security of $\mathcal{F}_{sgx}$. On the other side, the correct storage of {\fontfamily{cmss}\selectfont outp} on the blockchain is guaranteed by the security of blockchain functionality $\mathcal{F}_{blockchain}$. Once {\fontfamily{cmss}\selectfont outp} is verified and stored on the blockchain, it can not be tampered with. 

\subsection{Discussion of Robustness}
Voltran is resilient to a certain degree of a single point of failure, and FL tasks are not held up by a certain number of node failures. Malicious execution nodes or clients may intentionally delay the execution or even break down at any operation step. In Voltran, there are two execution modes for aggregation: a single SGX or multiple ones. In either case, a single point of failure of one SGX will crash the entire task. Voltran should be able to handle both sides of the fault separately to ensure the execution of the task. In Section III-E, we describe how we achieve high robustness.

\subsection{Discussion of TEE challenges}

\begin{itemize}
\item \textbf{TEE data leakage:} Adversaries may conduct external attacks such as side channel attacks to lead to data leakage in TEE. To overcome this threat, our system provides two mitigation measures. Due to our FL-oriented system, the first is to integrate secure aggregation algorithms into TEEs against side channel attacks such as [26]. The second is our secret key mechanism. Due to our committee, each time a new task starts, the keys are updated, thus avoiding the compromise of previous data confidentiality. Regarding the means of thoroughly resisting hardware attacks, we rely on hardware providers and application developers.
\item \textbf{TEE failures:} A sudden failure may happen on the TEE resulting in the loss of running enclaves. TEE lacks the ability to distinguish the correct current state for recovery. In our FL aggregation scenarios, each aggregation computation performed by TEE is independent and does not depend on the previous TEE state. Therefore, we do not need to consider the issues of TEE state failure and atomic delivery proposed by [24]. Even if a failure occurs in a TEE during the aggregation process, the committee will promptly detect the anomaly due to the heartbeat mechanism and re-schedule to ensure the continued execution of tasks, without the need to restore TEE state. Furthermore, to guarantee the consistency and persistency of TEE-based execution, we upload each aggregation results binding with the current task and round number as the state and guide the control of the TEE’s execution, rather than relying on time. Once chaos occurs, errors can be clearly identified through this state. 


\item \textbf{TEE coordination:} In our FL scenarios, due to our strategy to utilize multiple TEEs to execute one task, the TEE coordination becomes a challenge. To address it, our committee conducts a well-organized task schedule to clearly split the task into pieces and allocate them into each TEE. The local model owners and their corresponding TEE have their own unique identify and establish connection under the guidance of the committee. Therefore, according to these preparations, TEEs are coordinated in an orderly manner. 

\end{itemize}

\section{Evaluation and Discussion}
\subsection{Evaluation Metrics}
Given that clients exclusively serve as users of our framework by providing data, we intentionally exclude considerations pertaining to the specifics of their local training processes and potential variations in training speeds due to individual disparities. As we preserve the original computational procedure for training and aggregation, we anticipate that the accuracy and convergence speed of FL models will remain unaffected by Voltran. Our focus lies solely on assessing the performance implications introduced by Voltran on FL. The impact brought by Voltran stems from the distributed aggregation paradigm, which leads to an increase in the number of communications and encryption/decryption operations. These overheads are present in every phase of Voltran execution. The specific numbers of these metrics are subject to variability and are contingent upon the particular FL tasks at hand. In addition, Voltran incorporates more security features. Secure aggregation algorithms can be implemented in SGX to fortify against potential attacks. Compared to prevalent ciphertext aggregation algorithms such as homomorphic encryption, Voltran’s utilization of plaintext aggregation may significantly alleviate time costs. Therefore, we characterize the performance of the framework by measuring the following metrics: 

\textbf{Model Performance.} We measure three metrics to assess the performance of the model and Voltran-related process:

\begin{itemize}
\item \textbf{Accuracy}: We assess and compare the classification accuracy of trained models in Voltran with the vanilla FL baseline to evaluate the  impact introduced by Voltran. 


\item \textbf{Time overhead}: Voltran incurs an additional time overhead due to its paradigm shift. We assess the time overhead in three primary phases within Voltran: \textit{SendModeltoSGX}, \textit{Aggregate}, and \textit{SendResulttoChain}. We also compare the aggregation time overhead with that of vanilla federated learning (FL) as well as comparable state-of-the-art privacy-preserving aggregation schemes to underscore the alterations we bring about in the aggregation process.




\item \textbf{Communication overhead} Communication overhead indicates the volume of communication generated in the execution of Voltran and vanilla FL. 



\end{itemize}

\textbf{Privacy Metrics.} We use security strength to measure confidentiality performance. We compare the efficiency of Voltran to that of existing privacy-preserving schemes by maintaining the same level of security strength.





\textbf{Attack Defense.} The high scalability of Voltran allows clients to expand their aggregation functionalities, which can effectively employ off-the-shelf schemes for the targeted defence to resist malicious behaviours from clients, such as \textit{Backdoor attack}. These attacks possess measurable indicators to assess their impact. By integrating the corresponding defence strategies and evaluating the performance of Voltran using these indicators, we illustrate the framework’s robust scalability and ability to provide comprehensive security support.






\begin{itemize}
\item \textbf{Backdoor Attack.} The backdoor attack considered by CRFL \cite{xie2021crfl} involves the use of the model replacement approach \cite{bagdasaryan2020backdoor} where the attackers train local models using poisoned datasets and then scale the malicious updates before sending them to the server. In this approach, each attacker only performs the attack once, and they coordinate their model replacement attacks at the same round, denoted as $t_{adv}$. Prior to $t_{adv}$, the adversarial clients train their local models using original benign datasets. However, when they reach the $t_{adv}$ round, each local iteration is trained on the backdoored local dataset. This distributed yet coordinated backdoor attack has been shown to be effective in previous work  \cite{xie2019dba}. For more detailed information, please refer to CRFL \cite{xie2021crfl}.

\item \textbf{Byzantine Attack.} Here come the definition and conduction of the Byzantine attack. The definition of Byzantine attacks is presented in Definition 4 below from \cite{LingchenZhao2021SEARSA}. 

\textbf{Definition 4} (\textit{Byzantine Attack}): The vectors of the Byzantine models from the adversaries are defined as 
$$
\left[\mathbf{V}_i\right]_j=\left\{\begin{array}{l}
{\left[\mathbf{V}_i\right]_j \sim[\mathbf{G}]_j, \text { correct } j^{\text {th }} \text { dimension, }} \\
\text { \textit{arbitrary}, abnormal dimensions }
\end{array}\right.
$$, where $\left[\mathbf{V}_i\right]_j$ denotes the $j^{\text {th }}$ dimension of vector $\mathbf{V}_i$. 

Byzantine adversaries may negotiate in advance to substitute the same dimension of the vectors with similar abnormal values to counteract Byzantine defense methods and amplify the impact on the aggregation result.

\end{itemize}

\begin{figure*}[!h]
\centering
\begin{subfigure}[b]{0.244\textwidth}
\includegraphics[width=\textwidth]{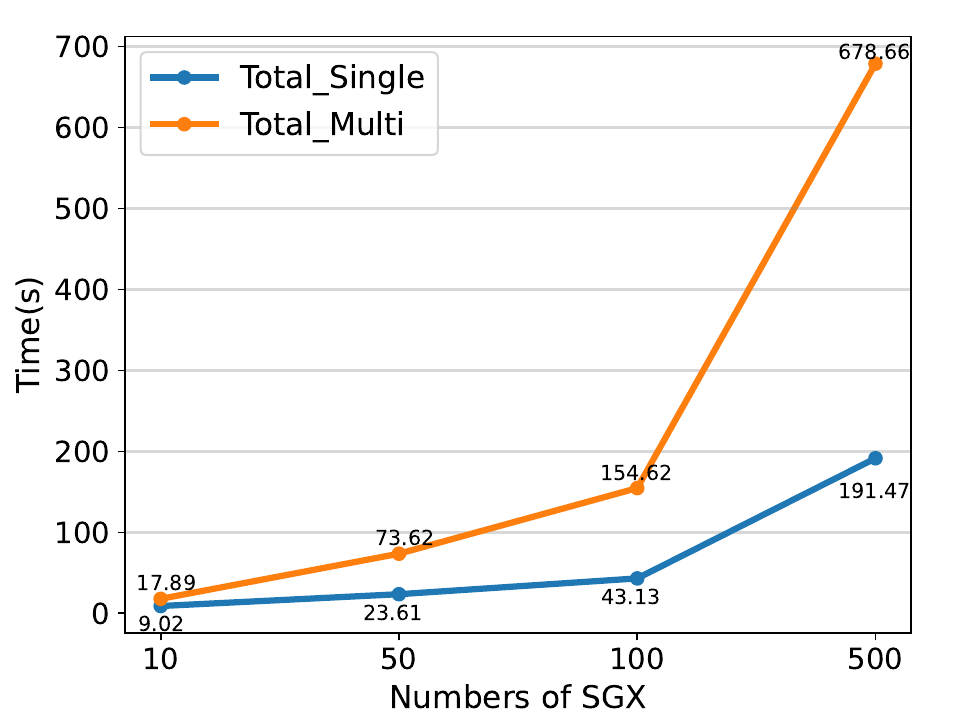}
\caption{Remote Attestation}
\label{RA}
\end{subfigure}
\hfill
\begin{subfigure}[b]{0.244\textwidth}
\includegraphics[width=\textwidth]{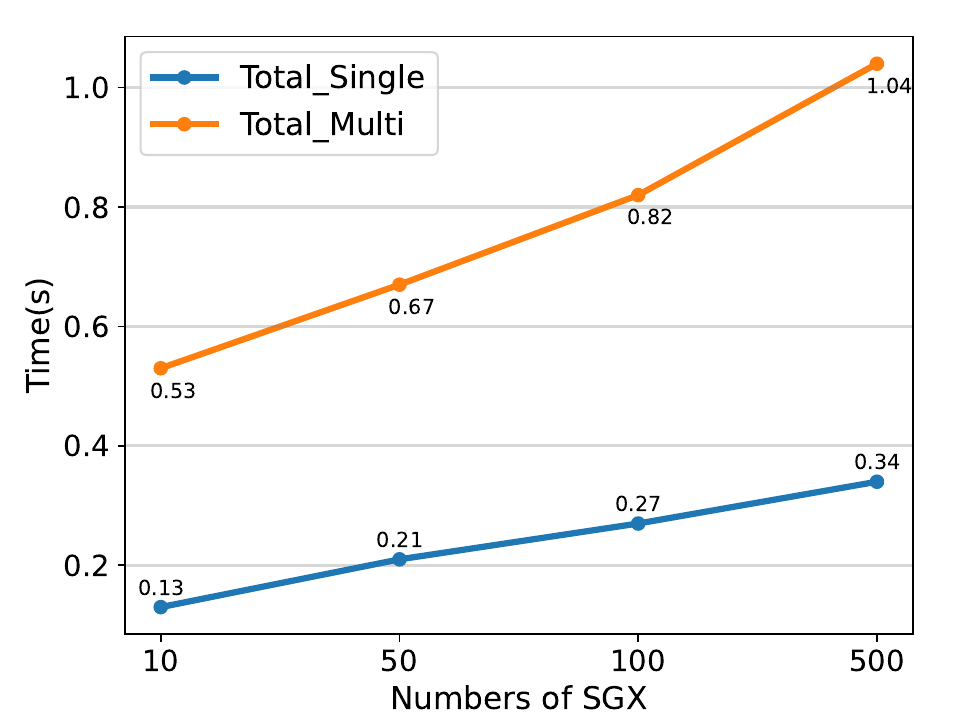}
\caption{Task Scheduling}
\label{Task Scheduling}
\end{subfigure}
\begin{subfigure}[b]{0.244\textwidth}
\includegraphics[width=\textwidth]{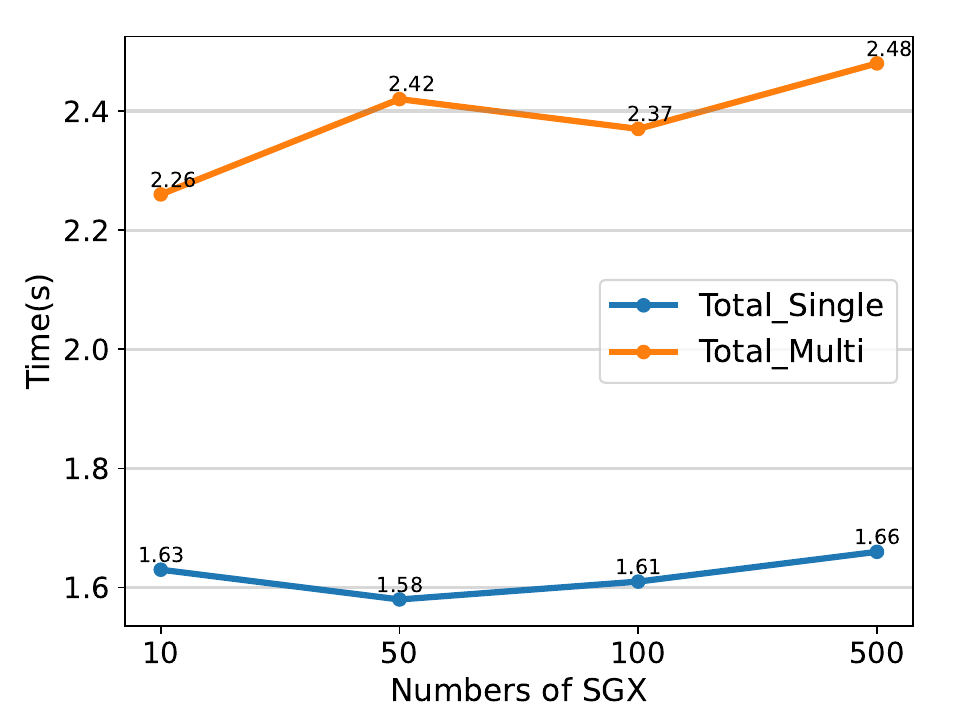}
\caption{Key Distribution}
\label{Key Distribution}
\end{subfigure}
\begin{subfigure}[b]{0.244\textwidth}
\includegraphics[width=\textwidth]{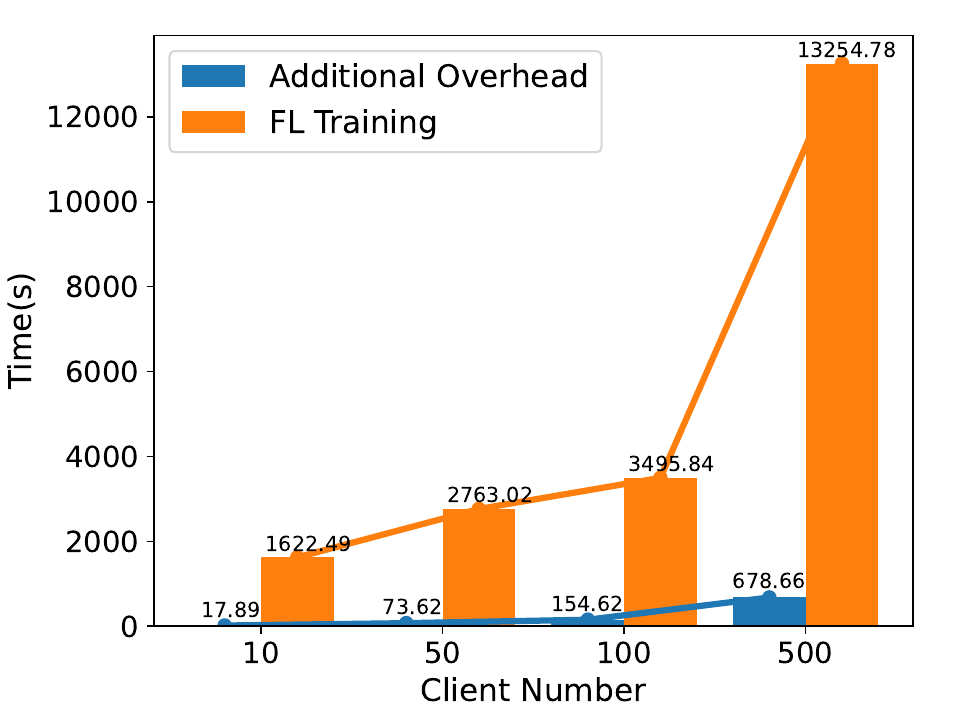}
\caption{Comparison}
\label{Comparison}
\end{subfigure}
\hfill
\caption{Time overhead of additional operations between single-SGX and multi-SGX and comparison with training time.}
\label{Addition overhead of multi-SGX}
\end{figure*}

\subsection{Comparisons with state-of-the-art FL frameworks}

To evaluate our practicality, we compare Voltran with two current state-of-the-art FL frameworks MNN \cite{280900}, an efficient and lightweight machine learning framework optimized for mobile devices, and \cite{bonawitz2017practical}, a pratical secure aggregation framework based on secret sharing proposed by Google. We conduct FL tasks on these two frameworks to evaluate our performance. 

First, since MNN is a framework for clients, we compare Voltran with it to evaluate the performance of our client side. Because MNN is a tensor compute engine along with the data processing and model execution libraries, we make our client side follow this framework with additional operations for security. We perform a recommendation task on the model of Linear Regression with the dataset Avazu and a natural language processing (NLP) task on the model of Stack LSTM with the dataset Sent140. Experimental results are shown in Table \ref{MNN}. It can be seen that our framework introduces almost no additional overhead on the client side in terms of runtime, communication volume, and energy consumption. The slight additional gap arises from the encryption and decryption operations brought by our approach to safeguard the privacy of the models.

\begin{table}[t]
\centering
\caption{Performance comparison between Voltran and MNN on two tasks.}
\scalebox{0.815}{
\begin{tabular}{cccccccc}
\toprule
\multirow{2}{*}{Task}       & \multirow{2}{*}{Device Class} & \multicolumn{2}{c}{Runtime(min)} & \multicolumn{2}{c}{Traffic(Mb)} & \multicolumn{2}{c}{Energy(mAh)} \\\cmidrule(r){3-4}  \cmidrule(r){5-6} \cmidrule(r){7-8}
&                               & MNN            & Voltran         & MNN           & Voltran         & MNN                 & Voltran               \\ \midrule
\multirow{2}{*}{LR}         & high-end                      & 6.72           &     6.81            & 0.75          &    0.75             & 0.13                &  0.13                     \\
& mid-end                       & 11.49          &    11.60             & 0.75          &    0.75             & 0.28                &    0.29                   \\\midrule
\multirow{2}{*}{Stack LSTM} & high-end                      & 2.61           &    2.72             & 3.34          &  3.34               & 3.05                &         3.06              \\
& mid-end                       & 4.13           &  4.24               & 3.34          &    3.34             & 7.78                & 7.80 \\
\bottomrule                    
\end{tabular}}
\label{MNN}
\end{table}

%
%
%
%
%

Second, we perform the same FL task compared with Google's classical secure framework \cite{bonawitz2017practical}, where the data vector size is fixed to 100K entries with 62 bits per entry, to evaluate the end-to-end running time. In this task, we utilize our multi-SGX strategy with two SGX uints. Results in Table \ref{GoogleAgg} demonstrate that our framework is more efficient than \cite{bonawitz2017practical} at both the client and server end (we take our Computational Layer combined with Consensus Layer seen as the decentralized FL server).

%
\begin{table}[!t]
\caption{Runtime comparison between Voltran and Scheme \cite{bonawitz2017practical} on the client and server end.}
\centering
\scalebox{1.18}{
\begin{tabular}{ccccc}
\toprule
\multirow{2}{*}{Clients} & \multicolumn{2}{l}{Client runtime(ms)} & \multicolumn{2}{l}{Server runtime(ms)} \\ \cmidrule(r){2-3}  \cmidrule(r){4-5}  
& Scheme {[}2{]}          & Voltran          & Scheme {[}2{]}        & Voltran        \\ \midrule
500 & 13159 & 4072 & 14670 & 3268    \\ 
1000 & 23497 & 4264 & 27855 & 4821           \\ \bottomrule
\end{tabular}}
\label{GoogleAgg}
\end{table}

\subsection{Additional overhead brought by multi-SGX mode}

We present the specific overhead of additional operations by multi-SGX. We showcase the cost of each operation individually in Fig. \ref{Addition overhead of multi-SGX}. Results on three operations indicate the multi-SGX mode requires more additional overhead than the single-SGX mode, especially the Remote Attestation operation. This is because multi-SGX mode asks each client to send its model to multiple SGXs, thus requiring more times of Remote Attestation. However, since these preparations only need to be executed once, subsequent multi-round large-scale FL training takes more time (because only large-scale tasks need multi-SGX mode to enhance efficiency), whereas pre-processing time can be seen less significant in comparison. We take the ResNet18 task as an instance in Fig. \ref{Addition overhead of multi-SGX}. We take 4 SGX units to perform the task. The total FL training time for the client number of 10, 50, 100, 500 during 50 rounds is 1622.49, 2763.02, 3495.84 and 13254.78 seconds, which are all significantly greater than their respective pre-processing times.

\subsection{Node Dropout Recovery}

\begin{table*} [!ht]
\caption{Time overhead of node dropout recovery on two FL tasks with different dropout ratios on various node scales. Time is measured in seconds. The Connect phase includes the remote attestation and key distribution processes.}  
\label{dropout}  
\resizebox{\textwidth}{15mm}{  
\begin{tabular}{cccccccccccccc}  
\toprule  
\multirow{2}{*}{Task} & \multirow{2}{*}{Phase} & \multicolumn{3}{c}{10 Nodes} & \multicolumn{3}{c}{20 Nodes} & \multicolumn{3}{c}{40 Nodes} & \multicolumn{3}{c}{80 Nodes}\\  
&  & 10\% & 20\% & 30\% & 10\% & 20\% & 30\% &10\% & 20\% & 30\% &10\% & 20\% & 30\%   \\  
\midrule

\multirow{2}{*}{CNN} &  Re-schedule &  0 & 0 & 0.31& 0 & 0 & 0.32 &  0 &  0.29 & 0.35 &  0.34 & 0.42 & 0.41 \\  
& Connect & 0 & 0 & 44.68 & 0 & 0 & 42.39 & 0 & 44.12& 45.19  & 45.55 & 67.42 & 66.98 \\  			\midrule  

\multirow{2}{*}{ResNet18} & Re-schedule & 0.89 & 0.93 & 1.24 & 0.86 & 1.09 & 1.37 & 1.12 & 1.26 & 1.44 & 1.27 & 1.52 & 1.72  \\  
& Connect & 45.07 & 43.69 & 68.64 & 42.62 & 70.87 & 97.59 & 73.20 & 96.24 & 102.31 & 93.51 & 108.55 & 148.89   \\

\bottomrule  

\end{tabular}  
}  
\end{table*}  

Due to the need to schedule FL tasks and allocate them to SGX nodes, our approach considers introducing a committee mechanism to achieve this purpose. However, the implementation and optimization of the committee mechanism are not the main contribution of our work. The specific implementation of the committee can be referred to schemes [1] or [2]. We provide simulated experiments for SGX node recovery referring to the committee implemented referred to [1]. Specifically, we measure the total recovery time of our mechanism against SGX node dropout in different network sizes with various dropout situations. We take the ResNet18 and ResNet50 tasks and set the number of clients as 100. We set the total number of SGX nodes in the system as 10, 20, 40, 80 and vary the random node dropout ratio from 10\% to 30\% to measure the mechanism execution time overhead in Table \ref{dropout}. The results indicate that recovery is only necessary when SGX utilization within the system is high, and the operations performed during the recovery process are the same as those in the initialization phase, with similar overheads.

\begin{table}[t]
\caption{Time overhead when the number of client n = 10000 on CNN.}
\centering
\scalebox{1.011}{
\begin{tabular}{cccc}
\toprule

Num of SGX & Preparation(s) & Communication(s) & Computation(s)\\

\midrule

1 & 1254.3 & 65.22 &  17.43  \\ 
4 & 4824.7 & 162.98 &  2.68  \\ 
\bottomrule
\end{tabular}}
\label{Google}
\end{table}

\subsection{Overhead with larger number of clients}

We conduct the CNN task on the dataset MNIST with 10000 clients. Table \ref{Google} presents the performance data. We take one and four SGX nodes to execute this task. The preparation time includes the remote attestation phase between clients and SGX nodes, the key distribution phase and other pre-execution operations. Results demonstrate that our system can support the large-scale tasks when the number of client n = 10000.

\subsection{Confidentiality}

\begin{table}[t]
\caption{Performance comparisons of confidentiality attacks on Voltran and vanilla FL.}
\centering
\scalebox{0.81}{
\begin{tabular}{ccccc}
\toprule \multirow{2}{*}{Learning Method}& \multirow{2}{*}{Model} & \multicolumn{3}{c}{ Performance under Three Attacks} \\
\cmidrule(r){3-5} & & \textit{MSE} under DRA & \textit{AUC} under PIA  & \textit{Prec}. under MIA  \\	\midrule
\multirow{2}{*}{Vanilla FL} & AlexNet & $0.017 $ & $0.930 $ & $0.874 $ \\
& VGG9 & $0.008$ & $0.862$ & $0.765$ \\
\midrule \multirow{2}{*}{Voltran} & AlexNet &  1.28 & 0.511 & 0.509 \\  & VGG9 & 1.27 & 0.513 & 0.510 \\
\bottomrule
\end{tabular}	}
\label{privacy}
\end{table}

We take the following three attacks: Data Reconstruction Attack (DRA) \cite{NEURIPS2019_60a6c400}, Property Inference Attack (PIA) \cite{ nasr2019comprehensive} and Membership Inference Attack (MIA) \cite{melis2019exploiting} and conduct experiments on AlexNet and VGG9 models on CIFAR10 in an IID setting.

We compare Voltran’s resistance to these attacks with vanilla end-to-end FL. Table \ref{privacy} shows the indicators of each attack measured in these schemes: Mean-Square-Error (MSE) under DRA, Area-Under-Curve (AUC) under the PIA, and Precision under MIA. MSE measures the difference between constructed images and target images and its range is $\left[0, \infty\right.$). The lower MSE is, the more privacy loss. AUC refers to the area under receiver operating curve. Prec. refers to Precision. The range of both AUC and Prec. is $[0.5,1]$. The value 0.5 is for random guesses. The higher AUC or Prec. is, the more privacy loss.

Results demonstrate that our framework can successfully defend against these attacks. Voltran can fully provide protection against DRA and PIA because our encryption on them. The DRA can only reconstruct a fully noised image for any target image (i.e., an MSE of $\sim1.3$ for the specific dataset), while the PIA always reports a random guess on confidentiality properties. Regarding the MIA on full-trained models, as Voltran keeps the global model in ciphertext, which significantly drops the MIA’s advantage. Thus, Voltran fully addresses confidentiality issues raised by these three attacks.

\subsection{Network Bandwidth}

\begin{figure}[t]
\centering
\includegraphics[width=0.6\linewidth]{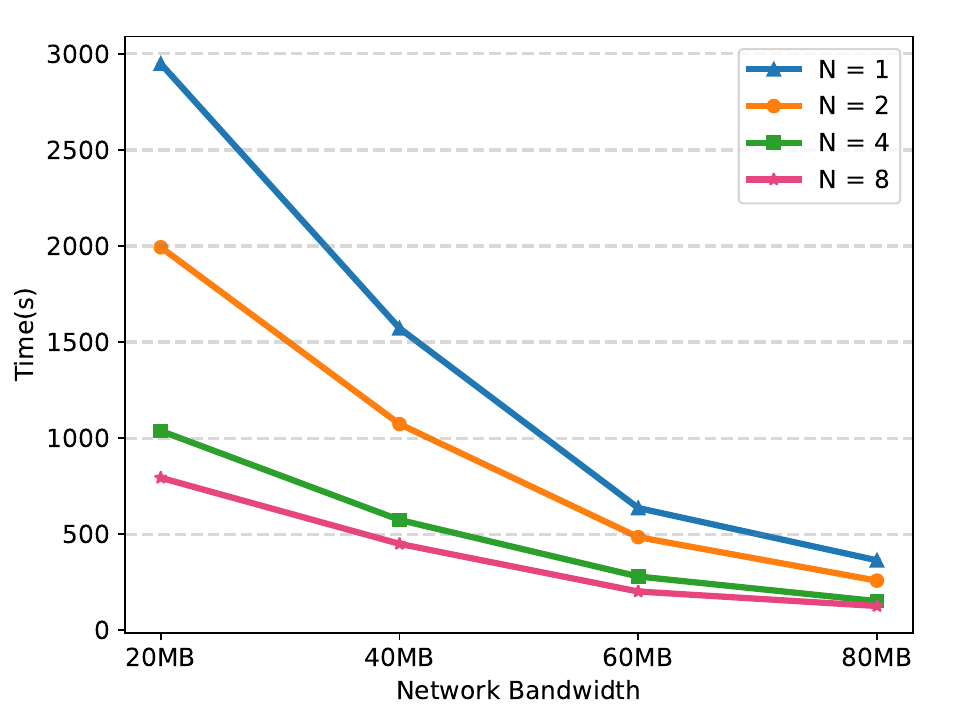}
\caption{Time overhead on different network bandwidth of one round on the ResNet18 task.}
\label{bandwidth}
\end{figure}

We also consider the influence caused by different network conditions. Fig. \ref{bandwidth} gives the performance of Voltran on different network bandwidths on the ResNet18 task. Results demonstrate that when the bandwidth becomes larger, our efficiency can be better.

\vfill

\end{document}